\documentclass[10pt]{elsarticle}

\usepackage{latexsym}
\usepackage{bm}
\usepackage{upgreek}
\usepackage{mathrsfs}
\usepackage{times}
\usepackage{amsthm}
\usepackage{amssymb}
\usepackage{epsfig}
\usepackage{graphicx}
\usepackage{amsmath}
\usepackage{color}
\usepackage{mathrsfs}         
\usepackage{epsfig}
\usepackage{graphicx}
\usepackage{color}
\newcommand{\rev}[1]{{#1}}
\newcommand{\rv}[1]{{#1}}

\newcommand{\rvv}[1]{{#1}}
\newcommand{\add}[1]{{#1}}

\definecolor{purple}{rgb}{0.8,0,0.6}
\newcommand{\new}[1]{{#1}}
\newcommand{\txt}[1]{{#1}}

\theoremstyle{plain}

\theoremstyle{definition}

\theoremstyle{remark}

\begin{document}

\title{Influence of interactions on Integer Quantum Hall Effect}

\author[ARIEL]{C.X.~Zhang}

\author[ARIEL]{M.A.~Zubkov
\footnote{
e-mail: mikhailzu@ariel.ac.il} }

\address[ARIEL]{Physics Department, Ariel University, Ariel 40700, Israel}

\begin{abstract}
Conductivity of Integer Quantum Hall Effect (IQHE) may be expressed as the topological invariant composed of the two - point Green function. Such a topological invariant is known both for the case of homogeneous systems with intrinsic Anomalous Quantum Hall Effect (AQHE) and for the case of IQHE in the inhomogeneous systems. In the latter case we may speak, for example, of the AQHE in the presence of elastic deformations and of the IQHE in presence of magnetic field. The topological invariant for the general case of inhomogeneous systems is expressed through the Wigner transformed Green functions and contains Moyal product. When it is reduced to the expression for the IQHE in the homogeneous systems the Moyal product is reduced to the ordinary one while the Wigner transformed Green function (defined in phase space) is reduced to the Green function in momentum space. Originally the mentioned above topological representation has been derived for the non - interacting systems. We demonstrate that in a wide range of different cases in the presence of interactions the Hall conductivity is given by the same expression, in which the noninteracting two - point Green function is substituted by the complete two - point Green function with the interactions taken into account. Several types of interactions are considered including the contact four - fermion interactions, Yukawa and Coulomb interactions. \new{We present the complete  proof of this statement up to the two loops, and argue that the similar result remains to all orders of perturbation theory.} It is based on the incorporation of Wigner - Weyl calculus to the perturbation theory. We, therefore, formulate Feynmann rules of diagram technique in terms of the Wigner transformed propagators.
\end{abstract}




\maketitle
\tableofcontents

\section{Introduction}
\label{SectIntro}

Generally, the introduction of an external field breaks translational symmetry,
which may lead to new physics, especially when the field is strong.
Take magnetic field as an example, quantum mechanics tells us that
a uniform magnetic field $\cal B$ changes the spectrum of a two-space-dimensional (2d)
charged particle from paraboloid into discrete Landau levels, with equal
spacing $\hbar\omega_c=q{\cal B} \hbar /mc$
($q$ is the charge and $m$ is the mass {of the particle}).
This is the case for a single free particle, while for the more complicated systems
 the interplay of magnetic field and interaction between the particles
is more interesting. For example, in the hydrogen atom (with Coulomb
interaction between proton and electron) magnetic field lifts its
energy degeneracy, which leads to the so-called "Zeeman effect".
If magnetic field is so strong that $\hbar\omega_c \gg e^2/a_0$ ($a_0$ is Bohr radius),
the atom will be significantly elongated along the direction of field $B$ and squeezed in the other directions,
which also drastically changes its energy spectrum \cite{Lai2015, Popov2014}.

In condensed matter physics magnetic field causes interesting phenomena,
such as quantum Hall effect (QHE), deHaas - van Alphen effect, and colossal
magneto - resistance effects. Among them, the most popular one is the QHE \cite{vonKlitzing1980}:
the Hall resistance $R_H$ as a function of ${\cal B}$ possesses plateaus in the
presence of electron-electron interactions and impurities.
Mechanism of IQHE (integer QHE:  the  $R_H$-plateaus are integer
multiples of $e^2/\hbar$) is more or less known \cite{Prange+Girvin1990};
while  mechanism of FQHE (fractional QHE) is still
under intensive investigations\cite{Murthy+Shankar2003}. In addition to magnetic field,
elastic deformations give another source of external field,
which also changes the electric transport behavior of materials
\cite{Guinea_2010,Arias_2015,Castro-Villarreal_2017,Amorim_2016}.


After the discovery of the QHE \cite{vonKlitzing1980} theorists took extensive efforts in order to understand
why there are plateaus in the $R_H -{\cal B}$ graph, i.e.
the "quantization" of Hall conductivity $\sigma_H$.
IQHE can be explained without taking into account electron-electron interactions.
The simple models of the IQHE include the 
free spin-polarized electrons that move within the plane in the presence of magnetic field orthogonal to this plane \cite{R1,R2,R3,R4}. The electronic states are described by  wave function that has the form of  Vandermonde determinant \cite{R5}. Say, Eq. (14)-(16) of \cite{R5} express such a wave function for the filling factor one.

Under magnetic field, the electrons in a disordered system follow the equi-potentials of
the disorder potential. Each Landau level broadens into a band,
and extented states are located  near the central energy of  the band
(otherwise they are  {localized states}).
When chemical potential (tuned by ${\cal B}$) crosses this  { energy region},
the Hall conductivity changes by an interger. When  chemical potential
moves within the {localized states},
the Hall conductivity remains on the same plateau \cite{Prange+Girvin1990}.
 In addition to this explanation, the universal integer values of the Hall plateaus prompt that $\sigma_H$ at the plateaus has the topological reason, i.e. it may be related to some topological invariant, which is robust to the smooth modification of the system. The seminal paper \cite{TKNN} shows that $\sigma_H$ may be expressed through the integral of Berry curvature over the occupied electronic states (this is the so - called TKNN invariant \cite{Fradkin,Tong:2016kpv,Miransky:2015ava,Hatsugai,Hall3DTI}).
Therefore, in the absence of inter - electron interactions, the $\sigma_H$ can be expressed by a
 topological invariant, and
{  its value is not changed when the system is modified smoothly
(e.g. a certain change of $B$, of chemical potential, etc)
\cite{Niu+Thouless1985,Shiba+Kanada1971,Zheng+Ando2002}.}
However, the question remains: what if one takes into account the inter-electron interactions? Can we express  $\sigma_H$ through the topological invariant when interactions are taken into account,
 and prove that $\sigma_H$ is robust to these interactions?
It is wildely believed that $\sigma_H$ can be expressed by the
same topological invariant written in terms of the interacting Green functions
\cite{Matsuyama1987,Volovik1988,Volovik2003,
Gurarie2011,Essin+Gurarie2011,Zubkov2016}.
Theoreticians made several attempts to prove such a statement
\cite{Imai+Ishikawa1990,Ishikawa+Aoyama2003}.
However, to the best of our knowledge no rigorous proof
has been given until recently, except for the case of
anomalous quantum Hall effect (AQHE) in $2+1$ D QED \cite{parity_anomaly}.
This is a very special case, when QHE exists in the absence of magnetic field in the $2+1$ D system with relativistic invariance and an exchange by $2+1$D photons.
{ The common lore typically extends the results of \cite{parity_anomaly} to all  $2d$ systems with integer QHE including those with magnetic field and disorder. }
However, as it has been mentioned above, no proof has been presented for the general case. Moreover, an expression for the Hall conductivity through Green functions in the presence of inhomogeneous magnetic field (and, more general, - the inhomogeneity of general type) has been given for the first time only in \cite{Zubkov+Wu_2019}.
In the present paper we review recent results on the general proof of the mentioned above statement in a wide range of systems using Wigner - Weyl formalism and ordinary perturbation theory.

Green function technique is a powerful tool in condensed matter physics.
The most commonly used Green function is the two-point one: $G(x_1,x_2)$.
\txt{In the homogeneous systems the two point Green function depends on the distance $x_1 - x_2$. Therefore, its Fourier transform depends on one momentum. At the same time in presence of external fields depending on coordinates $G(x_1,x_2)$ can not be
expressed as a function of $x_1-x_2$. Then after the Fourier transformation it
depends on two momenta: $\tilde{G}(p_1,p_2)$.
Instead of $\tilde{G}(p_1,p_2)$ the
Wigner-Weyl transform \cite{Wigner_1932,Groenewold_1946,Moyal_1949}
 $G_W(R,p)$ may be used. We demonstrate that the corresponding diagram technique
for scattering amplitudes contains the same amount of integrations over momenta
as conventional theory with the translational symmetry. However, in the corresponding expressions the complicated Moyal products appear in place of the ordinary multiplications.
In spite of the complication caused by the Moyal products it is more useful to express certain physical quantities throw $G_W(R,p)$ rather than $\tilde{G}(p_1,p_2)$. It is especially important if the topological quantites are considered, for example the Hall coonductivity or the conductivity of the chiral separation effect. Originally expression for Hall conductivity in terms of $G_W(R,p)$ was proposed within Keldysh technique of non - equilibrium theory \cite{Shitade,Sugimoto,Sugimoto2006,Sugimoto2007,Sugimoto2008}.} The corresponding expression is the topological invariant in phase space in equilibrium at zero temperature, i.e. its value is not changed under the smooth deformation of the system \cite{Zubkov+Wu_2019} (the similar analysis in the framework of Keldysh formalism has been given in \cite{Mokrousov,Mokrousov2}).
The similar (but simpler) constructions were used earlier to consider the intrinsic AQHE and chiral magnetic effect \cite{Zhang_2019_JETPL,Zubkov2016}. It has been shown that the corresponding currents are proportional to the topological invariants in momentum space. This method allows to reproduce the conventional expressions  for Hall conductivity \cite{TKNN}, and to prove the absence of equilibrium chiral magnetic effect.

In the present article, we report our results on the influence of interactions on QHE considered using technique of Wigner transformation. This approach has been developed in our earlier papers
\cite{Zubkov+Wu_2019,Zhang+Zubkov2019,Zhang_2019_JETPL,ZZ2019_FeynRule}.
We will show that
one can express $\sigma_H$ for the interacting systems through the
same topological invariant as for the non - interacting ones. This invariant is written
in terms of Wigner-transformed Green functions,
 but the Green functions here are not the free Green functions, but the renormalized ones (dressed by interactions).
We will consider explicitly the tight-binding model of electrons in 2d with Yukawa interactions (generalizations to the case of Coulomb interactions and to the case of the other interactions provided by an exchange by bosonic excitations
are staightforward).
Perturbation expansion will be applied to the calculation of $\sigma_H$ through the $G_W$'s.
{ Because of the presence of the Moyal products between the $G_W$'s
(instead of the ordinary multiplication), a new set of Feynman rules for $G_W$'s is necessary.}
This set of rules will be described as well.

In the present paper we concentrate on the  field theoretical approach to the description of the IQHE.
The paper is organized as follows.
In Sect. II we describe briefly how Wigner transformation may be applied to
ordinary non-relativistic quantum mechanics.
In Sect. III we discuss the particular tight - binding model of 2d topological insulator
in the presence of uniform external electric field. The case of Yukawa interactions between electrons is considered. We show that $\sigma_H$ is still given by the same expression of the topological invariant as in the non - interacting model,
in which the two-point Green function is substituted by the one with interactions.
In Sect. IV, we describe Feynman rules of the diagram technique with the Wigner-transformed Green functions
in external fields.
In Sect. V, we study systems in the presence of an inhomogeneity, when Hall conductivity has to be expressed through the Wigner transformed Green function. As well as in the homogeneous case it has been proven that the interactions affect the final expression for the Hall conductivity through the substitution of bare two - point Green function by the dressed one.
In Section VI we end with Conclusions.

{\it Notice that in order to simplify expressions we will widely use in the paper notation $\int_q = \int d^3 q/(2\pi)^3 $ .}

\section{Wigner-transformed Green functions}
\label{Schrodinger}

In this section we consider the Wigner transformed Green function both for the one - particle quantum mechanical system, and for the system of many identical particles in thermal equilibrium. We will take the simple well - known examples of the corresponding systems and demonstrate how the Wigner - Weyl calculus works in these cases. In particular, we calculate Hall conductivity in the simple field system using Wigner - Weyl formalism.

In quantum mechanics the wave function of a non-relativistic particle
satisfies Schrodinger equation $(i\partial_t-H)\psi=0$.
We define operator $\hat{Q}=i\partial_t-\hat{H}$, with Hamiltonian
$H=(i\partial_x)^2/2m + V(x)$.
Green function $G(t_1,{\bf x}_1|t_2,{\bf x}_2)$ is defined as
\begin{eqnarray}\label{Green_G}
(i\partial_t-\hat{H})G(t_1,x_1|t_2,x_2) =
\delta(x_1-x_2)\delta(t_1-t_2),
\end{eqnarray}
with boundary/initial condition
\begin{eqnarray}\label{Green_G_a}
 G(t_1x_1|t_2,x_2) =0, \quad \rm{when} \quad t_1<t_2.
\end{eqnarray}
This Green function determines evolution in time of the one - particle wave function $\Psi(t,x)$ according to the general theory of differential equations with partial derivatives:
{
$$
\Psi(t,x) =  \int d x' G(t,x|t',x') \Psi(t',x')
$$  }
Its Wigner transformation is defined as
\begin {eqnarray}\label{WignerTrans}
G_W (R,p)=\int dr G(R+r/2, R-r/2) e^{-ipr},
\end{eqnarray}
where $G(X_1,X_2) =G(t_1,x_1|t_2,x_2) $ and $X=(t,x)$. The corresponding
Wigner-transformed Green function $G_W$ satisfies the so - called Groenewold equation $Q_W(X,P)\star G_W(X,P)=1$, with the Moyal product given by
$\star=exp[i(\overleftarrow{\partial}_X \overrightarrow\partial_P
-\overleftarrow\partial_P\overrightarrow\partial_X)/2]$. Here $X$ includes
time-space coordinates $(t, x)$, and $P$ includes energy-momentum
variables $(E, p)$. The derivative with the arrow pointing to the left acts to the left from the star
while the derivative with the arrow pointing to the right act as usual.
In the following, for simplicity we will not use capital letters ($X$ and $P$)
to denote vector of energy and momentum as well as vector of time and coordinates.

\subsection{Ordinary quantum mechanics}

In this subsection, we present several examples of single-particle quantum - mechanical systems, where the Wigner transformed Green function may be calculated explicitly. These examples are themselves trivial. We use them to demonstrate how  the Wigner - Weyl calculus works in principle.

Our first example is the system with the only energy level $E_0$. In this case there are no space coordinates and phase space contains the pair of time $t$ and energy $E_0$. The corresponding Groenewold equation reads
\begin{eqnarray}\label{example1_a}
(E - E_0) \star G_W(t,E) = 1.
\end{eqnarray}
It is equavalent to the following equation
\begin{eqnarray}\label{example1_b}
(E - E_0) G_W(t,E) -\frac{i}{2}\partial_t G_W(t,E)= 1.
\end{eqnarray}
which is an ordinary differential equation (ODE).
Multiplying both sides by $e^{2i(E-E_0)t}$, we transform the equation into
$ -\frac{i}{2}\partial_t (e^{2i(E-E_0)t}G_W(t,E))=e^{2i(E-E_0)t}$.
Then it's easy to find general solution
\begin{eqnarray}\label{example1_c}
G_W(t,E)=\frac{1}{E-E_0 \pm i\epsilon}+Ce^{-2i(E-E_0)t},
\end{eqnarray}
where $\epsilon$ is a small positive number \cite{Vladimirov_book}, and
$C$ is an arbitary number (integration constant for the ODE).
Furthermore, in order to fulfil  the boundary conditions of  Eq.(\ref{Green_G_a}),
we choose $+i\epsilon$ and $C=0$, which leads to the final result
\begin{eqnarray}\label{example1_d}
G_W(t,E)=\frac{1}{E-E_0 + i\epsilon}.
\end{eqnarray}
We will omit such a $C$-term below in the consideration of the other examples if the Hamiltonian $H$
doesn't depend on time $t$.

In the next example we consider the case of a free particle with
$\hat{Q}=i\partial_t-(-i\partial_x)^2/2m$.
With $Q_W=E-p^2/2m$ the corresponding equation for $G_W(E;x,p) $ is
\begin{eqnarray}\label{example2_a}
(E - \frac{p^2}{2m}) G_W +\frac{ip}{2m} \partial_x G_W+\frac{1}{8m}\partial^2_x G_W= 1.
\end{eqnarray}
Applying transformation $F=  e^{2ipx} G_W$, we bring this equation to the following form:
\begin{eqnarray}\label{example2_b}
E F + \frac{1}{8m}\partial^2_x F=    e^{2ipx}.
\end{eqnarray}
Its general solution is given by the sum of a particular solution and the general solution of the
corresponding homogeneous equation. We take the mentioned particular solution in the form $F^*=A   e^{2ipx}$ with constant $A$. Inserting $F^*$ into Eq.(\ref{example2_b}) we get $(E-p^2)A=1$, from which
one obtains $A$. The general solution for the  homogeneous equation
\begin{eqnarray}\label{example2_c}
E F + \frac{1}{8m}\partial^2_x F=0.
\end{eqnarray}
is $F_h=C e^{2i\sqrt{2mE}x}$.
Summing up $F^*$ and $F_h$, we obtain general solution for the inhomogeneous equation
Eq.(\ref{example2_b}). The corresponding result for $G_W$ is
\begin{eqnarray}\label{example2_d}
G_W=\frac{1}{E-p^2/2m \pm i\epsilon}+ C e^{-2ipx+2i\sqrt{2mE}x}.
\end{eqnarray}
Then from the condition of Eq.(\ref{Green_G_a}), we obtain $C=0$,
and the final result is
\begin{eqnarray}\label{example2_e}
G_W=\frac{1}{E-p^2/2m + i\epsilon}.
\end{eqnarray}
This form corresponds to an ordinary retarded Green function.

Our third example is harmonic oscillator
(which paves the way to particle in uniform magnetic field).
The corresponding operator $\hat{Q}$ has the form:
$\hat{Q}=i\partial_t-(-i\partial_x)^2/(2m) - m \omega^2 x^2/2$.
Its Wigner transformation is $Q_W = E-p^2/(2m) - m \omega^2 x^2/2$. The Groenewold equation receives the form
\begin{eqnarray}\label{example4_a}
(E - \frac{i}{2}\partial_t) G_W
-\frac{1}{2m}(-i\partial_x/2+p)^2 G_W -
 \frac{m\omega^2}{2}(+i\partial_p/2+x)^2 G_W
=1
\end{eqnarray}
Instead of direct solution of this equation we consider first the expression for Green function in
coordinate space (for example, Eq. (8.1) of \cite{FeynmannBook}):
\begin{eqnarray}\label{example4a_a}
 G(t,x_1| 0, x_2)
 =-i\sqrt{ \frac{m\omega}{2\pi i  {\rm sin}(\omega t)} }
                     exp \Big(  \frac{i m\omega}{2  {\rm sin}(\omega t)}
 [(x_1^2+x_2^2){\rm cos}(\omega t)- 2x_1 x_2]    \Big).
\end{eqnarray}
After substitution $R=(x_1+x_2)/2$ and $r=x_1-x_2$, and Fourier transform $r \rightarrow p$, we obtain
\begin{eqnarray}\label{example4a_b}
G(t;R,p)=\frac{-i}{{\rm cos}(\omega t/2)} e^{-iW {\rm tan}(\omega t/2) },
\end{eqnarray}
with $W=\frac{1}{m\omega} p^2 +{m\omega}R^2-i\epsilon$. Small imaginary part of this variable $-i\epsilon$ provides that the resulting Green function is the retarded one.
Therefore, the final result for  $G_W(E;R,p)$ is
\begin{eqnarray}\label{example4a_c}
 G_W(E;R,p)=  
 \frac{-2i}{\omega}
   \int^{\infty}_{-\infty}  e^{i(2Eu/\omega - W \, {\rm tan}\,u) }
   \frac{du}{{\rm cos}\,u}.
\end{eqnarray}

The above expression may be used for the calculation of $G_W$ of particle moving in the presence of magnetic field. We consider motion in plane $O-xy$, in the presence of uniform magnetic field ${\cal B}$ directed along axis $z$. We choose the gauge $A_x=0$ and $A_y=-{\cal B}x$, and the
Hamitonian receives the form $H=p_x^2/2m+(p_y+{\cal B}x)^2/2m$.
The corresponding Groenewold equation is
\begin{eqnarray}\label{example5_a}
 EG_W
+\frac{1}{8m}(\partial_x+2p_x i)^2 G_W + 
 \frac{{\cal B}^2}{8m}(\partial_{p_x}-2xi-2ip_y/{\cal B})^2 G_W
=1,
\end{eqnarray}
We may omit dependence of $G_W$ on $t$,
and treat $E$ and $p_y$ as parameters,  i.e.
{$G_W=G_W(E; x, p_x, p_y)$.}
This PDE has the form of the above equation for harmonic
oscillator.
Therefore, $G_W(E; x, p_x, p_y)$ can be expressed similarly to
Eq.(\ref{example4a_c}) as
\begin{eqnarray}\label{example5_b}
 G_W(E; x, p_x, p_y)=  
  \frac{-2 i m}{\cal B}
                    \int e^{i(2E u m/{\cal B} - W {\rm tan}\,u) }
                    \frac{du}{{\rm cos}\,u}.
\end{eqnarray}
with $W= [p_x^2 +(p_y+{\cal B}x)^2]/{\cal B}-i\epsilon$.

\subsection{Systems of identical particles}

In this subsection we discuss multi - fermion system existing
in 2d plane in the presence of magnetic field orthogonal to this plane
(vector potential of external magnetic field is denoted by ${\bf A}$).
We do not obtain here any new results, and use the consideration of
this system to demonstrate how Wigner - Weyl calculus works in the description of conventional QHE.

On the language of path integrals the given  system is defined by Grassmann - valued  field $\psi$
with action
\begin{eqnarray}\label{lagrangian}
S_{0} = \int d\tau d^2{\bf x} \  \bar{\psi}
            \Big( - \partial_{\tau} -   
         \frac{(-i{\bf \nabla} - {\bf A})^2}{2m}+\mu \Big) \psi
\end{eqnarray}
where $\mu$ is chemical potential, $\psi, \bar{\psi}$ are Grassmann - valued functions of $({\bf x},\tau)$. We are considering the system in imaginary time $\tau = it$. Green function in spatial coordinates  $G_0 (x_1,x_2)$
is defined as
\begin{eqnarray}\label{Green_spatial}
G_{0}(x_1,x_2) &=&\add{ -}\frac{1}{Z_0}
\int {D\bar{\psi}D{\psi}}\,
       \psi(x_1)\bar{\psi}(x_2)
      e^{\add{S_0}}, \nonumber \\
\end{eqnarray}
which satisfies equation
\begin {eqnarray}\label{Green_NR_x_equ}
&&\Big( - \partial_{\tau_1} -\frac{(-i{\bf \nabla}_1 - {\bf A}(x_1))^2}{2m}+\mu \Big)
G_{0}(x_1,x_2)\nonumber \\
&&=\delta^3 (x_1-x_2).
\end{eqnarray}
Applying Wigner transformation, we obtain
\begin {eqnarray}\label{Green_NR_Wigner_equ}
\Big( i\omega -\frac{({\bf p} - {\bf A}(x))^2}{2m}+\mu \Big)
\star G_{W}(x,p)=1. \nonumber \\
\end{eqnarray}
In the absence of magnetic field, i.e. when ${\bf A}=0$,
the solution for $G_{W}(x,p)$ is
\begin {eqnarray}\label{Green_NR_Wigner_1}
G_{W}(x,p)=\frac{1}{i\omega -\frac{{\bf p} ^2}{2m}+\mu}.
\end{eqnarray}

Comparing the above equation with the single-particle case of Eq.(\ref{example2_d}),
we come to the following conclusion. Replacement of $E$ by $i\omega+\mu$ in the
single particle retarded Green function brings it to the form of Matsubara Green function of multi-particle system with
chemical potential $\mu$.
Applying this "golden rule" to the system in the presence of magnetic field, we will obtain
\begin{eqnarray}\label{example5_b}
 G_W(i\omega;x,p_x; p_y)= 
 \frac{-2i m}{\cal B}
 \int e^{-2m(\omega-i\mu) u/{\cal B} -i W\, {\rm tan}\, u }
                    \frac{du}{{\rm cos}\,u}.
\end{eqnarray}
with $W=[p_x^2 +(p_y-Bx)^2]/{\cal B}$.
As above we chose the gauge with $A_x=0, A_y=-{\cal B}x$.

In the next step, we consider the Hall current directed along the  $y$ - axis
corresponding to the external electric field $\cal E$ directed along the $x$ - axis.
Current density $J_y(x)$ is given by $\delta {\rm ln}\,Z/ \delta A(x)$,
which can be expressed through the Green function
\begin{eqnarray}\label{current_density_NR}
J_y(x)= -\int G_W \frac{\partial Q_W}{\partial p_y} \frac{d^3 p}{(2\pi)^3}
\end{eqnarray}
with $Q_W=i\omega -({\bf p} - {\bf A}(x))^2/2m + {\cal E}x+ \mu $.
 Here $G_W$ of Eq.(\ref{current_density_NR})
still can be expressed in the form of Eq.(\ref{example5_b}),
but $\mu$ and $W$ should be changed into
$\mu'=\mu - p_y {\cal E}/{\cal B} -m({\cal E}/{\cal B})^2/2$,
and $W'= [p_x^2 +(p_y+{\cal B}x+m {\cal E}/{\cal B})^2]/{\cal B}$
respectively.
We have
\begin{eqnarray}\label{current_density_NR__}
 J_y(x)=   
  \frac{-2i m}{\cal B}\int
                 e^{-2m( \omega-i\mu' ) u/{\cal B} -i W' {\rm tan}\, u }
                   \Big({\cal E}/{\cal B}-       
 p_y^\prime/m \Big) \frac{d^3 p}{(2\pi)^3} \frac{du}{{\rm cos}\,u}
\end{eqnarray}
where $p'_y=p_y+{\cal B}x+m{\cal E}/{\cal B}$.
After some algebra,
we found that the term linear in $\cal E$ in the current density had the form
\begin{eqnarray}\label{current_density_NR_2}
J_y    =  \frac{- m {\cal E}}{(2\pi)^2 {\cal B}} \int e^{-2m(\omega-i\mu)u/{\cal B}}
       \Big(\frac{1}{{\rm sin}\,u}-\frac{u\,{\rm cos}\,u}{{\rm sin}^2 u} \Big) du d\omega \nonumber
\end{eqnarray}
Note that $1/{\rm sin}u-u {\rm cos}u/{\rm sin}^2 u= (u/{\rm sin}u)'$, and then using
integration by parts, we obtain
\begin{eqnarray}\label{current_density_NR_3}
J_y= -\frac{ m {\cal E}}{(2\pi)^2 {\cal B}}
     \int \frac{2m(\omega-i\mu)}{\cal B} e^{-2m(\omega-i\mu)u/{\cal B}}
       \frac{udu }{{\rm sin}\,u}  d\omega  \nonumber
\end{eqnarray}
Let us expand here  $1/{\rm sin}\,u$ as
\rvv{
\begin{eqnarray}
1/{\rm sin}\,u = 2i\sum_{n=0}^{+\infty} e^{-i(u-i\epsilon)(2n+1)}  \nonumber
\end{eqnarray}   }
and integrate each term separately in $u$.
Thus we obtain
\begin{eqnarray}\label{current_density_NR_4}
J_y= - i \frac{\cal E}{(2\pi)^2}
    \int \sum^{\infty}_{n=0} \frac{1}{\omega-i(\mu-E_n)}  d\omega
\end{eqnarray}
where $E_n = (n+1/2) {\cal B}/m$.
Each term in the sum is formally divergent here at large $\omega$. However, we should recall that the theory to be defined properly is to be regularized. Actually, the standard regularization using the discretization of an interval of $\tau$ between $0$ and $1/T$ leads to the modification of propagator  $ \frac{1}{\omega-i(\mu-E_n)}  $ at large values of $\omega$. The general property of this modification is that in the integral over $\omega$ we may close the contour in the upper half of the complex plane. This gives us immediately the standard result for the Hall current:
\begin{eqnarray}\label{current_density_NR_4}
J_y=  \frac{\cal E}{2\pi}
     \sum^{\infty}_{n=0} \theta(\mu-E_n) =\sigma_H \cal E
\end{eqnarray}
One can see, that Hall conductivity $\sigma_H$ is equal to the number of occupied Landau levels $N$ (those with $E_n < \mu$) divided by $2\pi$. Recall, that we use here the relativistic system of units. In the conventional units we obtain the standard result $\sigma_H = N e^2/h$.

\section{IQHE in the non - interacting $2+1$ D tight - binding models}
\label{SectHall}
Starting from this section we deal with the lattice models. These models either represent the tight - binding models of solid state physics or the lattice regularized quantum field theory.

\subsection{Lattice models in coordinate space and in  momentum space}

Let us discuss first the 2+1 D lattice model of the non-interacting fermions with Euclidean action
\begin{eqnarray}\label{lagrangian}
S_{0} &=& \int d\tau \sum_{{\bf x,x'}}\bar{\psi}_{\bf x'}\Big(i(i \partial_{\tau} - A_3(-i\tau,{\bf x}))\delta_{\bf x,x'}- 
 i{\cal D}_{\bf x,x'}\Big)\psi_{\bf x}
\end{eqnarray}
\txt{Here $\psi$ and $\bar{\psi}$ are the independent Grassmann - valued fields.  $\tau = i t$ is the so - called imaginary time. $i{\cal D}_{\bf x,x'}$ is the lattice Hamiltonian. We suppose, that fermions are in the presence of external electromagnetic potential $A$. One of the possible Hamiltonians for the case of rectangular lattice is given by}
\begin{eqnarray}\label{lattice_difference}
{\cal D}_{\bf x,x'}= -\frac{i}{2} \sum_{i=1,2} [(1+\sigma^i)\delta_{x+e_i,x'}e^{iA_{x+e_i,x}} 
                           + (1-\sigma^i)\delta_{x-e_i,x'}e^{iA_{x-e_i,x}}] \sigma_3     
                   + i(m+2)\delta_{\bf x,x'}\sigma_3
\end{eqnarray}
where $A_{u,v}= \int^{u}_{v} A \cdot ds$.
\txt{In the following we will consider the case of the Hamiltonian of a more general form. The third Euclidean component $A_3$ of electromagnetic potential is expressed through the conventional real electric potential $\phi(t,{\bf x})$
as $A_3  =  -i \phi(-i \tau, {\bf x} )$.
We assume that spatial coordinates $\bf x$ are discrete, while "imaginary" time $\tau$ is continuous.
The Euclidean three-momentum is denoted as $p = (\omega, {\bf p})$. Here $\omega=-iE$ is the iMatsubara frequency taken at zero temperature.
In our three - dimensional space $x = (\tau, {\bf x})$, while $A = (-i \phi, {\bf A})$.
Since we are dealing with equilibrium theory all external fields as well as the Hamiltonian itself do not depend explicitly on time.}

\txt{Fourier transformed fermion field is defined as
\begin{eqnarray}\label{Fourier}
\psi(p)=\sum_{\bf x} e^{-ipx}\int \psi_{\bf x,\tau} e^{i\omega \tau} d\tau.
\end{eqnarray}   }
\txt{We can represent action in momentum space as}
\begin{eqnarray}\label{action@p}
S_{0} = \int \frac{d^3p}{\new{(2 \pi)^3}} \bar{\psi}(p)\Big(  i\omega -i\phi(i\partial_{\bf p}) 
                 - H(p-A(i\partial_{\bf p}))  \Big)   \psi(p),
\end{eqnarray}
\txt{where $\int d^3 p =\int^{\infty}_{-\infty} d\omega \int^{\pi}_{-\pi} d^2\bf{p}$ (we take rectangular lattice for simplicity). FOr example, for the lattice Hamiltonian given by Eq. (\ref{lattice_difference}) we obtain Hamiltonian in momentum space:
\begin{eqnarray}\label{Ham0}
H({\bf p}) = {\rm sin}\,p_1\, \sigma^2 - {\rm sin}\, p_2 \, \sigma^1 - 
  (m + \sum_{i=1,2}(1-{\rm cos}\,p_i)) \, \sigma^3,
\end{eqnarray}
We also define
$Q(p,x)=Q(\omega,{\bf p},\tau,{\bf x})=i(\omega - \phi(i\tau,{\bf x})) - H({\bf p} - {\bf A}(i\tau,{\bf x}))$.
As a result the momentum space  Green function $\tilde{G}_0 (p_1,p_2)$
may be expressed as
\begin{eqnarray}\label{Green_p}
\tilde{G}_{0}(p_1,p_2)
= \frac{1}{Z_0}
\int \frac{D\bar{\psi}D{\psi}}{(2\pi)^3}\,
       \bar{\psi}(p_2)\psi(p_1) 
      e^{\int \frac{d^3 p}{(2\pi)^3} \bar{\psi}(p) Q(p,i\partial_p)\psi(p)}.
\end{eqnarray}
It is a solution of
\begin {eqnarray}\label{Green_k_equ}
Q(p_1,i\partial_{p_1})\tilde{G}_0(p_1,p_2)=\delta^3 (p_1-p_2).
\end{eqnarray}
The coordinate space Green function $G_0(x_1,x_2)$
is obtained from $\tilde{G}_0$ by Fourier transformation
\begin {eqnarray}\label{Green_pp2xx}
G_0(x_1,x_2)= \int \frac{d^3 p_1}{(2\pi)^{3/2}} \int \frac{d^3 p_2}{(2\pi)^{3/2}}
 e^{ip_1 x_1}    
  \tilde{G}_0(p_1,p_2) e^{-ip_2 x_2}
\end{eqnarray}
Using this expression we may continue definition of the Green function to continuous values of coordinates $x_i$.
Then for discrete  $\bf x'_1$ and $\bf x'_2$ we have %
\begin {eqnarray}\label{Green_x_equ}
 Q(-i\partial_{x_1},x_1) G_0(x_1,x'_2) \Big|_{x_1=(\tau_1,\bf x'_1)}= 
 \delta (\tau_1-\tau_2)\delta_{\bf x'_1,\bf x'_2}.
\end{eqnarray}
The last equation may be proved easily:
\begin {eqnarray}\label{Green_x_equ_proofA}
& & Q(-i\partial_{x_1},x_1) G_0(x_1,x'_2)  \nonumber \\
&=& \int \frac{d^3 p_1 d^3 p_2}{(2\pi)^{3}}
 [Q(-i\partial_{x_1},x_1) e^{ip_1 x_1}]  
 \tilde{G}_0(p_1,p_2) e^{-ip_2 x'_2}      \nonumber\\
&=& \int \frac{d^3 p_1 d^3 p_2}{(2\pi)^{3}}
[Q(p_1,-i\partial_{p_1}) e^{ip_1 x_1}] 
  \tilde{G}_0(p_1,p_2) e^{-ip_2 x'_2}.  \nonumber
\end{eqnarray}
Using integration by parts and Eq.(\ref{Green_k_equ}) we come to
\begin {eqnarray}\label{Green_x_equ_proofB}
&& Q(-i\partial_{x_1},x_1) G_0(x_1,x'_2) \Big|_{x_1=(\tau_1,\bf x'_1)}   \nonumber \\
&=& \int \frac{d^3 p_1 d^3 p_2}{(2\pi)^{3}}
e^{ip_1 x'_1} [Q(p_1,i\partial_{p_1}) \tilde{G}_0(p_1,p_2) ]  e^{-ip_2 x'_2}    \nonumber  \\
&=& \int \frac{d^3 p_1 d^3 p_2}{(2\pi)^{3}}
e^{ip_1 x'_1}   \delta(p_1-p_2) e^{-ip_2 x'_2}    \nonumber  \\
&=& \delta (\tau_1-\tau_2)\delta_{\bf{x'_1},\bf{x'_2}}.
\end{eqnarray}}

\subsection{Wigner transformation}

\txt{Let us assume that field $A$ is slowly varying such that
 its variations on the distances of the order of the lattice spacing are negligible. Then Wigner transformation leads to the so - called Groenewold equation
\begin {eqnarray}\label{Green0_w_equ}
Q_W(x,p)\star G_{0,W}(x,p) =1
\end{eqnarray}
Here $Q_W(x,p)$ and $G_{0,W}$ are defined as
\begin{eqnarray}
{G}_{0,W}(x,p)& = & \int d^3q e^{ix q} \tilde{G}_0({p+q/2}, {p-q/2})\nonumber \\ \label{GWx}\\
{Q}_{W}(x,p)& = & \int d^3q e^{ix q} \tilde{Q}({p+q/2}, {p-q/2}), \nonumber
\end{eqnarray}
while
\begin{equation}
\tilde{Q}(p_1,p_2) \equiv  \int d^3k \delta^{(3)}(p_1-k) {Q}(k,i\partial_{k}) \delta^{(3)}(p_2 - k)\nonumber
\end{equation}
are the matrix elements of $\hat{Q}$. Moyal product $\star$ is defined as $\star=e^{i\tilde{\Delta}/2}$, where
 $\tilde{\Delta} =\overleftarrow{\partial}_x\overrightarrow{\partial}_p-\overleftarrow{\partial}_p\overrightarrow{\partial}_x$.
The gradient expansion gives $G_{0,W}=G^{(0)}_{0,W}+G^{(1)}_{0,W}+...$
where $G^{(n)}_{0,W} \sim O(\partial^n_x)$.
$G^{(0)}_{0,W}$ may be expressed as $G^{(0)}_{0,W}(x,p)=g(p-{\cal A}(x))$ \cite{FZ2019} with
$
g(p) =[i\omega  - H({\bf p} )]^{-1},
$
Here ($\mu=1,2$)
\begin{equation}\label{calAs}
{\cal A}_\mu({\bf x}) =  \int \big[\frac{\sin(k_{\mu}/2)}{k_{\mu}/2}\tilde A_\mu({\bf k})e^{i{\bf kx}}+c.c. \big]dk
\end{equation}
and
\begin{eqnarray}
{\cal A}_1({\bf x}) & = & \int_{{\bf x} - {\bf e}_1/2}^{{\bf x} + {\bf e}_1/2} A_1(y_1,x_2)dy_1 \nonumber\\
{\cal A}_2({\bf x}) & = & \int_{{\bf x} - {\bf e}_2/2}^{{\bf x} + {\bf e}_2/2} A_2(x_1,y_2)dy_2\label{calAs}
\end{eqnarray}
By ${\bf e}_{\mu}$  we denote the unit vector along the $\mu$ - th direction. At the same time we may represent
${A}_\mu({\bf x}) =  \int \big[\tilde A_\mu({\bf k})e^{i{\bf kx}}+c.c. \big]dk $, and for case of slowly varying fields we simply replace $\cal A$ by $A$.}

\subsection{Electric current and Hall conductivity}\label{Sec3C}

Here vector potential is divided into two contributions:
$A_\mu=A^{(m)}_{\mu} +A^{(e)}_{\mu}$, where $A^{(m)}_{\mu}$ corresponds to magnetic field,
and $A^{(e)}_{\mu}$ corresponds to electric field. Furthermore, we assume
the electric field in constant and small, and then $A^{(e)}_{\mu}$ is
denoted by $\delta A_{\mu}$.
Using the expansion of $Q_W$ in powers of $\delta A$, i.e.
${\cal Q}(p-A(R)-\delta A)=
{\cal Q}(p-A(R))-\partial^{\mu}{\cal Q}\delta A_{\mu}$,
we expand the function $G_{0,W}$ correspondingly:
 $G_{0,W}(R,p)=G^{(0)}_{0,W}+G^{(1)}_{0,W}+...$,
with $G^{(n)}_{0,W}\sim (\delta A)^n$.
From Eq.(\ref{Green0_w_equ}), the $G^{(n)}_{0,W}$'s
can be obtained iteratively.

At the leading order (zeroth order), $G^{(0)}_{W}$ satisfies
\begin {eqnarray}\label{WignerEqu_(0)}
G^{(0)}_{0,W}(R,p) \star {\cal Q}(p- A(R))= 1.
\end{eqnarray}
At the next leading order (the first order),
$G^{(1)}_{0,W}$ satisfies
\begin {eqnarray}\label{WignerEqu_(1)}
 G^{(1)}_{0,W}(R,p) \star {\cal Q}(p- A(R))-   
 G^{(0)}_{0,W}(R,p)\star (\partial_{\mu} {\cal Q}\delta A^{\mu})= 0.
\end{eqnarray}
One can solve the equation and find that
\begin {eqnarray}\label{WignerEqu_1}
G^{(1)}_{W}(R,p)=G^{(0)}_{W}(R,p)\star (\frac{\partial {\cal Q}}{\partial p_{\mu}} \delta A^{\mu}) \star G^{(0)}_{W}(R,p).\nonumber\\
\end{eqnarray}
$G^{(1)}_{0,W}$ can be expressed as \cite{Suleymanov2019,Zubkov+Wu_2019}
\begin{eqnarray}\label{G0(1)}
G_{0,W}^{(1)}(x,p) =  -\frac{\partial G_{0,W}^{(0)}}{\partial p_\mu} \delta A_{\mu} \new{-} 
                        \frac{i}{2}  G_{0,W}^{(0)} \star \frac{\partial Q_W}{\partial p_\mu}
                               \star \frac{\partial G_{0,W}^{(0)}}{\partial p_\nu} \delta F_{\mu\nu}
\end{eqnarray}
\add{Notice that in the similar expression of \cite{Z2016_1} there was mistake in sign (Eq. (24)). Correspondingly, several expressions of \cite{Z2016_1} (the journal version) are to be corrected \footnote{For more details see Corrigendum of  \cite{Zubkov+Wu_2019}. Namely, in the original journal version of \cite{Z2016_1}  there was mistake in the second row of Eq. (24): the correct sign is plus. As a result in \cite{Z2016_1} (the journal version) the sign is to be changed to the opposite in Eqs. (30), (35), (49), (74), (80), (86), (90), the first equation in (33).  Besides, there were misprints and the signs are to be changed to the opposite in \cite{Z2016_1} in Eqs. (17), (20), (27), (28), (29), (30), (43), (44); the second and the third rows of Eq. (26). Notice that the expressions given in Appendix A are valid for the systems with sufficiently weak inhomogeneity that may be neglected at the distance of the lattice spacing. In Appendix A the integration regions are not specified. Notice that the integral over $\bf P$ in Eq. (99) and over momenta in the further expressions of Appendix A are over the Brillouin zone (and not over the Brillouin zone extended twice along each axis of the reciprocal lattice). The mentioned mistakes/misprints are  corrected in ArXiV version of \cite{Z2016_1}. }.}
Electric current can be considered as the linear response to the external field,
i.e. $\delta {\rm log}\,Z=J^k(x)\,\delta A_k(x)$, with $Z$ the partition function.
Here, we consider a 2d system in $O-xy$ plane under a magnetic field in the z-direction
and an electric field along the x-direction, and
the electric current density along the $y$-axis is given by
\begin{eqnarray}\label{current_2D}
J_2(x)& = & - \int \frac{d^3 p}{(2\pi)^3}  Tr G_{0,W}(x,p) \frac{\partial Q_W}{\partial p_2}.\nonumber \\
\end{eqnarray}
Corresponding to expansion $G_0=G_0^{(0)}+G_0^{(1)}+...$
in powers of $\delta A$,
$J_k$ is expanded as $J_k=J_k^{(0)}+J_k^{(1)}+...$.
We find the current density, up to the order of $\delta A $ (the first power),
as follows
\begin{eqnarray}\label{current_b}
J_2(x) &=& -\int \frac{d^3 p}{(2\pi)^3}  Tr G^{(1)}_{0,W}(R,p)  \partial_2 {\cal Q} - 
                                     Tr G^{(0)}_{0,W}(R,p)  \partial_2   (\partial_{\mu}{\cal Q}\delta A^{\mu}) \nonumber\\
      &=&  \new{+}\frac{i}{2}\delta F_{13}   \int \frac{d^3 p}{(2\pi)^3}Tr [ \partial_1 G^{(0)}_{0,W}\star  
       \partial_3 Q^{(0)}_{W}\star G^{(0)}_{0,W}]  \cdot \partial_2 Q^{(0)}_{W} \nonumber\\
         & &  \new{+}\frac{i}{2}\delta F_{31}   \int \frac{d^3 p}{(2\pi)^3}Tr [ \partial_3 G^{(0)}_{0,W}\star 
          \partial_1 Q^{(0)}_{W}\star G^{(0)}_{0,W}] \cdot \partial_2 Q^{(0)}_{W}.
\end{eqnarray}
$\delta F_{lm}$'s are related to the constant electric field strength, and most of the components are zero
except $\delta F_{13}=iE_1$ and $\delta F_{31}=-iE_1$.
Here we denote $Q_{W}^{(0)} =  {\cal Q}(p- A(R))$.
Now let us consider the averaged total current
\begin{eqnarray}\label{current_c}
{\cal J}^{(1)}_2 &=& \frac{1}{\cal S} \int {d^2 x} J_2(x) \nonumber\\
      &=& \new{+}\frac{i}{2} \sum_{l,m \in \{1,3\} }\delta F_{lm}   \int  \frac{d^2 x}{\cal S} \frac{d^3 p}{(2\pi)^3}
            [ \partial_l G^{(0)}_{0,W}\star 
       \partial_m Q^{(0)}_{W}\star G^{(0)}_{0,W}]\star  \partial_2 Q^{(0)}_{W} \nonumber\\
    &=& \new{+}\frac{E_1}{2} \int  \frac{d^2 x}{\cal S} \frac{d^3 p}{(2\pi)^3}
                             Tr(W_1\star W_3 - 
     W_3\star W_1 ) \star W_2
\end{eqnarray}
where $\cal S$ is the overall area of the system, and $W_m=G^{(0)}_{0,W}\star \partial_m Q^{(0)} $.
In the second line of the above equation, we can substitute an ordinary product by the $\star$  product,
because all of the factors in the integrand do not depend on electric field
(they depend on $A^{(m)}_{\mu}$ only, not on $A^{(e)}_{\mu}$),
and then the periodic boundary condition in spatial coordinates can be satisfied.

Taking advantage of the equality
$Tr(U_1 U_2 - U_2 U_1)U_3=(1/3) \sum_{ijk}  \epsilon_{ijk} Tr U_i U_j U_k$,
with $U_i$'s arbitrary matrices,
the term linear in the field strength for Hall current
can be expressed as
\begin{eqnarray} \label{j2d}
{\cal J}_k^{(1)}  &= & \epsilon_{ijk} {\cal M}_{} \delta F_{ij}, \\
 {\cal M} &=& - \frac{i}{12}\,\epsilon_{abc}  \int \frac{d^2 x}{\cal S} \frac{d^3p}{(2\pi)^3} \,{\rm Tr}\,   \,
\Big( W_a \star W_b \star W_c \Big)  \nonumber
\end{eqnarray}
in which the Green function ${\cal G}$ satisfies ${\cal G}^{-1} = i \omega - H({\bf p})$,
with $H$ the one - particle Hamiltonian.
Here $\delta F_{ij}$ is the Euclidean field strength
$\delta F_{ij} = \partial_i \delta A_j  - \partial_j \delta A_i$. In Euclidean space (the theory in imaginary time) we do not distinguish between lower and upper indices, and define the components  $\delta A^k$ for $k = 1,2$  as equal to the space components of real external electromagnetic potential $\delta{\bf A}$ in Minkowski space - time. Correspondingly, $ \delta A^3 = -i \delta A^0$, where $\delta A^0$ is the external electric potential.
Therefore, $\delta F_{3k} = -i E_k$ with $k=1,2$,
corresponding to the external electric field
$ {\bf E} = (E_1,E_2)$.
The generalization to the case of the $3+1$ D models is straightforward.
It is worth mentioning, that the derivation of Eq. (\ref{j2d}) requires that
the field $A$ does not vary fast, i.e. its variation on the distance of the order of lattice spacing may be neglected.

We suppose, that the fermions are gapped and the Green function ${\cal G}({ p})$
depends on the three - vector ${p} = (p_1,p_2,p_3)$ of Euclidean momentum
(the third component of vector corresponds to imaginary time).
In order to express the Hall current and the Hall conductance,
let us introduce the electric field strength into Eq.(\ref{j2d}), which
leads to the following expression for the Hall current
\begin{equation}
{\cal J}^k_{Hall} = \frac{1}{2\pi}\,{\cal N}\,\epsilon^{ki}E_i,\label{HALLj}
\end{equation}
where the topological invariant denoted by ${\cal N}$ is to be calculated for the original system with vanishing component $\delta A$ of gauge field:
\begin{eqnarray}
{\cal N} =  \new{+}\rv{\frac{1}{24 \pi^2 {\cal S}}} {\rm Tr}\, \int_{} Q^{(0)}_{W}\star 
  d G^{(0)}_{0,W}\star \wedge d Q^{(0)}_{W}\star \wedge d G^{(0)}_{0,W}\label{N3A}
\end{eqnarray}
Eq. (\ref{N3A}) defines the topological invariant.

\subsection{Homogeneous systems}

In this subsection, we consider the Hall conductivity in homogeneous systems,
as a special case of the previous subsection.
In homogeneous systems, the translational symmetry is satisfied,
if electric potential $\delta A$ is not introduced, i.e. the only source of
 translational symmetry breaking comes from $\delta A$.
In this case, $\cal G$ is the Green function in momentum space,
i.e. the Fourier transformation of the two point Green function in coordinate space.
Thus we are speaking here about the intrinsic AQHE existing in the 2d topological insulators.

Because of the  translational symmetry,
 the above obtained expression for the current density  $J_k(x)$ does not depend on $x$, and we obtain
\begin{eqnarray} \label{j2d_}
J_k^{(1)}  &= & \frac{1}{4\pi}\epsilon_{ijk} {\cal M}_{}\delta  F_{ij},\\
 {\cal M} &=& \new{-}\rv{ \frac{i}{3!\,4\pi^2}\,\epsilon_{ijk} \int_{} \,{\rm Tr}\, d^3p  \, \Big[ {\cal G}^{-1} \partial_{p_i} {\cal G} \partial_{p_j} {\cal G}^{-1} \partial_{p_k} {\cal G}\Big]}  \nonumber
\end{eqnarray}
where Green function ${\cal G}$ satisfies ${\cal G}^{-1} = i \omega - H({\bf p})$,
and $H$ is the one - particle Hamiltonian while  $\delta F_{ij}$ is the Euclidean field strength
$\delta F_{ij} = \partial_i \delta A_j  - \partial_j \delta A_i$.
We suppose, that the fermions are gapped and the Green function ${\cal G}({ p})$
depends on the three - vector ${p} = (p_1,p_2,p_3)$ of Euclidean momentum
(the third component of vector corresponds to imaginary time).
\txt{As above we add the external electric field
${\bf E} = (E_1,E_2)$ as $\delta  F_{3k} = -i E_k$ to be substituted to  Eq.(\ref{j2d_}). The Hall current can be calculated as follows
\begin{equation}
{J}^k_{Hall} = \frac{1}{2\pi}\,{\cal N}\,\epsilon^{ki}E_i,\label{HALLj}
\end{equation}
Here in order to calculate ${\cal N}$ we may take the model with the external electromagnetic field switched off:
\begin{eqnarray}
{\cal N} &=&  \new{+}\rv{\frac{1}{24 \pi^2}} {\rm Tr}\, \int_{} {\cal G}^{-1} d {\cal G} \wedge d {\cal G}^{-1} \wedge d {\cal G}\label{N3A_}
\end{eqnarray}
Eq. (\ref{N3A_}) is the topological invariant composed of the momentum space Green function $\cal G$. To obtain the above expression we require that the system under consideration is homogeneous.}

In many cases the value of ${\cal N}$ may be computed directly.
Take the Hamiltonian in  Eq.(\ref{Ham0}) for example,  the corresponding Green function has the form ${\cal G}^{-1} = i \omega - H({\bf p})$ .
In the absence of external fields, for $m \in (-2,0)$ we have ${\cal N} = \new{-}1$,
while  ${\cal N}  = \new{+}1 $ for $m\in (-4,-2)$,
and ${\cal N}  = 0 $ for $m\in (-\infty,-4)\cup (0,\infty)$. This calculation is given in Appendix \ref{AppendixA}.

\section{Homogeneous systems with interactions}

{ In this section, we consider the interaction effect on the Hall conductivity
in homogeneous system (without magnetic field).
}

\subsection{Exchange by bosonic excitations}
\label{SectScalarA}

\txt{Here, we consider the two - dimensional model with the interactions resulted from the field $\varphi$.
Coulomb interactions also belong to this class. The action defined in imaginary time  is given by
\begin{eqnarray}\label{action_Yukawa3D}
 S_{\eta}=S_0
+ \int d\tau \Big(\sum_{{\bf x,x'}}\varphi_{\bf x'}(\tau)
\Big( \partial^2_{\tau}\delta_{\bf x,x'}+ 
 {\cal W}_{\bf x',x}  \Big) \varphi_{\bf x}(\tau)
-\eta \sum_{\bf x}\bar{\psi}(\tau,{\bf x})\psi(\tau, {\bf x}) \varphi_{\bf x}(\tau)\Big). \nonumber \\
\end{eqnarray}
Here matrix $\cal W$ is specific for the given type of excitations. In particular, for the \add{attractive} Yukawa interactions we have
\begin {eqnarray}\label{bosonic_lattice_diff}
{\cal W}_{\bf x',x}=\sum_{i=1,2}(\delta_{x',x+e_i}+\delta_{x',x-e_i})\new{-(M^2+4)}\delta_{x',x}
\end{eqnarray}
Parameter $M$ plays the role of mass (in lattice units). }

In the case of \add{repulsive} Coulomb interactions, instead of an additional field $\varphi$
we may consider the following modification of the action of $2+1$ D tight-binding model:
\begin{eqnarray}\label{action_Coulomb3D}
 S  = S_0-\alpha \int d\tau \sum_{{\bf x,x'}} 
 \bar{\psi}(\tau,{\bf x})\psi(\tau, {\bf x})V({\bf x-x'})\bar{\psi}(\tau,{\bf x'})\psi(\tau, {\bf x'}),
\end{eqnarray}
where $V$ is Coulomb potential $V({\bf x})=1/|{\bf x}|=1/\sqrt{x_1^2+x_2^2}$.
Now the role of the above parameter $\eta$ is played by $\alpha$, and
in the following, we may interchange $\eta^2$ and $\alpha$. The Green function is given by
\begin {eqnarray}\label{Green_W_express_Coulomb3D}
{\cal G}_{\alpha}(p) =[i\omega  - H({\bf p} )-\alpha \Sigma(p)  ]^{-1}+O(\alpha^2) \nonumber
\end{eqnarray}     
In the leading order, the self-energy function
\rev{\begin{eqnarray}\label{self-energy_b}
\Sigma(p) & = & \new{-} \int_q {\cal G}_{\alpha = 0}(q) \tilde{V}(p-q), 
\end{eqnarray}}
where $\int_q = \int d^3 q/(2\pi)^3 $ and
$\tilde{V}(p)$ is Coulomb potential in momentum space
$\tilde{V}(p) =\sum_{\bf x} e^{i{\bf p\cdot x}}/ \sqrt{x_1^2+x_2^2}$.
Therefore, $\Sigma(p)$ depends only on $p_1$ and $p_2$.

Electric current is given by
\begin {eqnarray}\label{current_3D_a}
J^k_\eta(x) =- \int_p  Tr G_{\eta,W}(x,p) \frac{\partial { Q_W    } }{\partial p_k}.
\end{eqnarray}
$G_{\eta,W}(x,p)$ is the full Green function with the interactions taken into account, and satisfies
\begin {eqnarray}\label{WignerEqu}
G^{(0)}_{0,W}(R,p) \star ({\cal Q}(p- A(R))-\Sigma)= 1.
\end{eqnarray}
$G_{\eta,W}(x,p)$ can be expanded into
$G_{\eta,W}(x,p)=G^{(0)}_{\eta,W}(x,p)+G^{(1)}_{\eta,W}(x,p)+...$, in which
the term $G^{(k)}_{\eta,W}(x,p)$ is proportional to the product of $k$ derivatives
$\frac{\partial}{\partial x}$.
In particular,
\begin {eqnarray}\label{Green_W_express_Yukawa3D}
G^{(0)}_{\eta,W}(x,p) = \Big[ i(\omega - A_3(x)) - H({\bf p} - {A}(x)) 
 -\eta^2 \Sigma (p,x)  \Big]^{-1},
\end{eqnarray}
where the self-energy function $\eta^2 \Sigma=\eta^2 \Sigma_1+\eta^4 \Sigma_2+...$.
At the leading order, $\Sigma_1(x,p)$ is given by
\begin{eqnarray}\label{self-energy}
\Sigma_1(x,p) & = & \add{+}\int_q G_{0,W}(x,q) D(p-q)  \nonumber\\
& =&\Sigma_1^{(0)}+\Sigma_1^{(1)}+...,
\end{eqnarray}
which is also expanded in powers of $\frac{\partial}{\partial x}$, according to the expansion of $G_{0,W}$.
More presisely, the Green function $G^{(0)}_{\eta,W}$ in
Eq.(\ref{Green_W_express_Yukawa3D})  should be written as
\begin {eqnarray}\label{Green_W_express_Yukawa3D_a}
G^{(0)}_{\eta,W}(x,p) = \Big[ i(\omega - A_3(x)) - H({\bf p} - {A}(x))  
 -\eta^2 \Sigma^{(0)} (p,x)  \Big]^{-1}.
\end{eqnarray}

The  bosonic Green function is
\begin {eqnarray}\label{bosonic_Green}
D(p) = \frac{1}{\omega^2 + {\rm sin}^2 p_1 + {\rm sin}^2 p_2 + M^2}.
\end{eqnarray}
\add{(For the case of Coulomb interactions we substitute $D \to - V$.)}
The contribution of Yukawa interactions to the self-energy depends both on momenta and space coordinates. \txt{
Below we will give an expression for  $\mathcal{K}=J^k_{\eta}-J^k_{\eta \eta}$. Here
$$
J_{\eta\eta} =\new{-}\int_p\, {\rm Tr}\, G^{}_{\eta,W}(x,p) \frac{\partial}{\partial p_k}({ Q_W - \Sigma    })
$$
The meaning of $\cal K$ is the difference between electric current and its modification calculated using the renormalized velocity of electrons.
We obtain
\begin {eqnarray}\label{current_change}
\mathcal{K}
&=& \new{-} \int_p \, {\rm Tr}\, G_{\eta,W}(x,p) \frac{\partial}{\partial p_k} \Sigma(x,p)  \nonumber \\
&=&  \new{-}\eta^2\int_p \, {\rm Tr}\, G_{0,W}(x,p) \frac{\partial}{\partial p_k} \Sigma_1(x,p) + O(\eta^4).\nonumber \\
\end{eqnarray}
Next, we expand in powers of derivatives $G_{0,W}= G^{(0)}_{0,W}(x,p)+ G^{(1)}_{0,W}(x,p)+...$
Here $G^{(n)}_{\eta,W} \sim O(\partial_x^n)$. In the same way we represent $\mathcal{K}= \mathcal{K}^{(0)}+ \mathcal{K}^{(1)}+...$
We start discussion from the zeroth order  $ \mathcal{K}^{(0)}$. For brevity below $G^{(0)}_{0,W}(x,p)$ is referred to as $g(p)$:
\begin {eqnarray}\label{current_change_LO}  
\mathcal{K}^{(0)}
&=&\add{-} \eta^2\int_p Tr \Big[ G^{(0)}_{0,W}(x,p)\cdot
       \frac{\partial}{\partial p_k} \int_q G^{(0)}_{0,W}(x,p-q)D(q)\Big]  \nonumber \\
&=& \add{-} \eta^2\int_p Tr \Big[ g(p)
\frac{\partial}{\partial p_k} \int_q g(p-q)D(q) \Big], \nonumber
\end{eqnarray}
Let us introduce integral $I$ through  $\mathcal{K}^{(0)}= \eta^2 I$:
\begin {eqnarray}\label{current_change_LO_2}  
I &=& \add{-} \int_p\int_q
{\rm Tr} \big[ g(p)\frac{\partial}{\partial p_k} g(p-q) \big] D(q)   \nonumber \\
&=& \int_{p,q}
{\rm Tr} \big[ \frac{\partial g(p)}{\partial p_k}  g(p-q)\big] D(q)   \nonumber \\
&=& \int_{p,q}
{\rm Tr} \big[ g(p-q) \frac{\partial g(p)}{\partial p_k}  \big] D(q)   \nonumber \\
&=& \int_{\rvv{s,q}}
{\rm Tr} \big[ g(s) \frac{\partial g(s+q)}{\partial s_k}  \big] D(q)   \nonumber \\
&=& \int_{s,q}
{\rm Tr} \big[ g(s) \frac{\partial g(s-t)}{\partial s_k}  \big] D(-t).   \nonumber
\end{eqnarray}
It is supposed that the bosonic Green function is even function of momentum $D(-t)=D(t)$, and we obtain $I=-I$. As a result we come to the conclusion that  $I=0$ and
$\mathcal{K}^{(0)} =0$.}

\txt{The first order term may be calculated as follows
\begin {eqnarray}\label{current_change_NLO}  
 \mathcal{K}^{(1)}&=&\new{-}
  \eta^2\int_p\,
  \Big({\rm Tr}\,  G^{(1)}_{0,W}(x,p)\frac{\partial }{\partial p_k} \Sigma^{(0)}(x,p)
+  {\rm Tr}\,  G_{0,W}^{(0)}(x,p)\frac{\partial }{\partial p_k} \Sigma^{(1)}(x,p)\Big)   \nonumber \\
&=&\add{-}\eta^2 \int_p \int_q
\, \Big({\rm Tr}\,  G_{0,W}^{(1)}(x,p) G_{0,W}^{(0)}(x,q)  \frac{\partial}{\partial p_k} D(p-q)
+  {\rm Tr}\,  G_{0,W}^{(0)}(x,p)G_{0,W}^{(1)}(x,q)\frac{\partial }{\partial p_k} D(p-q)   \Big)\nonumber \\
&=&\add{-}\eta^2\int_p \int_q
\, \Big({\rm Tr}\,  G_{0,W}^{(1)}(x,p) G_{0,W}^{(0)}(x,q) \frac{\partial}{\partial p_k} D(p-q)
+  {\rm Tr}\,  G_{0,W}^{(1)}(x,q)G_{0,W}^{(0)}(x,p)\frac{\partial }{\partial p_k} D(p-q)  \Big) \nonumber \\
&=&\add{-}\eta^2\int_p \int_q
\, \Big({\rm Tr}\,  G_{0,W}^{(1)}(x,q) G_{0,W}^{(0)}(x,p)  \frac{\partial}{\partial q_k} D(q-p)
+  {\rm Tr}\,  G_{0,W}^{(1)}(x,q)G_{0,W}^{(0)}(x,p)\frac{\partial }{\partial p_k} D(p-q)  \Big) \nonumber \\
&=& 0
\end{eqnarray}
Again, we take into account that bosonic propagator is an even function $D(-t) = D(t)$, and $D'(-t)=-D'(t)$. This allows to obtain $\delta J^{k,(1)} =\mathcal{K}^{(1)}=0$. }

\txt{Thus we prove that to one - loop order the Hall current is given by
\begin{equation}
{j}^k_{Hall} = \frac{1}{2\pi}\,{\cal N}\,\epsilon^{ki} E_i ,\label{HALLji}
\end{equation}
Here ${\cal N}$ may be expressed through the interacting Green function
${\cal G}_\eta = [i\omega  - H({\bf p})-\eta^2 \Sigma_1^{(0)}(p)]^{-1} $
(with $\Sigma_1^{(0)}(p) = -\int_q {\cal G}_{0}(q) D(p-q)$):
\begin{eqnarray}
{\cal N} &=& \new{+}\rv{ \frac{1}{24 \pi^2}} {\rm Tr}\, \int_{} {\cal G}_\eta^{-1} d {\cal G}_\eta \wedge d {\cal G}_\eta^{-1} \wedge d {\cal G}_\eta\label{N3Ai}
\end{eqnarray}}

\subsection{Higher-order corrections}

\txt{Here we extend our discussion to the higher orders of perturbation theory (see also \cite{Zhang_2019_JETPL}).}

\begin{figure}[h]
	\centering  %
	\includegraphics[width=0.4\linewidth]{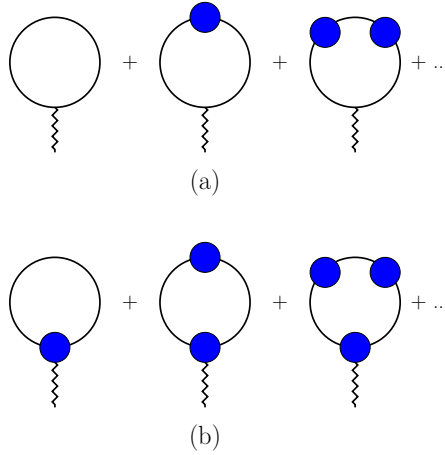}  %
	\caption{Tadpole diagrams. The black solid lines are the propagators of fermions,
while the black zigzag represents an external field. The shaded blue circles correspond to the self-energy functions.
{ (a) Diagrams related to the current density $J^k_{\eta}$.
(b)Diagrams related to the difference $J^k_{\eta} - J^k_{\eta\eta}$.    }   }  %
	\label{fig_tadpole_general}   %
\end{figure}

\txt{In the following we assume that $G_W(x,p)$ and $\Sigma_W(x,p)$ depend on the combination $p-A(x)$. In order to prove this we should remember that the external electric field is homogeneous.  Therefore, we obtain for the current density
\begin {eqnarray}\label{current_star}
J_k(x) &=& \new{-}\int_p\, Tr G_W(x,p) \partial_{p_k} Q \nonumber \\
      &=& \new{-}\int_p \, Tr G_W(x,p)\star \partial_{p_k} Q
\end{eqnarray}
Here the star product is inserted. This may be done because we consider response to constant external electric field (see also Appendix \ref{AppB}).
We come to
$J^k_{\eta} = \new{-}\int_p \,Tr G_{\eta} \star \partial_{p_k} Q$,
while $\mathcal{K}=J^k_{\eta}-J^k_{\eta \eta}=\new{-} \int_p  \, Tr G_{\eta} \star \partial_{p_k} \Sigma$.
Using the approach of  \cite{Zhang+Zubkov2019}
we represent $J^k_{\eta}=\sum_{n=0}^{\infty} \mathcal{J}[n]$ with
\begin {eqnarray}\label{J_component}
\mathcal{J}[n] =\new{-} \int_p \, Tr (G_{0}\star \Sigma \star)^n G_0  \star \partial_{p_k} Q,
\end{eqnarray}
We represent this expansion in Fig.\ref{fig_tadpole_general}(a).
  $\mathcal{K}$ is represented in Fig.\ref{fig_tadpole_general}(b). The latter is given by  $\mathcal{K}=\sum_{n=0}^{\infty} \mathcal{K}[n]$, where
\begin {eqnarray}\label{deltaJ_component}
\mathcal{K}[n] = \new{-}\int_p \,  Tr (\Sigma \star G_0\star)^n G_{0} \star  \partial_{p_k} \Sigma.
\end{eqnarray}
In  \cite{Zhang_2019_JETPL} we obtained relation between $J^k_{\eta} $ and  $\mathcal{K}$ (see also below Sect. 6.2.):
\begin {eqnarray}\label{deltaJ_and_J}
\mathcal{K}[n] = \mathcal{J}[n+1] .
\end{eqnarray}
In the other words, the sum of the radiative corrections to electric current is equal to the total value of $\mathcal{K}$: $$\sum_{n\ge 1} \mathcal{J}[n] = \mathcal{K}$$}

\begin{figure}[h]
	\centering  %
	\includegraphics[width=0.4\linewidth]{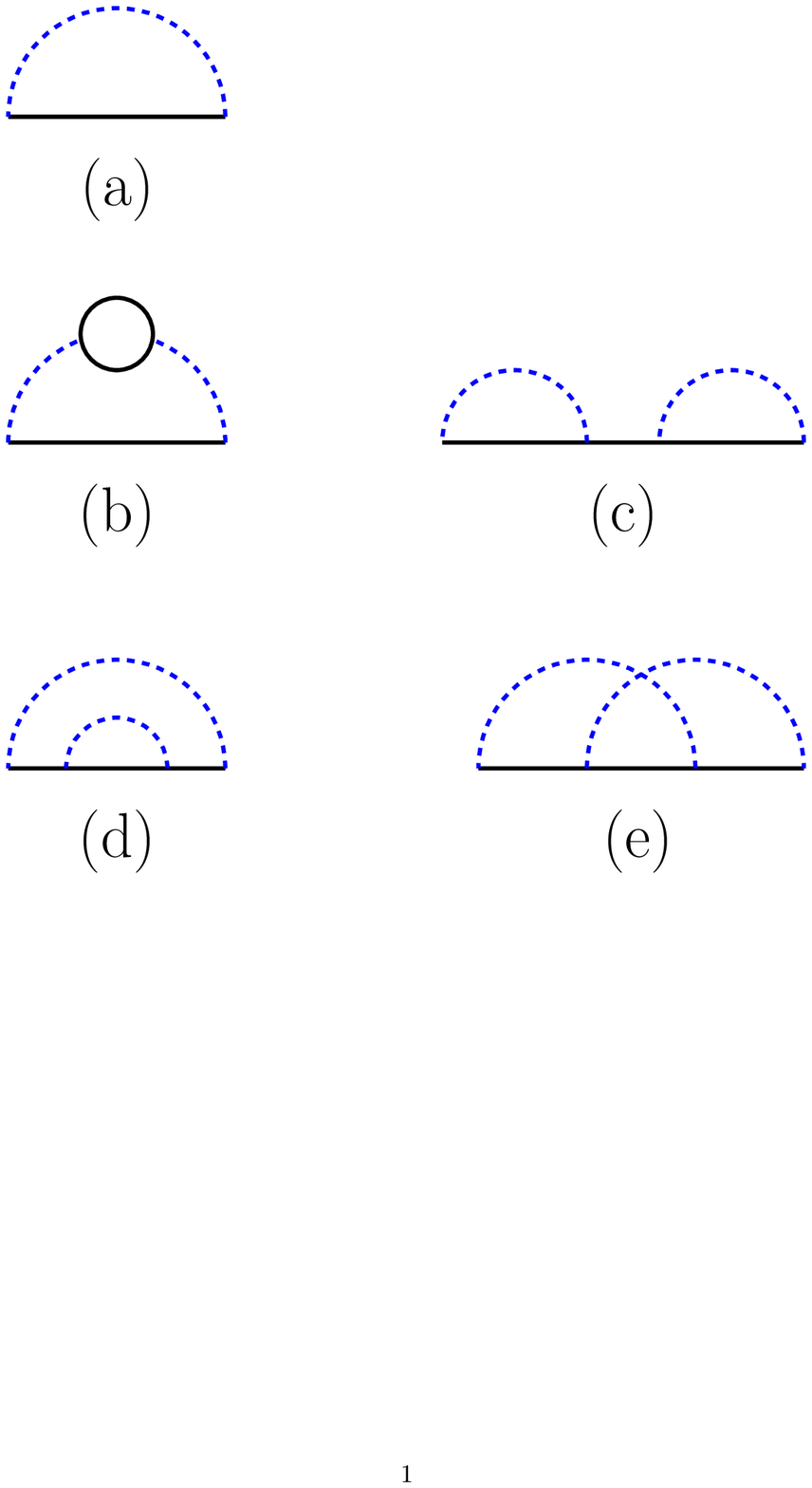}  %
	\caption{Self-energy functions. The dashed blue lines correspond to
the bosons responsible for the Yukawa interaction.}  %
	\label{fig_self-energy}   %
\end{figure}

\begin{figure}[h]
	\centering  %
	\includegraphics[width=0.4\linewidth]{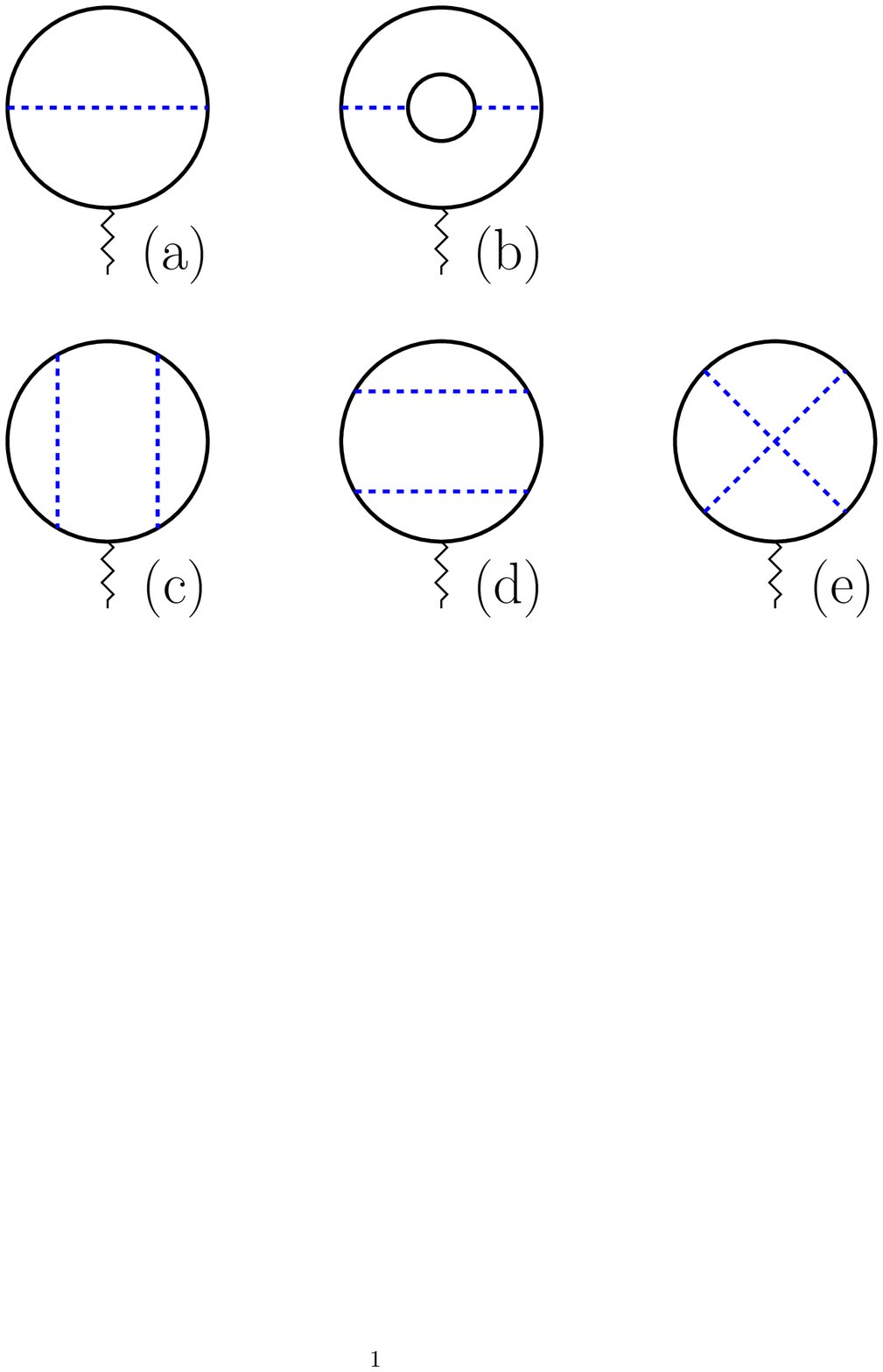}  %
	\caption{Tadpole graphs in the first and the second order.}  %
	\label{fig_tadpole}   %
\end{figure}

\txt{Below we prove that $\mathcal{K}=0$
up to two loops, i.e. to order $\eta^4$. The consideration of the higher order in perturbation theory is similar.
First, we discuss $\mathcal{K}_1$ (this is the contribution to $\mathcal{K}$ proportional to $\eta^2$):
\begin {eqnarray}\label{K1(0)}
\mathcal{K}_1
&=& \new{+}\int_p {\rm Tr} \Big[ \Sigma_{1,W} (x,p) \star \partial_{p_k} G_{0,W}(x,p)\Big] \nonumber \\
&=& \add{+}\int_{p,q}
       {\rm Tr}\Big[  G_{0,W}(x,p-q)D(q) \star   
     \partial_{p_k} G_{0,W}(x,p)\Big] ,
\end{eqnarray}
Self - energy function $\Sigma_{1,W}$ is represented in Fig.\ref{fig_self-energy}(a).
$\mathcal{K}_1$ is given by Fig.\ref{fig_tadpole}(a).}

\begin{figure}[h]
	\centering  %
	\includegraphics[width=0.3\linewidth]{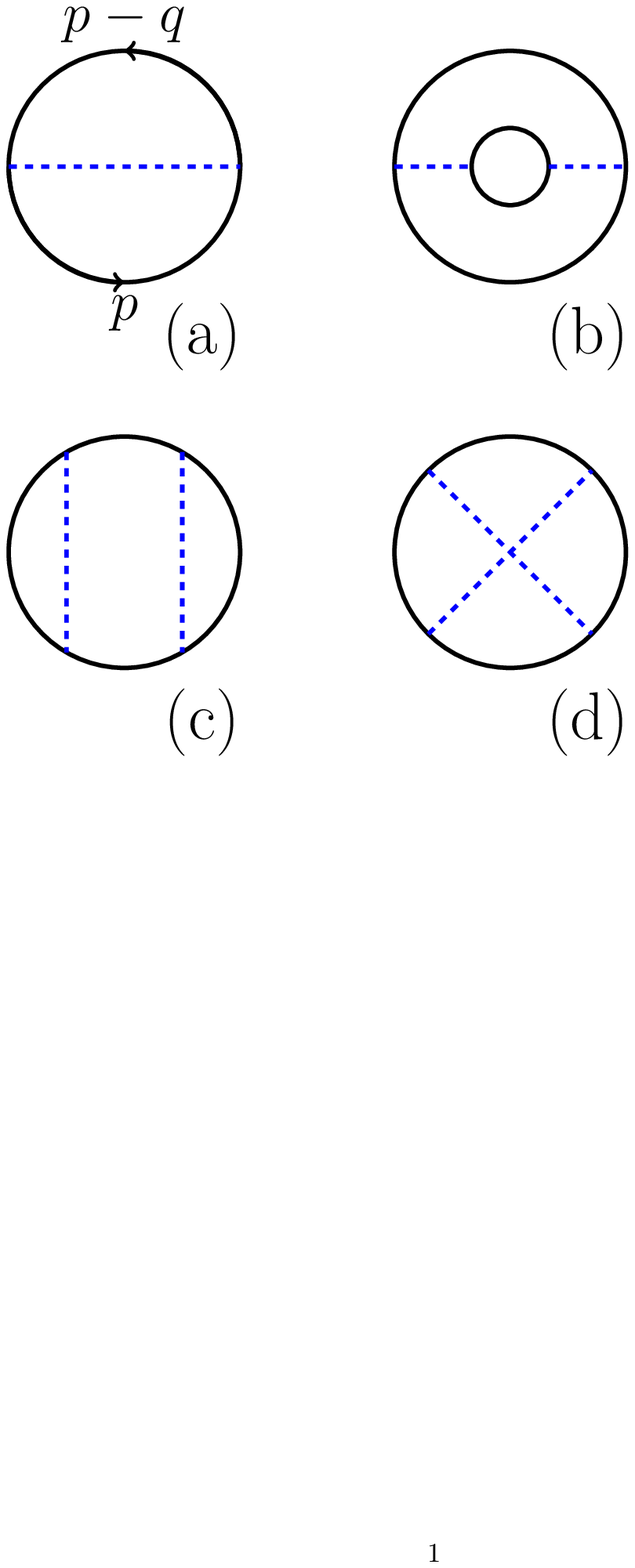}  %
	\caption{Graphs of bubble-like "progenitors" in the first and the second order.}  %
	\label{fig_bubbles}   %
\end{figure}

\begin{figure}[h]
	\centering  %
	\includegraphics[width=0.3\linewidth]{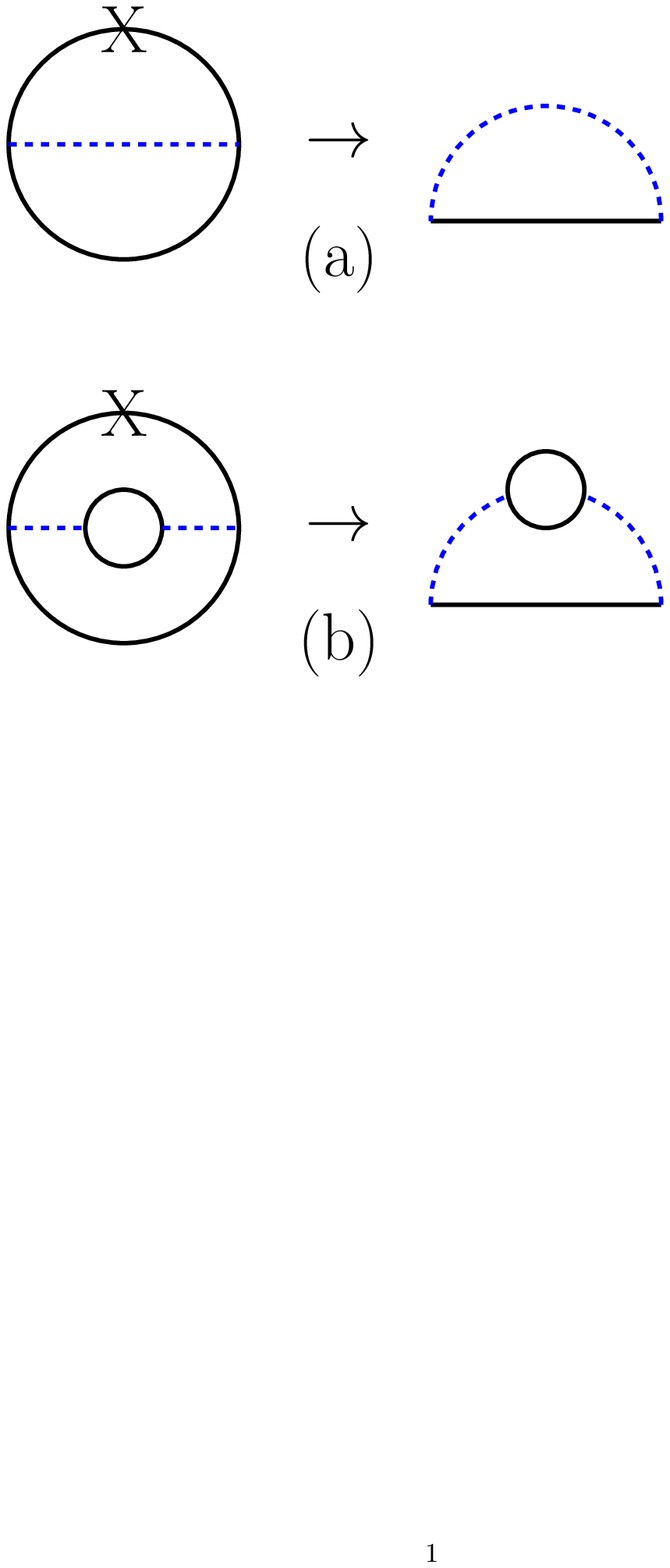}  %
	\caption{"Progenitors" and the corresponding self-energies, in the first and the second order.}  %
	\label{fig_cut-glue_1}   %
\end{figure}

\txt{Here expression without $\partial_{p_k}$
given by the bubble diagram of Fig.\ref{fig_bubbles}(a)
is
\begin {eqnarray}\label{bubble_1}
\mathcal{B}_1
=\add{+}\int_{p,q} {\rm Tr}  [G_{0,W}(x,p-q) \star G_{0,W}(x,p)] D(q), \nonumber \\
\end{eqnarray}
Following \cite{Zhang_2019_JETPL} we call this diagram the "progenitor", which may be
understood as follows.
If we insert into the integration over $p$ the derivative $\partial_{p_k}$, then the integration gives vanishing result because of the periodicity in momentum space:
\begin {eqnarray}\label{bubble_1_partial}
 \int_{p,q}
   {\rm Tr}  \partial_{p_k} [G_{0,W}(x,p-q) \star G_{0,W}(x,p)]D(q) 
 =0
\end{eqnarray}
Insertion of $\partial_{p_k}$ gives rise to the two terms, which have equal values
(see also subsection \ref{SectScalarA}).
Each of those terms is given by the
 diagram of Fig.\ref{fig_tadpole} (a).
One can see that the derivative $\partial_{p_k}$ being added to the integrand of $\mathcal{B}_1$ "generates"  expression for $\mathcal{K}_1$. As a result the latter appears to be vanishing.
On the language of the diagrams we if we cut the fermion line marked by cross "X" in Fig.\ref{fig_cut-glue_1} (a), we
come to the self-energy diagram of  Fig.\ref{fig_self-energy} (a). Thus  bubble-like "progenitor" is transformed to
to one of the self-energy diagrams.}

\begin{figure}[h]
	\centering  %
	\includegraphics[width=0.3\linewidth]{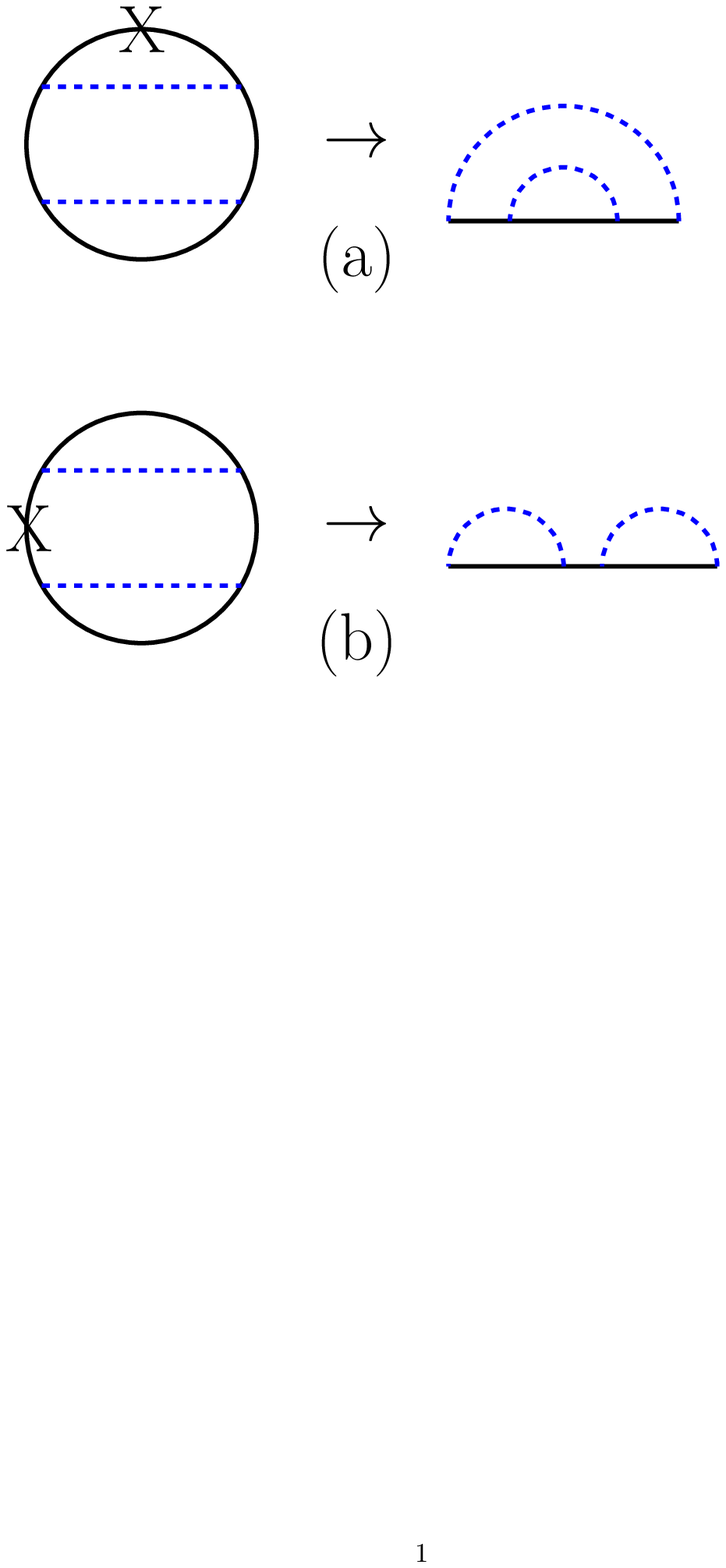}  %
	\caption{"Progenitors" and the corresponding self-energies, in the second order (non-entangled case).}  %
	\label{fig_cut-glue_2}   %
\end{figure}

\begin{figure}[h]
	\centering  %
	\includegraphics[width=0.3\linewidth]{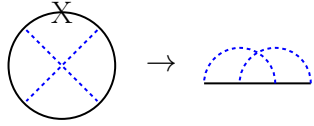}  %
	\caption{"Progenitor" and the corresponding self-energy, in the second order (the entangled case).}  %
	\label{fig_cut-glue_3}   %
\end{figure}

\txt{Eq.(\ref{bubble_1_partial}) gives   $\mathcal{K}_1=0$. As a result, using
 Eq.(\ref{deltaJ_and_J}) we obtain $\mathcal{J}_1=0$, and the interaction contribution vanishes to the order  $O(\eta^2)$. This is an alternative proof of the result obtained already in the previous subsection. Now we will extend this result to the consideration of
 $\mathcal{K}_2$ (which is the order $O(\eta^4)$):
\begin {eqnarray}\label{K2(0)}
\mathcal{K}_2 = \new{+}\int_p {\rm Tr} \Xi_{2,W} (x,p) \star \partial_{p_k} G_{0,W}(x,p)
\end{eqnarray}
Here we denote $\Xi_{2,W} = \Sigma_{2,W} + \Sigma_{1,W} \star G_{0,W} \star \Sigma_{1,W}$. It is represented
by the diagrams of Fig.\ref{fig_self-energy}(b-e),
while $\mathcal{K}_2$ is given by
Fig.\ref{fig_tadpole}(b-e).
Fig.\ref{fig_tadpole}(b) represents the  simplest case. It may be shown that similar to Fig.\ref{fig_tadpole}(a),
 Fig.\ref{fig_tadpole}(b) gives zero. Here we should only replace
$D(q)$ in Eq. (\ref{K1(0)}) by "dressed" propagator $D(q)\Pi(q^2)D(q)$, where $\Pi(q^2)$ represents vacuum
polarization.
The same may be obtained through the progenitor of Fig.\ref{fig_bubbles}(b).
Using the rule of Fig.\ref{fig_cut-glue_1}(b) we obtain that adding the cross to the progenitor gives rise to the needed self - energy diagrams.
As a result the contribution to the current of Fig.\ref{fig_self-energy}(b), i.e. Fig.\ref{fig_tadpole}(b), is zero.}

\txt{In the so - called Rainbow approximation the self energy is given by  Fig.\ref{fig_self-energy} (c) and (d). The corresponding contribution to $\cal K$ is
\begin {eqnarray}\label{K2(0)_rainbow}
 \mathcal{K}_{2,r.b.}= 
 \new{+}\int_{p,q,k}
 {\rm Tr}  && \Big[ G_{W}(x,p-q)D(q) \star G_{W}(x,p) 
   \star G_{W}(x,p-k)D(k) \star \partial_{p_k}G_{W}(x,p)+ \nonumber \\
&&  G_{W}(x,p-q)D(q) \star G_{W}(x,p-q-k)D(k)
  \star G_{W}(x,p-q) \star \partial_{p_k}G_{W}(x,p)  \Big], 
\end{eqnarray}
It is represented in  Fig.\ref{fig_tadpole}(c) and (d), and may be obtained from the progenitor of Fig.\ref{fig_bubbles}(c) (see
Fig.\ref{fig_cut-glue_2}). As a result the  contribution of the self-energy diagrams of Fig.\ref{fig_self-energy}(c) and (d)
is zero. In the other words, the sum of the diagrams Fig.\ref{fig_tadpole}(c) and (d)
vanishes. }

\txt{The cross diagram of Fig.\ref{fig_self-energy}(e),
(i.e. Fig.\ref{fig_tadpole}(e) ) may be considered in a similar way. One should add
 crosses to the diagram of  Fig.\ref{fig_bubbles} (d), and this will give 4 diagrams represented in
Fig.\ref{fig_cut-glue_3}.
This calculation proves that the contribution of Fig.\ref{fig_tadpole}(e) to $\cal K$ is zero.}

\txt{
Thus we come to conclusion that  $\mathcal{K}_1=\mathcal{K}_2=0$.
Using Eq.(\ref{deltaJ_and_J}) we obtain that the radiative corrections to the current density vanish in the orders $O(\eta^2)$ and $O(\eta^4)$. }

\txt{The consideration of the higher orders is similar.}


\section{Feynman rules for  Wigner-transformed Green functions}
\label{sectbubble}
\label{SectIV}

\txt{In the previous section we encountered the particular cases of diagram technique that deals with the Wigner-transformed propagators. Here we describe systematically construction of such a technique. For definiteness we consider the particular relativistic model. However,  the extension of this formalism to the systems of general type is straightforward.}

\subsection{Model under consideration}

\txt{In the $D$ - dimensional homogeneous systems the Green function in momentum space depends on $D$ components of momentum. In the non - homogeneous systems the Green functions $\tilde{G}(p_1,p_2)$ depend on $2D$ components of incoming and outgoing momenta. Therefore, the usual Feynman diagrams contain extra integrations over momenta. Alternatively, one may  express all physical quantities through the Wigner-transformed Green functions.
As a result we reduce the number of integrations. However, the new diagrams contain Moyal products, which complicate calculations. In certain such a diagram technique is more useful than the conventional one. One of the cases is the integer quantum Hall effect.}

\txt{Let us start from the Dirac fermion interacting with the scalar field. It is supposed that the inhomogeneous background is provided by external gauge field. \add{In Euclidean space - time (after Wick rotation) the lagrangian has the form:}
\begin {eqnarray}\label{Yukawa}
\mathcal{L}=\bar{\psi} (( i \partial_{\mu}-A_{\mu})\gamma^{\mu} - m) \psi 
         \add{-} (i \partial_{\mu} - B_{\mu})\phi (i \partial^{\mu} - B^{\mu})\phi - 
           m^2_{\phi}\phi^2   -g \bar{\psi} \psi \phi
\end{eqnarray}
Here $A_{\mu}$ and $B_{\mu}$ are two different external vector potentials. }

\txt{The two - point Green function $G$ satisfies equation $\hat{Q}(x_1) G(x_1,x_2)=\delta(x_1-x_2)$. Here
 $\hat{Q}(x)=( i \partial_{\mu}-A_{\mu}(x))\gamma^{\mu} - m $.
We define Wigner - Weyl transform of $G$ as
\begin {eqnarray}\label{Wigner}
G_W (R,p)=\int dr G(R+r/2, R-r/2) e^{-ipr}. \nonumber \\
\end{eqnarray}
The Groenewold equation reads $$Q_W(R,p)\star G_W (R,p)=1$$ (see \cite{Zubkov2016,Suleymanov2019}).
Here $Q_W$ is Weyl symbol of
operator $\hat{Q}$, while
$\star=e^{i(\overleftarrow\partial_R\overrightarrow\partial_p - \overleftarrow\partial_p\overrightarrow\partial_R) /2}$ is the Moyal product.}

\txt{Inverse propagator of the field $\phi$ is
$\hat{U}(x)=( i\partial_{\mu}-B_{\mu}(x)) ( i\partial^{\mu}-B^{\mu}(x))- m^2_{\phi} $.
Similarly, the Wigner - Weyl transform of  $D_W$ gives $U_W(R,p) \star D_W (R,p)=1$.
The standard Wigner - Weyl calculus  \cite{Suleymanov2019,Bayen_1978,Littlejohn_1986} results in
\begin{eqnarray}\label{product}
C(x_1,x_2)=\int A(x_1,y)B(y,x_2)dy \Rightarrow \nonumber \\
C_W(R,p)= A_W(R,p)\star B_W(R,p).
\end{eqnarray}}

\subsection{Feynman rules for the self energy and the fermion bubbles}

\label{SectRules0}
\txt{We start construction of Feynman rules from those self energy diagrams  \cite{Peskin} that do not contain internal fermion loops.
Wigner-transformed propagators are denoted by $D^{(j)}$ and $G_{a}$, where indices $j$ and $a$ mark boson and fermion lines correspondingly.
Let us list several auxiliary results.}
\begin{enumerate}
\item{}
\txt{Associativity of Moyal product
\begin{eqnarray}
	\Big(A(R,p) \star B(R,p)\Big) \star C(R,p) = 
	A(R,p) \star \Big( B(R,p) \star C(R,p)\Big)   \nonumber
\end{eqnarray}
As a result in the product of several functions in phase space we may omit the brackets.}

\item{}
\begin{eqnarray}
C(x_1,x_2)&=&\int A(x_1,y)H(y)B(y,x_2)dy \Rightarrow   \nonumber \\
C_W(R,p)&=& A(R,p)\star H(R)\star B(R,p)\label{f1}
\end{eqnarray}
\txt{In order to prove this formula one should use  Eq. (\ref{product}).}

\item{}
\begin {eqnarray}\label{lemma1}
e^{ikR}\star G_{W}(R,p)= e^{ikR} G_{W}(R,p-k/2),    \nonumber\\
G_{W}(R,p)\star e^{ikR} = e^{ikR} G_{W}(R,p+k/2) \nonumber
\end{eqnarray}
\begin {eqnarray}\label{lemma2}
&&(A(R,p)e^{ikR})\star B(R,p)=   
 [A(R,p)\star B(R,p-k/2)] e^{ikR},  \nonumber  \\
&&A(R,p)\star (e^{ikR} B(R,p))= 
 [A(R,p+k/2)\star B(R,p)] e^{ikR}. \nonumber
\end{eqnarray}

\item{}
\txt{The above results allow to prove that
\begin {eqnarray}\label{Theorem}
&&G_1(R,p)\star e^{ik_1R}\star G_2(R,p)\star ...\star 
  e^{ik_n R}\star G_{n+1}(R,p)\nonumber \\
&&\overset{def}{=} G_1(R,p)\prod^{n}_{i=1} {}^{\star} (e^{ik_i R}\star G_{i+1}(R,p))  \nonumber \\
&&= [\prod^{n}_{i=1} {}^{\star} G_{i}(R,p+p_i/2)] \quad e^{i\sum_j^n k_j R},  \nonumber \\
\end{eqnarray}
Here $p_m=-\sum_{j=1}^{m-1} k_j +\sum_{j=m}^{n} k_j$.}
\end{enumerate}

\begin{figure}[h]
	\centering  %
     \includegraphics[width=3cm]{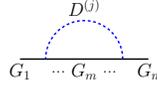} \vspace{0.5cm}
	\caption{The schematic representation of the diagrams of the fermionic self-energy without internal fermion loops. The solid line represents the fermion while
the dashed line represents the scalar. $G_i$ and $D^{(j)}$ are the fermionic and the bosonic Green functions
respectively. Dots stand for the additional ejections and absorbtions of the scalar by the fermion (those that are not shown explicitly).}  %
	\label{fig.1}   %
\end{figure}

\txt{On the Feynmann diagrams fermion propagators are typically
represented by the solid lines, while the boson propagators are represented by the dashed lines. Fourier transformation makes bosonic propagator $\tilde D(k_{a},k_{b})$ the function of the two momenta. Consider an example of $\tilde D^{(j)}(k_{ja},k_{jb})$ from Fig. \ref{fig.1}. As a result of Fourier transform Green function $D^{(j)}$ gives rise to factor $e^{ik_{ja}R}$ standing left to it, while $e^{ik_{jb}R}$ is to the right.}

\txt{The fermionic Green function $G_m$ may take different places with respect to a dash (corresponding to the Green function of boson):}
\begin{enumerate}

\item{} The dashed line is completely right to $G_m$.

\item{} The dashed line is completely left to $G_m$.

\item{} The dashed line begins left to $G_m$, and
ends right to $G_m$.

\end{enumerate}

\txt{Suppose, that the given diagram is composed of the product of $n$ fermion propagators. The given dashed line connects the right end of the $s$ - th fermion propagator with left end of the $t$ -th fermion propagator. There are also the other dashed line in the diagram, but we temporarily disregard their influence on the Wigner transformed fermion Green function, and consider only the influence of the given dash as if there would be no other dashed lines at all. One can see, that the product of the fermion Green functions receives the form
\begin {eqnarray}\label{GreenStar1}
&&G_1(R,p+q_j/2)\star ... \star G_s(R,p+q_j/2)\star         \nonumber\\
&&G_{s+1}(R,p-k_j)\star ... \star G_{t-1}(R,p-k_j)\star    \nonumber\\
&&G_{t}(R,p-q_j/2)\star ... \star G_{n}(R,p-q_j/2)  \nonumber
\end{eqnarray}
Here $q_j=k_{ja}-k_{jb}$, while $k_j=(k_{ja}+k_{jb})/2$. Symbols of the Wigner transformation are omitted.
The diagram with the integrations over momenta receives the following form.
(Here the exponential factors
that are due to the other dashed lines are not written explicitly.)
\begin {eqnarray}\label{GreenStar2}
&& \int [G_1(R,p) ... \star G_s(R,p)\circ_j\star  
 G_{s+1}(R,p-k_j)\star   ...  \nonumber\\
&& \star G_{t-1}(R,p-k_j)\star_j\circ
G_{t}(R,p)\star ...  G_{n}(R,p)] D^{(j)}(R,k_j)dk_1 ... dk_j ...  
\end{eqnarray}
We denote $\circ_j=e^{-i\overleftarrow\partial_p\partial^{(j)}_R /2}$ and
${}_j\circ =e^{i \partial^{(j)}_R \overrightarrow\partial_p /2}$.
$\partial^{(j)}_R$ acts on $D^{(j)}$ only. Operator  $\overrightarrow\partial_p$ acts on all fermion Green functions standing right to the symbol ${}_j\circ$. Operator   $\overleftarrow\partial_p$ acts on all fermio Green functions standing left to the symbol $\circ_j$.}

\txt{The diagram rules will be better understood if we will consider the particular examples of Fig.\ref{fig.2}.
Fig.\ref{fig.2}(a) represents the one - loop interaction contribution to the
fermion Green function.
The direct expression is
\begin {eqnarray}\label{Green_1st}
 \int \Big[ G_1(R,p) \circ_D \star G_2(R,p-k)\circ_D \star G_1(R,p)\Big] 
 D_W(R,k) dk \nonumber
\end{eqnarray}
Fig.\ref{fig.2}(b) is a two - loop contribution, which is out of rainbow approximation. Here solid line consists of 5 fermion propagators.
The second and the third fermion propagators are "parallel" with $D^{(1)}$,
and their momenta include $k_1$. Operators $\circ_1$
and ${}_j \circ$ stand before and after these fermion propagators. In the similar way, $D^{(2)}$ affects the third and the fourth fermion propagators.}
\txt{Feynmann diagram of Fig.\ref{fig.2}(b) may be written as
\begin {eqnarray}\label{GreenStar_eg}
\int\int && [G_1(R,p) \circ_1 \star G_2(R,p-k_1)\circ_2 \star 
 G_3(R,p-k_1-k_2)\star {}_{1}\circ    \nonumber\\
&&G_4(R,p-k_2)\star {}_{1}\circ G_5(R,p)]          
  D^{(1)}_W(R,k_1) D^{(2)}_W(R,k_2) dk_1 dk_2.  \nonumber
\end{eqnarray}
\begin{figure}[h]
	\centering  %
	\includegraphics[width=7cm]{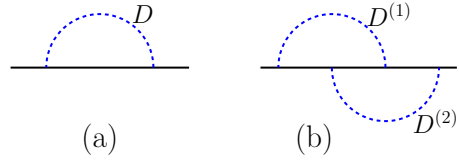}\vspace{1cm}  %
	\caption{(a)One-loop Feynmann diagram for the self energy.
     (b)An entangled two-loop Feynmann diagram for the self energy.}  %
	\label{fig.2}   %
\end{figure}}

\begin{figure}[h]
	\centering  %
	\includegraphics[width=5cm]{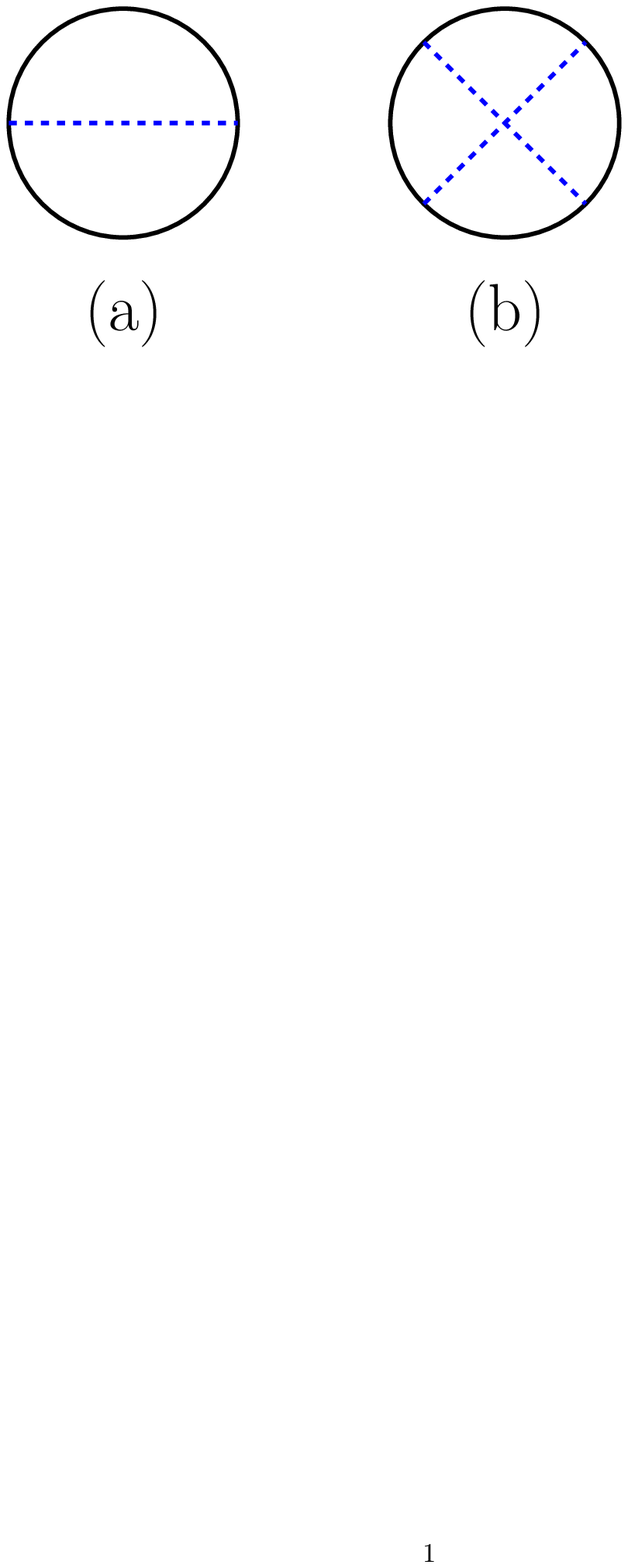}  
	\caption{Fermionic bubles.}  %
	\label{fig.3}   %
\end{figure}

\txt{Fermionic bubbles of Figs. \ref{fig.3} typically do not appear directly in the scattering amplitudes. But such bubbles are needed for the calculation of various  thermodynamical potentials. Besides, they represent the progenitors needed to prove the non - renormalization of Hall conductivity by interactions \cite{Zhang_2019_JETPL}.}

\txt{As an example we consider the Feynman diagram of Fig. \ref{fig.3}. The first (a) bubble reads
\begin {eqnarray}\label{bubble_a}
 \frac{1}{2}\int Tr \Big[ G_W(R,p-k) \star {}_{1}\circ G_W(R,p)\Big]
  \quad D^{(1)}_W(R,k) dk. \nonumber
\end{eqnarray}
Due to the trace we may also represent it as
\begin {eqnarray}\label{bubble_a'}
\frac{1}{2}\int Tr [G_W(R,p) \circ_{1}\star G_W(R,p-k)] D^{(1)}_W(R,k) dk,  \nonumber
\end{eqnarray}
In the similar way the diagram (b) results in
\begin {eqnarray}\label{bubble_b}
&&\frac{1}{4}\int Tr \Big[ G_W(R,p-k_1) \circ_2 \star   
   G_W(R,p-k_1-k_2)     \star {}_1\circ G_W(R,p-k_2)\star {}_2\circ   \nonumber\\
&& G_W(R,p)\Big] D^{(1)}_{W}(R,k_1) D^{(2)}_{W}(R,k_2) dk_1 dk_2.
\end{eqnarray}
It is worth mentioning that glueing the end points of each diagram of Fig.\ref{fig.2}, one obtains the
diagram of Fig.\ref{fig.3}.}

%

\txt{Let us now formulate the direct rules for the calculation of Feynman diagrams considered above, i.e. for the corrections to the fermion propagator in quenched approximation (without internal closed fermion lines), and for the fermion bubbles with only one closed fermion line. Later we will extend these rules to the more general case. }
\begin{enumerate}

\item{}

\txt{Taking the given graph we label momenta $p$, $p-k_j$ ... in accordance with  "momentum conservation", just as in the conventional Feynmann diagram. One should add to each diagram a combinatorial symmetry factor identical to that of the conventional diagram technique.}

\item{}

\txt{Instead of the simple product of fermion propagators of conventional technique we write down the sequence $G_W(R,p)\star G_W(R,p-k_j)\star ...$ with the Moyal products along the
fermion line. The  result of the  inhomogeneity is the appearance of dependence on $R$ and the Moyal products.}

\item{}

\txt{The more specific feature of our diagram technique is the insertion of operators $\circ_j$ and ${}_j\circ$ at the beginning and the and end
points of each dashed line corresponding to boson propagator $D^{(j)}(R,k_j)$. Finally, when the fermion line is closed (bubble) we add the trace, while the very first $\circ_{D}\star$
 is removed.}

\end{enumerate}

\subsection{The cases when the internal fermion loop is present}
\label{SectIL}

\begin{figure}[h]
	\centering  %
	\includegraphics[width=5cm]{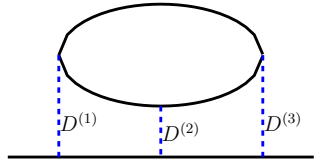} 
	\caption{An example of the  Feynmann diagram in self-energy, which contains two fermion lines. One of the fermion lines forms  an internal loop. }  %
	\label{fig.4}   %
\end{figure}

\txt{Let us consider the complication of the diagrams considered in the previous section, in which the internal closed fermion loop is present. We still consider the case when the number of external fermion legs is two or zero. \add{Notice that with our sign conventions the diagram of any physical amplitude containing $n$ fermion propagators should be supplemented by factor $(-1)^n$ coming from the minus sign in expression $G(x,y) = - \langle \psi_x \bar{\psi}_y\rangle$.} An example of such a diagram is given in Fig. \ref{fig.4}.
It may be written as follows
\begin {eqnarray}\label{Green_2line(1)}
{\cal F}(x_1|x_2) &=&  
 \int G(x_1,y_1)G(y_1,y_2)G(y_2,y_3)G(y_3,x_2)
 Tr[G(y_4,y_5)G(y_5,y_6)G(y_6,y_4)]    \nonumber\\
&& \quad  D(y_1,y_4)D(y_2,y_5)D(y_3,y_6) dy_1 ... dy_6
\end{eqnarray}
Let us use relation $D(x,y)=\int e^{ik_a x}\tilde{D}(k_a,k_b)e^{-ik_b y} dk$,
and perform Wigner transformation:
\begin {eqnarray}\label{Green_2line(2)}
{\cal F}_W(R_1|p_1) &=&
\int G_W(R_1,p_1+\frac{k_{1a}}{2}+\frac{k_{2a}}{2}+\frac{k_{3a}}{2})  \star  
 G_W(R_1,p_1-\frac{k_{1a}}{2}+\frac{k_{2a}}{2}+\frac{k_{3a}}{2}) \star  \nonumber\\
&&G_W(R_1,p_1-\frac{k_{1a}}{2}-\frac{k_{2a}}{2}+\frac{k_{3a}}{2}) \star  
  G_W(R_1,p_1-\frac{k_{1a}}{2}-\frac{k_{2a}}{2}-\frac{k_{3a}}{2})         \nonumber\\
&&Tr[G_W(R_2,p_2-\frac{k_{1b}}{2}-\frac{k_{2b}}{2}-\frac{k_{3b}}{2})  \star 
 G_W(R_2,p_2+\frac{k_{1b}}{2}+\frac{k_{2b}}{2}-\frac{k_{3b}}{2})  \nonumber\\
&&  \star G_W(R_2,p_2+\frac{k_{1b}}{2}+\frac{k_{2b}}{2}+\frac{k_{3b}}{2})]    
  e^{ik_{1a} R_1}\tilde{D}(k_{1a},k_{1b})e^{-ik_{1b} R_2} \nonumber\\
&&e^{ik_{2a} R_1}\tilde{D}(k_{2a},k_{2b})e^{-ik_{2b} R_2}
  e^{ik_{3a} R_1}\tilde{D}(k_{3a},k_{3b})e^{-ik_{3b} R_2}\nonumber\\
&& dk_{1a} dk_{2a} dk_{3a} dk_{1b} dk_{2b} dk_{3b} dR_2 dp_2
\end{eqnarray}
Here we encounter the two groups of variables: $(R_1,p_1)$ and $(R_2,p_2)$. The first corresponds to the fermion line that passes through the diagram, while the second corresponds to internal fermion loop. In order to simplify this expression we introduce Moyal product $\circ$ between the fermion propagator and bosonic propagator:
\begin {eqnarray}\label{Green_2line(3)}
 {\cal F}_W(R_1|p_1) &=&
 G_W(R_1,p_1) \star \overset{(1,1)}{\circ} G_W(R_1,p_1) \star  
 \overset{(2,1)}{\circ} G_W(R_1,p_1) \star \overset{(3,1)}{\circ} G_W(R_1,p_1)  \nonumber\\
&& \int Tr [ \overset{(1,2)}{\circ} G_W(R_2,p_2) \star \overset{(2,2)}{\circ} G_W(R_2,p_2) \star  
  \overset{(3,2)}{\circ} G_W(R_2,p_2)]  \nonumber\\
&& \quad D^{(1)}(R_1,R_2)D^{(2)}(R_1,R_2)
   D^{(3)}(R_1,R_2) dR_2 dp_2
\end{eqnarray}
Here $\overset{(i,j)}{\circ}=exp(\frac{i}{2}({\partial}^{(i)}_{R_j} \overrightarrow{\partial}_{p_j}-\overleftarrow{\partial}_{p_j}{\partial}^{(i)}_{R_j}))$,
and ${\partial}^{(i)}_{R_j} $ acts on $D^{(i)}(R_1,R_2)$ only. The operator   $\overrightarrow\partial_{p_j}$ acts on all fermion propagators standing right to the symbol $\overset{(i,j)}{\circ}$. The operator $\overleftarrow\partial_{p_j}$ acts on all propagators standing left to the symbol $\overset{(i,j)}{\circ}$.  }

\txt{The generalization of this typical case to the diagram of general type (i.e. with any number of internal fermion loops)  is straigntforward.}

\subsection{Diagrams with more than two legs}

\begin{figure}[h]
	\centering  %
	\includegraphics[width=7cm]{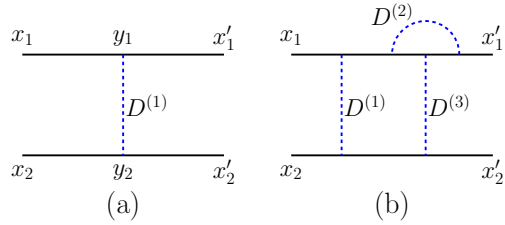} 
	\caption{(a) The simplest diagram with  four external fermion lines.
            (b) A more complicated example of the diagram with four external fermion lines.}  %
	\label{fig.5}   %
\end{figure}

\txt{The last generalization needed to formulate the diagram rules for any possible diagrams consists of adding the extra fermion lines crossing the whole diagram. As a result we will come to the consideration of an arbitrary diagram with any even number of external fermion legs.
As a typical example of such a generalization let us consider diagram of Fig.\ref{fig.5}(a).
In coordinate space it reads
\begin {eqnarray}\label{Green_4legs_a}
{\cal F}(x_1,x'_1|x_2,x'_2)= 
 \int G(x_1,y_1)G(y_1,x'_1)D(y_1,y_2) 
     G(x_2,y_2)G(y_2,x'_2) dy_1 dy_2.
\end{eqnarray}
We define Wigner transformation of this  diagram corresponding to the pairs $(x_1,x'_1)\rightarrow (R_1,p_1)$,
$(x_2,x'_2)\rightarrow (R_2,p_2)$. We obtain the final result for the Wigner transformation
${\cal F}_W(R_1,R_2|p_1,p_2)$:
\begin {eqnarray}\label{Green_4legs_b}
{\cal F}_W(R_1,R_2|p_1,p_2)=
 [G_W(R_1,p_1)\overset{(1,1)}{\circ} \star G_W(R_1,p_1)] 
 [G_W(R_2,p_2)\overset{(1,2)}{\circ} \star G_W(R_2,p_2)] D^{(1)}(R_1,R_2)
\end{eqnarray}}

\txt{A slightly more complicated case is represented in Fig.\ref{fig.5}(b). The
Wigner-transformed diagram is
\begin {eqnarray}\label{Green_4legs_c}
&& {\cal F}_W(R_1,R_2|p_1,p_2)=  \nonumber\\
&& \int dk  [G_W(R_1,p_1)\overset{(1,1)}{\circ} 
 \star G_W(R_1,p_1)\circ_2 \star G_W(R_1,p_1-k)   
 \overset{(3,1)}{\circ}\star G_W(R_1,p_1-k) \star{}_2\circ G_W(R_1,p_1)]  \nonumber\\
&&\quad [G_W(R_2,p_2)\overset{(1,2)}{\circ} \star G_W(R_2,p_2)\overset{(3,2)}{\circ} \star G_W(R_2,p_2)]
 D^{(1)}(R_1,R_2)D_W^{(2)}(R_1,k)D^{(3)}(R_1,R_2) \nonumber
\end{eqnarray}}
\txt{Obviously, the obtained expressions may easily be extended to the more general case.}

\txt{Now we are in a position to summarize the general rules of the diagram technique.
\begin{enumerate}
\item{}
Each fermion line $L_i$ (either open or closed) acquires spatial coordinate $R_i$. Each propagator of this line carries momentum $p_a$ that obeys the "conseration low". One should write down the sequence $G\star G...$ according to the rules presented at the end of
section \ref{SectRules0}. $R$ of those rules is to be replaced by $R_i$. \add{The diagram of any physical amplitude containing $n$ fermion propagators should be supplemented by factor $(-1)^n$. Besides, each internal fermionic loop produces an additional factor $-1$.}
\item{}
The bosonic dashed lines, which begin and end at the same fermion line result in the same operators $\circ_j$ and ${}_j\circ$ as in section \ref{SectRules0}.
\item{}
Dashed line connecting different fermion lines $L_i$ and $L_j$ corresponds to boson Green function $D(R_i,R_j)$ that is written in coordinate space. We do not propose here to write the Wigner ransformed propagator $D_W$ instead. The operators $\overset{(i,j)}{\circ}$ are added to the sequence $G\star G...$ at the positions of the ejection/absorbtion of the dashed line connecting $L_i$ and $L_j$.
\item{} Finally, it is necessary to add combinatorial symmetry factors that are equal to those of the conventional diagram technique.
\end{enumerate}}


\section{Non-uniform systems with interactions}

\subsection{Hall conductivity for the systems in the presence of (varying) magnetic field}

\txt{It has been shown above that for the general case of the two - dimensional non - homogeneous system the Hall conductivity has the form
$\sigma_{xy} = {\cal N} / (2 \pi)$,
where ${\cal N}$ is the topological invariant in phase space,
which is the generalization of the  classical TKNN invariant \cite{TKNN}:
\begin{eqnarray}\label{calM2d23c}
 {\cal N}=   \new{-}\frac{\epsilon_{ijk}}{24\pi^2}
\int  \frac{d^2 x } {S} \, \int d^3p \, Tr
  {G}_W(p,x)\star   
  \frac{\partial {Q}_W(p,x)}{\partial p_i}
  \star \frac{\partial  {G}_W( p,x)}{\partial p_j} \star \frac{\partial  {Q}_W(p,x)}{\partial p_k} \nonumber\\
\end{eqnarray}
Here the inhomogeneity may be caused by varying external magnetic field, by electric potential of impurities or by other reasons. The given expression has been derived for the system with the  interactions neglected. It may be shown  \cite{Zubkov+Wu_2019} that Eq. (\ref{calM2d23c}) is reduced to the TKNN invariant, when constant magnetic field is the only source of inhomogeneity.}

\txt{It is natural to suppose that in the presence of interaction one simply has to substitute to Eq. (\ref{calM2d23c}) the fermion propagator with radiative corrections. Below we prove this conjecture. }

\subsection{$2+1$ D tight - binding model in the presence of Coulomb interactions. Setup of the {\it Gedankenexperiment}.}

\label{SectCoulomb}


\txt{For definiteness let us consider the $2+1$ D tight-binding model with
Coulomb interactions. We place our system into a large torus, which gives rise to periodic boundary conditions. Equivalently, one may think that the system belongs to the surface of cylinder closed through the spatial infinity. We introduce the coordinate axes as follows. The x- axis is  is along the cylinder axis, while the y-axis is along the circle, and $y\in (-L,L]$. $L$ is assumed very large. }

\txt{ We  divide the cylinder into the two pieces:
(I) For $y\in [0,L]$ the effective coupling constant $\alpha$ is not zero, and gives rise to Coulomb interactions;
(II) For $y\in (-L,0)$ the value of effective coupling constant $\alpha^\prime$ is different. We discuss tha case $\alpha^\prime \to 0$, when Coulomb interactions are off.}

\txt{Magnetic field is normal to the surface of the mentioned cylinder, and may vary around a certain value. Besides, there may be the inhomogeneous electric potential. We require that the dependence of electrommagnetic field on coordinates in the the part (II) repeats its dependence on coordinates inside the part (I).  }

\txt{Vector potential $A_{\mu}$ is represented as the sum of the two terms: $A_{\mu}=A^{(m)}_{\mu} + A^{(e)}_{\mu}$. Here $A^{(m)}_{\mu}$ is responsible for the magnetic field and for the electric field of impurities, as well as for the inhomogeneities of another type. $A^{(e)}_{\mu}$ corresponds to external electric field. The latter term is assumed  to be small, and the external electric field is uniform within each region, but is directed oppositely. Inside (I) for $y\in [0,L]$ the electric field is positive and is along the y-axis. Inside (II) for $y\in [-L,0]$ it is along the same axis, but is negative. The Euclidean action of the model is
\begin{eqnarray}
S &=& \int d\tau \sum_{{\bf x,x'}}\Big[\bar{\psi}_{\bf x'}\Big(i(i \partial_{\tau} - A_3(i\tau,{\bf x}))\delta_{\bf x,x'} 
 - i{\cal B}_{\bf x,x'}\Big)\psi_{\bf x} \new{-}  \nonumber\\
&& \alpha \bar{\psi}(\tau,{\bf x})\psi(\tau, {\bf x})
 \theta(y)V({\bf x-x'})\theta(y')\bar{\psi}(\tau,{\bf x'})\psi(\tau, {\bf x'})\Big] 
\end{eqnarray}
Here ${\cal B}_{\bf x,x'}$ is specific for the electrons in the given material.  $A_{x,y}=\int^y_x A^{\mu} ds_{\mu} $.
$V$ is Coulomb potential $V({\bf x})=1/|{\bf x}|=1/\sqrt{x_1^2+x_2^2}$,
for ${\bf x}\not= {\bf 0}$.
Deep inside (I) one may drop to momentum space:
$$S= \int  dp \bar{\psi}_{p}\hat{Q}(p,i\partial_p)\psi_{p}
\new{-}\alpha \int  dp dq dk \bar{\psi}_{p+q}\psi_{p}\tilde{V}({\bf q})\bar{\psi}_{k}\psi_{q+k},
$$
where $\hat{Q}(p,i\partial_p)=i\Big(\sum_{k=1,2,3}\sigma^k g_k(p-A(i\partial_p))-im(p-A(i\partial_p))\Big)\sigma^3$
 \cite{Z2016_1}, and
%
$\tilde{V}({\bf q}) =\sum_{\bf x} e^{i{\bf q\cdot x}}/\sqrt{x_1^2+x_2^2}$.
%
Coulomb interaction results in the non - trivial self-energy of the fermions.}

\txt{Dressed fermion propagator may be calculated using Feynman diagrams:
\begin {eqnarray}\label{Green_a}
G_{\alpha}(x,y) &=&
  G_{0}(x,y) + 
  \int G_{0}(x,z_1) \Sigma(z_1,z_2)G_{0}(z_2,y)dz_1 dz_2 +   \nonumber\\
& & \int G_{0}(x,z_1) \Sigma(z_1,z_2)G_{0}(z_2,z_3)\Sigma(z_3,z_4)  
 G_{0}(z_4,y)dz_1 dz_2 dz_3 dz_4 +...
\end{eqnarray}
with
$\Sigma(z_1,z_2)=\add{-}\alpha G_{0}(z_1,z_2) \theta(y_1)V({\bf z}_1-{\bf z}_2)\theta(y_2)+O(\alpha^2)$,
with $z_i=({\bf z}_i,\tau_i)$, and $\bf z_i=(x_i,y_i)$.
Applying Wigner transformation we come to
\begin {eqnarray}\label{Green_b_Wigner}
G_{\alpha,W}(R,p) = G_{0,W}(R,p)  
 +  G_{0,W}(R,p)\star \Sigma_W(R,p)\star G_{0,W}(R,p) + ...,  
\end{eqnarray}
Here $G_{0,W}(R,p)$ is solution of Groenewold equation $Q_{0,W}(R,p)\star G_{0, W}(R,p) = 1$. $\Sigma_W$ is the Wigner transformed self - energy $\Sigma$.}

\subsection{Expression for the electric current through the interacting Green function}

\txt{We expand $G_{\alpha,W}(R,p)$ in powers of $\alpha$: $G_{\alpha,W}= {\cal G}_0 + \alpha {\cal G}_1 +\alpha^2 {\cal G}_2 +... $
In the same way
$G_{\alpha,W}=  G^{(0)}_{\alpha,W}+  G^{(1)}_{\alpha,W}+ ...$
and $G^{(l)}_{\alpha,W}=\sum_k \alpha^k {\cal G}^{(l)}_k $.
We expand the total electric current in powers of $\alpha$. It is expressed as
\begin {eqnarray} \label{current_3D_a}
I^k(\alpha) &=& \new{-} \int \frac{d^2 R}{S} \int_p  Tr G_{\alpha,W}(R,p) \star 
   \frac{\partial}{\partial p_k} Q_{0,W}(R,p)\nonumber\\
           &=&  \new{-}\int \frac{d^2 R}{S} \int_p
         Tr  \sum_{n=0}^{\infty} G_{0,W} 
   (\star \Sigma_W\star G_{0,W})^n
         \star \frac{\partial}{\partial p_k} Q_{0,W}(R,p) 
\end{eqnarray}
We rewrite $\Sigma_W = \alpha \Sigma_{1,W} +\alpha^2 \Sigma_{2,W} +... $. Then
 the current is equal to
 $I^{\mu}= I^{\mu}_0 + \alpha I^{\mu}_1 +\alpha^2 I^{\mu}_2 +... $,
where
$I^{k}_0= \int \frac{d^2 R}{S} \int_p Tr G_{0,W}\star \frac{\partial}{\partial p_k} Q_{0,W}$,
and
\begin {eqnarray}\label{current_i}
I^k_r = \new{-}\int \frac{d^2 R }{S} \int_p
          Tr \sum_{k_1+...+k_n=r}\Big[\prod_{i = 1...n} \Sigma_{k_i,W} 
       \star G_{0,W}\Big] \star \frac{\partial}{\partial p_k} Q_{0,W}, 
\end{eqnarray}
with $r\geq 1$.}

\txt{If we substitute in expression for electric current the "velocity" $\partial_{p_k} Q_0$ by the corresponding renormalized expression, we will obtain
\begin{eqnarray}\label{tildeI}
\tilde{I}^k(\alpha) = \new{-}\int \frac{d^2 R}{S} \int_p Tr G_{\alpha,W}(R,p) \star  
                      \frac{\partial}{\partial p_k} \Big( Q_{0,W}(R,p)-\Sigma  \Big).
\end{eqnarray}
Here $G_{\alpha,W}(R,p)$ satisfies equation
\begin{equation}
 G_{\alpha,W}(R,p)\star \Big( Q_{0,W}(R,p)-\Sigma \Big) = 1 \label{groenewold}
\end{equation}
Difference between this "modified" current and the original one is denoted
$$\Delta I^k(\alpha) = {I}^k(\alpha)-\tilde{I}^k(\alpha)$$
It is given by
\begin {eqnarray}\label{current_3D_a}
&& \new{\Delta I^k(\alpha)=-}\int \frac{d^2 R}{S} \int_p Tr G_{\alpha,W}(R,p) \star \frac{\partial}{\partial p_k} \Sigma_{W}(R,p)\nonumber\\
        &&= \new{-}\int \frac{d^2 R}{S} \int_p
     Tr \Big( G_{0,W} + \sum_{n=1}^{\infty} G_{0,W} 
       (\star\Sigma_W\star G_{0,W})^n  \Big)
     \star \frac{\partial}{\partial p_k} \Sigma_{\alpha,W}(R,p)\nonumber\\
         &&= \new{-}\alpha \int \frac{d^2 R}{S} \int_p  Tr G_{0,W}\star
                \frac{\partial}{\partial p_k} \Sigma_{1,W}(R,p) \nonumber\\
         & &  \new{-}\alpha^2 \int \frac{d^2 R}{S} \int_p \Big(Tr \Sigma_{1,W} \star G_{0,W} \star  
            \frac{\partial}{\partial p_k}\Sigma_{1,W}(R,p) \star G_{0,W}+ 
            Tr  G_{0,W}\star \frac{\partial}{\partial p_k}\Sigma_{2,W}(R,p)\Big)+... \nonumber
\end{eqnarray}
$\Delta I_k$ is expressed through the diagrams in Fig. \ref{fig_tadpole_general} (b).
The diagram with $n$ insertion of self energy $\Sigma_W$ gives
\begin{eqnarray}
\Delta I^{(n)}_k
 &=&-(n+1)\int \frac{d^2 R }{S} \int_p Tr G_{0,W}\star \new{\Sigma_{W}}\star 
  \partial_{p_k} Q_{0,W}\star \new{\Sigma_{W}}\star
     ...\star \Sigma_{W} \nonumber\\
 &&\new{+}n\int \frac{d^2 R}{S}\int_p Tr G_{0,W}\star \partial_{p_k}\Sigma_W\star ...
    \star \Sigma_{W}\star G_{0,W}\star \Sigma_{W}\nonumber
\end{eqnarray}
We obtain an interesting relation
\begin{eqnarray}
& & (n+1)\Delta I^{(n)}_k
= \new{-}(n+1)\int \frac{d^2 R}{S} \int_p Tr G_{0,W}\star \new{\Sigma_{W}}\star 
 \partial_{p_k} Q_{0,W}
   \star \new{\Sigma_{W}}\star G_{0,W}\star...\star \Sigma_{W}\star G_{0,W}\star \Sigma_{W},  \nonumber
\end{eqnarray}
It gives
$
\Delta I^{(n)}_k = I^{(n+1)}_k
$.
Here $I^{(n+1)}$ is the radiative correction to electric current caused by  $n+1$ insertions of $\Sigma_W$. It is drawn
 in Fig. \ref{fig_tadpole_general} (a) (the $n+2$ -th term). As a result, we obtain:
$$
\new{ I_k(\alpha) - \tilde{I}_k(\alpha) = I_k(\alpha)-I_k(0)}
$$}

\new{We find that the total current is robust to the introduction of interactions as long as the electric current calculated using the renormalized velocity, i.e. $\tilde{I}_k(\alpha)$, is equal to the one calculated with bare velocity.   }

\subsection{Non - renormalization of Hall conductance by interactions}

\subsubsection{First order}

\txt{In the proposed {\it Gedankenexperiment} without interactions the electric current may be calculated as
$$
I^k_0 =  \new{-}\int d^2 R/S \int_p  Tr G_{0,W}(R,p) \star \frac{\partial}{\partial p_k} Q_{0,W}(R,p)
$$
Here we give the proof that this expression is not renormalized by interactions, i.e. for  $j \geq 1$, $I^k_j=0$.
Let us start from $I^k_1$:
\begin {eqnarray}\label{current_1st}
I^k_1 &=&\add{+}\int \frac{d^2 R}{S} \int_{p,q} Tr  \Big( G_{0,W}(R,p-q)                    
         D_W(R,q)\Big) \star \frac{\partial}{\partial p_k} G_{0,W}(R,p) \nonumber\\
         &=&\add{+}\int \frac{d^2 R}{S} \int_{p,q} Tr  \Big( G_{0,W}(R,p-q)                   
            D_W(R,q)\Big) \frac{\partial}{\partial p_k} G_{0,W}(R,p)
\end{eqnarray}
where $D_W$ is Wigner transform of
\begin {eqnarray}\label{Sigma_1}
D(z_1,z_2)&=&\add{-}\alpha \theta(y_1)V({\bf z}_1-{\bf z}_2)\theta(y_2),\label{DV}
\end{eqnarray}
\add{The minus sign here means that we deal with repulsive interaction.}
For any $R$ this expression is proportional to
\begin {eqnarray}\label{lemma_1}
\int\int {\cal F}_R(p-q){\cal D}_R(q){\cal F}_R'(p)dpdq=0,
\end{eqnarray}
with ${\cal D}_R(q) = D_W(R,q)$, which is an even function of momentum $q$. At the same time ${\cal F}_R(q) = G_{0,W}(R,q)$,  ${\cal F}'$ is the derivative of $\cal F$.
Such an expression leads to conclusion that $I^k_1 =0$. Namely, we integrate by parts and obtain $I_1^k = -I^k_1$.}

\subsubsection{Second order}

\txt{For the second order contribution $I^k_2$ we obtain
\begin{eqnarray}\label{current_i2}
 I^k_2 = \new{+}\int \frac{d^2 R }{S}\int_p Tr \Sigma_{2,W} \star\partial_{p_k} G_{0,W}\new{+} 
 \int \frac{d^2 R}{S} \int_p Tr \Sigma_{1,W} \star G_{0,W} \star \Sigma_{1,W}\star
     \partial_{p_k} G_{0,W}\nonumber
\end{eqnarray}
If one would restrict to the rainbow approximation of  $\Sigma_2$, he would obtain (see Fig. \ref{fig_cut-glue_2})
\begin{eqnarray}\label{current_i2}
&& I^{k}_2 \approx   \nonumber\\
&&\add{+} \int  \frac{d^2 R }{S}  \int_{p,k,q} \,Tr \Big[G_{0,W}(R,p-k)
 \star G_{0,W}(R,p-k-q)D_W(R,q) \star 
 G_{0,W}(R,p-k)\Big] D_W(R,k)\star \partial_{p_k}G_{0,W}(R,p)  \nonumber\\
&&\add{+} \int  \frac{d^2 R }{S}  \int_{p,k,q} \,Tr G_{0,W}(R,p-q)D_W(R,q)
 \star G_{0,W}(R,p)\star G_{0,W}(R,p-k)D_W(R,k)\star   %
 \partial_{p_k}G_{0,W}(R,p) \nonumber\\
\end{eqnarray}
Here the star before $\partial_{p_k}$ may be removed. It will be inserted after that before the last boson propagator $D_W$:
\begin{eqnarray}\label{current_i2}
 I^{k}_2 &\approx & \add{+}\int  \frac{d^2 R }{S}\int_{p,k,q} \,Tr \Big[G_{0,W}(R,p-k) \star 
   G_{0,W}(R,p-k-q)D_W(R,q)\star  
   G_{0,W}(R,p-k)\Big]\star \nonumber\\
&& \quad\quad   D_W(R,k)\partial_{p_k}G_{0,W}(R,p)  \nonumber\\
&&  \add{+}\int \frac{d^2 R}{S}\int_{\rvv{p,k,q}} \,Tr G_{0,W}(R,p-q)D_W(R,q)\star   
   G_{0,W}(R,p)\star G_{0,W}(R,p-k)D_W(R,k)
  \star \partial_{p_k}G_{0,W}(R,p)\nonumber\\
&=& \add{+}\frac{1}{2}\int  \frac{d^2 R }{S}\int_{p,k,q}\,\partial_{p_k} \,Tr \Big[G_{0,W}(R,p-k)\star  
   G_{0,W}(R,p-k-q)D_W(R,q)\star 
   G_{0,W}(R,p-k)\Big]\star \nonumber\\
&& \quad\quad  D_W(R,k)G_{0,W}(R,p), \nonumber
\end{eqnarray}
The result is zero.
Here we again encounter the "progenitors" discussed in the previous sections. The  corresponding Feynmann diagram is represented in Fig. \ref{fig_bubbles} (c). This is the progenitor for the diagrams of Fig. \ref{fig_cut-glue_2}. }

\txt{The other two loop diagrams of  Fig. \ref{fig_cut-glue_3}  result in
\begin{eqnarray}\label{current_i2}
I^{k(cross)}_2 &=& \add{+}\int \frac{ d^2 R}{S} \int_{p,k,q}\,Tr \Big[G_{0,W}(R,p-k)\circ_{2}\star  
 G_{0,W}(R,p-k-q)\star \ _{1}\circ G_{0,W}(R,p-q)\star 
 \partial_{p_k}G_{0,W}(R,p) \Big]\nonumber\\
 & & D_{W(1)}(R,k)D_{W(2)}(R,q)   \nonumber\\
&=&  \add{+}\frac{1}{4}\int \frac{ d^2 R}{S} \int_{p,k,q} \partial_{p_k} \,Tr \Big[G_{0,W}(R,p-k)\circ_{2}\star  
   G_{0,W}(R,p-k-q)\star \ _{1}\circ  G_{0,W}(R,p-q)\star  
  G_{0,W}(R,p) \Big] \nonumber\\
& & D_{W(1)}(R,k) D_{W(2)}(R,q)   \nonumber
\end{eqnarray}
The result is vanishing as well.
Here
$\star=e^{i\overleftarrow{\partial}_R\overrightarrow{\partial}_p/2-i\overleftarrow{\partial}_p\overrightarrow{\partial}_R/2}$
acts only on $G$ and does not act on $D$.  As in the previous section we use notation
$\circ_{i} = e^{-i\overleftarrow{\partial}_p\overrightarrow{\partial}_R/2}$
with the derivatives over $p$ and $R$. Here the derivatives with the right arrow act on $D_{W(i)}$, while the derivatives with the left arrow act on the  fermion propagator that is placed left to this symbol.
In  $_{i}\circ=e^{i\overleftarrow{\partial}_R\overrightarrow{\partial}_p/2}$
the derivatives with the right arrow
act on expression following this symbol while the derivatives with the left arrow act on $D_{W(i)}$.
It is worth mentioning that $D_{W(i)}$ does not contain $p$. Therefore, operator $\circ$ acts only on $D_{W(i)}$, and does not affect $G$.
The remaining row of the above formula  corresponds to Fig. \ref{fig_bubbles} (d).}

\add{Thus we prove that $I_2^k=0$. We checked this statement manually via direct consideration of all possible diagrams also for the three - loop order. The consideration of the higher orders may be performed in the similar way. }

\subsubsection{Higher orders}

{ Let us consider the contribution $I_j^k$ of the order $j$ to electric current $I^k$ (averaged over the whole volume of the system). We may express this contribution as follows:
\begin{equation}
I_j^k = -\frac{1}{(2j)! S\beta}\int d^3 w d^3 z_1 ... d^3 z_{2j}  \langle {\rm Tr} \,   (\bar{\psi}_{z_1} \psi_{z_1} \phi_{z_1}) ... (\bar{\psi}_{z_{2j}} \psi_{z_{2j}} \phi_{z_{2j}}) (\bar{\psi}_w \partial_{\hat{p}_k} \hat{Q}_{0}(\hat{p},w) {\psi}_w)\rangle 	
\end{equation}
By $\partial_{\hat{p}_k} \hat{Q}_{0}(\hat{p},w)$ we understand here $-\partial_{{A}_k} \hat{Q}_{0}(\hat{p}-A,w)|_{A_k=0}$, where $\hat{Q}_{0}(\hat{p},w)$ is Dirac operator in the presence of external magnetic field, electric potential and, possibly, the other source of inhomogeneity. Here vacuum average $\langle ... \rangle$ is with respect to the system with the interactions turned off. Field $\phi$ gives rise to bosonic propagator
$$
D(x_1,x_2) =  \langle \phi(x_1) \phi(x_2) \rangle
$$	
(Recall that for the case of Coulomb interactions $D$ is given by Eq. (\ref{DV}),   and it is negative, which reflects the repulsive nature of Coulomb interactions.)
 The above written expression with the aid of Wick theorem may be represented as the sum over the Feynmann diagrams $I_{j(\alpha)}^k$. Factor $(2j)!$ in the denominator is cancelled by the combinatorial factor coming from  reordering of the set $\{z_1, ... , z_{2j}\}$:
 \begin{equation}
 I_j^k = \sum_{\alpha \in {\cal I}_j } [\alpha]^k\label{ExpI}
 \end{equation}
 Here we denote by ${\cal I}_j$ the set of the diagrams of the order $j$. By $[\alpha]$ we denote numerical value of the given diagram $\alpha$.
 For each configuration of the fermion propagators (i.e. for each way to glue them) the number of diagrams with different contractions of $\phi$ (i.e. with the dashed lines drawn in different ways) is $(2j-1)!!$. However, we exclude from the sum the disconnected diagrams since they are cancelled in the final expression for each physical quantity.
 Here each diagram containing $N(\alpha)$ internal fermionic loops, and one "external" one (with insertion of $\partial Q$) has the form
 \begin{eqnarray}
 	[\alpha]^k &=& - (-1)^{N(\alpha)} \frac{1}{S \beta} \int  \frac{d^3 R d^3 p}{(2\pi)^3} {\rm Tr} \Bigl(G(R,p-q_1) \circ_{...} \star ... G(R,p) \partial_{p^k} Q(R,p) G(R,p)\Bigr)\nonumber\\&&  \Pi_{i=1,...,N(\alpha)} \frac{d^3 R_i d^3 p_i}{(2\pi)^3} {\rm Tr} \Bigl(G(R_i,p_i-q_{i1}) \circ_{...} \star ... G(R_i,p_i)\Bigr)\nonumber\\&& D(R,q_{m})... D(R_i,q_{im})... D(R,R_j)...
 \end{eqnarray}
Here $D$ are the bosonic propagators, their arguments as well as the circle products are to be inserted into the above expressions according to the rules listed above in Sect. \ref{SectRules0}, Sect. \ref{SectIL} and in \cite{Zhang+Zubkov2019}. The general set of the rules is given at the end of Sect. \ref{SectIV}. It is worth mentioning that  the circle products do not appear between all fermion propagators. For the dashes connecting points of the same loop the circle product is to be added only at one of the ends. Depending on the position of the dash the circle product acts either on the fermion propagators standing left to its position or on the nearest fermion propagator standing to the right; the circle product always acts on the corresponding boson propagator.}

 { The corresponding progenitor diagrams are obtained in the similar expansion of
 \begin{equation}
 	\hat{B}_j = -\frac{1}{(2j)! }\int d^3 z_1 ... d^3 z_{2j}  \langle {\rm tr} \,   (\bar{\psi}_{z_1} \psi_{z_1} \phi_{z_1}) ... (\bar{\psi}_{z_{2j}} \psi_{z_{2j}} \phi_{z_{2j}}) \rangle \label{Bsource}	
 \end{equation}
Namely, we obtain  the sum over all possible Feynmann diagrams with the bubbles containing $j$ bosonic lines (the bubble set is denoted by $\cal B$)
\begin{equation}
\hat{B}_j = \sum_{\lambda \in {\cal B}_j} \gamma(\lambda)[\lambda]\label{ExpB}
\end{equation}
Here we denote by ${\cal B}_j$ the set of the bubble diagram of order $j$. $[\lambda]$ is the numerical value of diagram $\lambda$. $\gamma(\lambda)$ is the number of ways to produce the given topologically unique diagram $\lambda$ via Wick theorem applied to Eq. (\ref{Bsource}). For example, for the diagram containing the single fermion loop with $2j$ vertices this factor is equal to $\gamma(\lambda) = \frac{2j}{s(\lambda)}$, where $s(\lambda)$ is the symmetry factor of the given diagram equal to the number of rotations of the fermion loop that lead to the identical diagrams (see also \ref{SectAppD}) plus one. In the following we will also use the sum over the topologically different bubble diagrams that differs from $\hat{B}_j$:
$$
{B}_j = \sum_{\lambda \in {\cal B}_j} [\lambda]
$$
We will see that this is $B_j$ rather than $\hat{B}_j$ that is related directly to $I^k_j$. 
  Each diagram contains $N(\lambda)$ internal fermionic loops and has the value
\begin{eqnarray}
	[\lambda] &=&  -{(-1)^{N(\lambda)}}  \int \Pi_{i=1,...,N(\lambda)} \frac{d^3 R_i d^3 p_i}{(2\pi)^3} {\rm Tr} \Bigl(G(R_i,p_i-q_{i1}) \circ_{...} \star ... G(R_i,p_i)\Bigr)\nonumber\\&& D(R_1,q_{i1})... D(R_i,q_{im})... D(R_1,R_j)...
\end{eqnarray}}

\begin{figure}[h]\centering  	\includegraphics[width=0.7\linewidth]{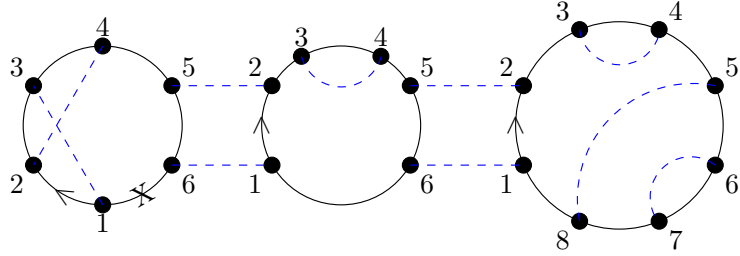}  	\caption{Diagram that illustrates the proof for arbitrary order.}  	\label{proof_Z}   \end{figure}

We would like to comment here on the difference between the expansions of Eqs. (\ref{ExpI}) and (\ref{ExpB}). Actually, the only but important difference between the two is the presence of crosses in one of the fermion lines at the diagrams of Eq. (\ref{ExpI}) and their absence in the diagrams of Eq. (\ref{ExpB}). The presence of cross also causes the absence of symmetry factor in Eq. (\ref{ExpI}). The set of the bubble diagrams of Eq. (\ref{ExpB}) is composed of the diagrams with arbitrary numbers of fermion loops. At each fermion loop there are vertices. Their total number is $2j$. The configurations of fermion propagators may be described by the sets $\{i_1, ..., i_N\}$ with $i_1 \le i_2 \le ... \le i_N$; $i_1 + i_2 + ... + i_N = 2j$. Here $N$ is the number of loops, while $i_k$ is the number of vertices from the $k$-th fermion loop. The configurations are different if the sets $\{i_1, ..., i_N\}$ are different. We illustrate this by Fig. \ref{proof_Z} (where the cross is to be disregarded). It corresponds to the configuration $\{6,6,8\}$.

Next, vertices are connected in pairs in all possible ways by the bosonic dashed lines. We represent the given choice of the dashed lines as  mapping  ${\cal P}:\{a^{(k)}|k=1,...,N; a=1,...,i_k\} \to \{b^{(l)}|l=1,...,N; b=1,...,i_l \} $ such that ${\cal P}({\cal P}(a^{(k)})) = a^{(k)} $, i.e. ${\cal P}^2 =1$. Here $a^{(k)}$  labels the $a$ - th vertex of the $k$ - th fermion loop. The adjacent vertices are represented by numbers that differ by $1$.
We denote by $\tilde{\cal B}$ the set of pairs
	$$
\lambda = \left(\begin{array}{c}
	\{i_1, ..., i_N\}\\ {\cal P}
\end{array} \right)
$$
However different elements of $\tilde{\cal B}$ may represent the identical bubble Feynmann diagrams. Namely, we define the equivalence relation $\cal J$ as follows. 	
	 The two sets ${\cal P}$, ${\cal P}^{\prime}$ of the dashed lines are thought of as equivalent  if they may be obtained one from another by cyclic rearrangement of vertices inside each fermion loop like $1,2,...,i_k \rightarrow i_k,1,...,i_k-1$ and by any relabeling of fermion loops of the same length. We may write this formally as ${\cal J}: {\cal P}(([a+M]_{{\rm mod}\, i_k})^{(k)})\sim {\cal P}(a^{(k)})$ for $M\in Z$ and ${\cal P}(a^{(j[k])})\sim {\cal P}(a^{(k)})$ if $j[k]$ is an invertible function such that $i_{j[k]} = i_{k}$. Thus the set of  Feynmann bubble diagrams is equal to
	$$
{\cal B}_j = \tilde{\cal B}_j/{\cal J}
$$
(The subset of $\cal B$ with the $j$ - th order diagrams is denoted by ${\cal B}_j$.) Below it will be useful to rewrite the sum over all these diagrams through the sum over elements of $\tilde{\cal B}_j$:
\begin{equation}
	B_j = \sum_{\bar{\lambda} \in \tilde{\cal B}_j} \frac{s(\bar{\lambda})}{n_1(\bar{\lambda})! ... n_{M(\bar{\lambda})}(\bar{\lambda})! i_1(\bar{\lambda}) i_2(\bar{\lambda})  ... i_{N(\bar{\lambda})}(\bar{\lambda})}[\bar{\lambda}]\label{ExpB2}
\end{equation}
Here the combinatorial factor takes into account the number of elements of $\tilde{\cal B} $ that are equivalent from the point of view of ${\cal B}$.
We suppose that there are $M(\lambda)$ groups of loops. Within the $k$ - th group all loops are of the same length, and the number of those loops is $n_k$. For example, if $i_1 = i_2< i_3 < i_4$ then the product of $n_k!$ reads $2! 1! 1!$.   $s(\lambda)$ is the number of the above mentioned rearrangements of vertices that do not lead at all to any change of function ${\cal P}(a^{(k)})$.

On the diagram of Fig. \ref{proof_Z} we represent a possible way to draw the dashed lines connecting vertices. The mentioned equivalence $\cal J$, for example, contains relabeling vertices of each of the three circles. Say, we may relabel them as follows. First loop: $1,2,3,4,5,6 \to 2,3,4,5,6,1$; second loop: $1,2,3,4,5,6 \to 3,4,5,6,1,2$; third loop: $1,2,3,4,5,6,7,8 \to 2,3,4,5,6,7,8,1$. We may also interchange the labels of the first and the second loops. The number of such transformations is $2! 1! 6 \times 6 \times 8 = n_1!\times n3! \times i_1 \times i_2 \times i_3$. Some of those transformations do not lead to modification of function $\cal P$, their number is $s(\lambda)$.

 To compose the set of the diagrams of Eq. (\ref{ExpI}), contributing electric current, we draw all possible closed fermionic loops with vertices on them (with the total number of vertices equal to $2j$). As it was mentioned above, possible configurations of the fermion loops are described by the sets $\{i_1, ..., i_N\}$.
Next, we draw a cross on one of the loops (say, the $k$-th loop). Thus the given set of fermionic loops with one cross is represented by the triad
$$
\left(\begin{array}{c}
	\{i_1, ..., i_N\}\\ k\\m
\end{array} \right)
$$	
Here $k$ is the number of the fermionic loop with the cross, and $m$ is the number of segment within this loop, where the cross is placed.	
In addition we should define the equivalence relation $\cal H$. According to it the configurations with crosses at different fermionic segments of the same $k$ - th loop are considered as equivalent. Moreover,
 the configurations with crosses drawn on the $k$ - th and $l$ - th loops are equivalent if $i_k=i_l$.
In case of Fig. \ref{proof_Z} we may put the cross, for example, to one of the segments of the first loop. The six configurations of fermionic propagators are equivalent (those with the cross at one of the segments). Also the six configurations with the cross on one of the segments of the second loop are equivalent to those with the cross on the first loop. Overall we have $6 + 6 = 12 $ equivalent diagrams (we still did not add the dashed lines).

	   We supplement the obtained configurations of the fermionic loops with crosses by the ways to connect vertices by dashes. The set of bosonic dashed lines is described by mapping ${\cal P}:\{a^{(m)}|m=1,...,N; a=1,...,i_m\} \to \{b^{(l)}|l=1,...,N; b=1,...,i_l \} $. This mapping satisfies ${\cal P}^2=1$. We denote by $\tilde{\cal I}$ the set of tetrads
	   $$
	   \alpha = \left(\begin{array}{c}
	   	\{i_1, ..., i_N\}\\ k\\ m\\ {\cal P}
	   \end{array} \right)
	   $$
	    Next, we define equivalence relation ${\cal J}^\prime$. According to it different configurations of dashes are equivalent if they differ by cyclic rearrangements of vertices along any fermionic loop (except for the one with the cross), and by relabeling of fermion loops with equal lengths (except for the one with the cross). 

For the example of Fig. \ref{proof_Z} we place the cross between the 6-th and the first points of the first loop, i.e. $k=1,m=6$.
Equivalence ${\cal J}^\prime$  in this example contains relabeling of the second and the third loops. The total number of equivalent elements is $6\times 8 = i_2 \times i_3$. (The symmetry factor does not appear in the presence of cross.) 

Overall, the set of diagrams contributing electric current is
    $$
{\cal I} = \tilde{\cal I}/({\cal J}^\prime\otimes {\cal H})
$$
Here by $\cal H$ we understand moving of cross to the other loops of the same length followed by the corresponding change of $\cal P$.
Now we can write
\begin{equation}
	I^i_j = \sum_{\bar{\alpha} \in \tilde{\cal I}_j}\frac{1}{n_1(\bar{\alpha})!...n_{M(\bar{\alpha})}(\bar{\alpha})!i_1(\bar{\alpha}) ... i_{N(\bar{\alpha})}(\bar{\alpha})}[\bar{\alpha}]^i\label{ExpI2}
\end{equation}
(The subset of $\cal I$ with the $j$ - th order diagrams is denoted by ${\cal I}_j$.) We denote by $u(\alpha)$ the number of fermion loops (of the given diagram $\alpha$) with the number of vertices equal to $i_{k(\alpha)}$.
For the case of Fig. \ref{proof_Z} we have $k(\alpha)=1, m(\alpha)=6$ (we put the cross to the first loop), $u(\alpha) = 2$ (the first and the second loops have the same numbers of vertices), and the total number of elements in $({\cal J}^\prime\otimes {\cal H})$ is $2 \times 6 \times 6 \times 8 = u(\alpha)! \times i_1(\alpha) \times i_2(\alpha) \times i_3(\alpha)$.

{Now let us define the following transformation of bubble $\lambda$:
\begin{eqnarray}
&&	[X[\lambda]]^k =  -\sum_{l=1...N(\lambda)}(-1)^{N(\lambda)} \int \Pi_{i=1,...,N(\lambda)}\nonumber\\&&\Big( \frac{d^3 R_i d^3 p_i}{(2\pi)^3}\Big)\partial_{p_l^k} \Pi_{i=1,...,N(\lambda)}{\rm Tr} \Bigl(G(R_i,p_i-q_{i1}) \circ_{...} \star ... G(R_i,p_i)\Bigr)\nonumber\\&& D_W(R_1,q_{11})... D_W(R_i,q_{im})... D(R_1,R_j)...\nonumber\\ &&=
	-\sum_{l=1...N(\lambda)}(-1)^{N(\lambda)} \int \Pi_{i=1,...,N(\lambda)}\Big( \frac{d^3 R_i d^3 p_i}{(2\pi)^3}\Big){\rm Tr} \Bigl(\partial_{p_l^k}G(R_l,p_l-q_{l1}) \circ_{...} \star ... G(R_i,p_i)\nonumber\\ &&+ \, ... \,
	 +G(R_l,p_l-q_{l1}) \circ_{...} \star ... \partial_{p_l^k}G(R_i,p_i)\Bigr)\nonumber\\&& \Pi_{i=1,...,l-1,l+1,...,N(\lambda)}{\rm Tr} \Bigl(G(R_i,p_i-q_{i1}) \circ_{...} \star ... G(R_i,p_i)\Bigr)\nonumber\\&& D_W(R_1,q_{11})... D_W(R_i,q_i)... D_W(R_1,R_j)...
\end{eqnarray}
  Next, we compose the transformation of the diagrams of $\tilde{\cal B}$ weighted with weight function $\frac{1}{s(\lambda)}$: 
$$
{\cal X}^q[B_j] = \sum_{\lambda\in {\cal B}}\frac{1}{ s(\lambda)} [X[\lambda]]^q = \sum_{\bar{\lambda}\in \tilde{\cal B}}\frac{1}{ s(\bar{\lambda})} \frac{s(\bar{\lambda})}{n_1(\bar{\lambda})! ... n_{M(\bar{\lambda})}(\bar{\lambda})! i_1(\bar{\lambda}) i_2(\bar{\lambda})  ... i_{N(\bar{\lambda})}(\bar{\lambda})}[X^q[\bar{\lambda}]]
$$
The sum symmetrizes contributions of $\partial_p G$. As a result we can represent the above sum through the diagrams of $\tilde{\cal I}$:
$$
{\cal X}^q[B_j] = \sum_{\bar{\alpha} \in X[\tilde{\cal B}_j]} \frac{ S\beta}{n_1(\bar{\alpha})! ... n_{M(\bar{\alpha})}(\bar{\alpha})!i_1(\bar{\alpha})...  i_{N(\bar{\alpha})}(\bar{\alpha})}[\bar{\alpha}]^q = {\beta S} I^q_j
$$
Here $X[\tilde{\cal B}_j]$ is the set of diagrams obtained from ${\cal B}_j$ adding a cross to each diagram. We used that
$ \tilde{\cal I}_j = X[\tilde{\cal B}_j]$, which means that set $\tilde{\cal I}_j$ contains diagrams that are obtained from $\tilde{\cal B}_j$ adding cross to each diagram. (At the same time bubble topologically unique diagram $\lambda \in {\cal B}_j$ with symmetry factor $s(\lambda)$ produces several topologically different contributions to $X[\lambda]$, each of these contributions appears with factor $s(\lambda)$.) The set $\tilde{\cal I}_j$ is larger than $\tilde{\cal B}_j$ because its definition contains the numbers $k$ and $m$. The latter are to be chosen in $u(\alpha) i_{k(\alpha)}$ ways.  One can see that  ${\cal X}^k[B_j]$ appears to be equal to
the $j$ - th order correction to electric current multiplied by $\beta S $:
 \begin{equation}
{\cal X}^k[B_j] =  {S\beta}  I_j^k
 \end{equation}
 The alternative derivation of this result has been given in \ref{SectAppD}.
 Since inside each term of $X[B_j]$ there is derivative with respect to momentum under the integral, we have
 $$
 {\cal X}^k[B_j] = I_j^k  =0
 $$
In the other words, the sum of all diagrams contributing to the electric current in the given order in $\alpha$ is equal to the integral over $p$ of the sum of expressions given by the corresponding progenitor diagrams. The integrals are equal to zero since the integration is over the closed Brillouin zone without boundary.}


\subsubsection{Symmetry of bosonic propagator}

{Above we used silently the important property of bosonic propagators (more precisely, of their Wigner transformations): the direction of momentum inside bosonic propagator is irrelevant. Due to this  property, for example, the diagram of Fig. \ref{fig_cut-glue_1} (a) is identical to the similar diagram with the cross placed at the lower segment of fermionic line instead of the upper one.}

\begin{figure}[h]
	\centering  %
	\includegraphics[width=0.7\linewidth]{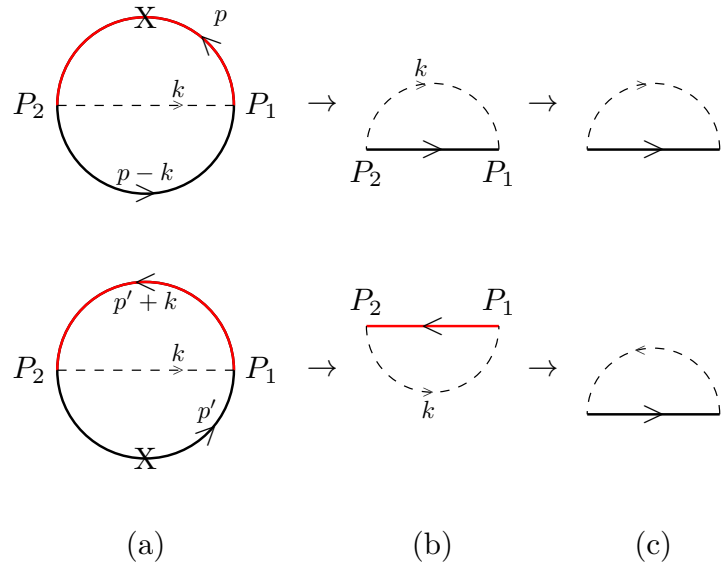}  %
	\caption{The two ways to add the cross to the progenitor diagram with one dashed line. On the right - hand side we represent the self - energy diagram $\Sigma$ contributing to electric current $\frac{1}{S \beta}\int d^3 x \int_p \Sigma(x,p) \partial_{p^k} G(x,p)$.}  %
	\label{proof_N_2}   %
\end{figure}

{Unlike the bosonic lines (dashes) the fermionic lines are oriented, and are to be supplied by the arrows. Those arrows are typically omitted (we may imply that all arrows point out the clockwise direction) if direction of bosonic dashes are irrelevant.
If direction of momentum along the dashed line would be important, then arrows are to be pointed out on both types of lines. For example, we compose two diagrams adding the cross to one of the two segments of Fig. \ref{fig.3} (a). This is illustrated by Fig. \ref{proof_N_2}.
The two diagrams are different if direction of the dash is important. Both are identical if the bosonic propagator is an even function of momentum, i.e.
$D_W(R,k)=D_W(R,-k)$. The same refers of Fig. \ref{fig.3} (b): if bosonic propagator is even function of momentum, adding the cross to one of the four fermionic segments, we produce identical diagrams contributing to electric current. At the leading order it is trivial for the majority of possible pair interactions: $D_0(k)=D_0(-k)$. However,
in general case (including the higher-order corrections) equality
$D_W(R,-k)=D_W(R,k)$ is to be proven.}

{We start the proof from the following}

{\bf Lemma}

{If function $F(x,y)$ satisfies
$F(x,y)=F(y,x)$, then its Wigner transformation
\begin{eqnarray}
	F_W(R,k)=\int_{-\infty}^{+\infty} F(R+r/2, R-r/2) e^{-ikr} dr
\end{eqnarray}
is even as a function of momentum: $F_W(R,-k)=F_W(R,k)$.}

{The proof of this lemma is straightforward:
\begin{eqnarray}
	F_W(R,-k) &=& \int_{-\infty}^{+\infty} F(R+r/2, R-r/2) e^{ikr} dr \nonumber\\
	&=& \int_{+\infty}^{-\infty} F(R-s/2, R+s/2) e^{-iks} (-1)ds \nonumber\\
	&=& \int_{-\infty}^{+\infty} F(R-s/2, R+s/2) e^{-iks} ds \nonumber\\
	&=& \int_{-\infty}^{+\infty} F(R+s/2, R-s/2) e^{-iks} ds \nonumber\\
	&=& F_W(R,k). \nonumber
\end{eqnarray}}

{In general case bosonic propagator can be expressed as
\begin{eqnarray}\label{D0_to_D}
	D(x,y)=D_0(x,y)+\int D_0(x,x')\Pi(x',y')D_0(y',y) dx'dy'
\end{eqnarray}
where $\Pi(x',y')$ is polarization operator written in coordinate space (it contains  not only the simply - connected diagrams, but also those that may be disconneced cutting one fermionic line).
At the leading order, $D_0(x,y)=D_0(y,x)$ and
$D_{0,W}(R,-k)=D_{0,W}(R,k)$.
In the following, we will prove that $\Pi(x',y')=\Pi(y',x')$.}

\begin{figure}[h]
	\centering  %
	\includegraphics[width=9cm]{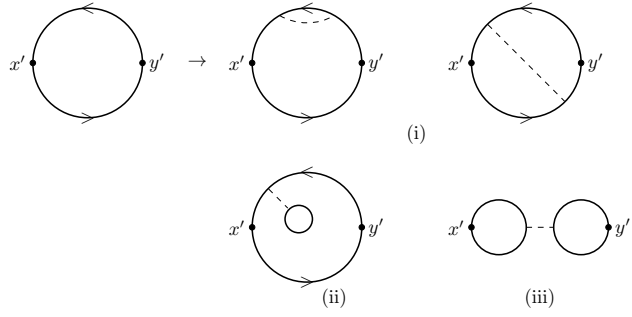}  %
	\caption{Feynman diagrams for the polarization operator $\Pi$ at the leading orders. Notice that in addition to the irreducible diagrams we include here those that may be divided cutting only one fermionic segment. }  %
	\label{proof_N_polar}   %
\end{figure}

{Suppose that at the order $n$ (when  each Feynman diagram contains n dashes)
$\Pi_n(x',y')=\Pi_n(y',x')$. Diagrammatically, this means that
if we consider $\Pi_n(x',y')$ as a set of the diagrams, then
$a \in \Pi_n(x',y')$ implies $\mathcal{I}(a)\in \Pi_n(x',y')$,
in which $\mathcal{I}$ is the operator interchanging the
labels $x'$ and $y'$ in a diagram.
One can construct the set $\Pi_{n+1}(x',y')$ from the set $\Pi_n(x',y')$
in the following ways (see Fig. \ref{proof_N_polar}):}

(i)  adding one dash to the diagrams in $\Pi_n(x',y')$;

(ii) adding one tadpole to a certain fermionic segment of the diagram;

(iii) adding to the polarization operator diagram an extra fermionic bubble.

{For arbitrary $\alpha \in \Pi_{n+1}(x',y')$, there is a diagram $a\in \Pi_{n}(x',y')$,
such that $\alpha\in a^*$.
Here $a^*$ is the set of diagrams formed from the diagram $a$
through the above mentioned three ways (i)-(iii).
Because $\Pi_n(x',y')=\Pi_n(y',x')$, $a\in \Pi_{n}(y',x')$.
Therefore,  $a^* \subseteq \Pi_{n+1}(y',x')$.
Noticing that $\alpha\in a^*$, we found $\alpha\in \Pi_{n+1}(y',x')$.
Now we have shown that for arbitrary $\alpha \in \Pi_{n+1}(x',y')$,
$\alpha\in \Pi_{n+1}(y',x')$, which implies
$\Pi_{n+1}(x',y')\subseteq \Pi_{n+1}(y',x')$.
In a similar way, $\Pi_{n+1}(y',x')\subseteq \Pi_{n+1}(x',y')$. Therefore, $\Pi_{n+1}(y',x')=\Pi_{n+1}(x',y')$.}

{Thus, we applied mathematical induction to show that $\Pi(x',y')=\Pi(y',x')$.
Furthermore, from Eq.(\ref{D0_to_D}), one finds that
$D(x,y)=D(y,x)$, which implies
$D_W(R,-k)=D_W(R,k)$.}

\subsubsection{Absence of corrections to Hall conductivity}

In the previous subsections we have proven that the electric current (averaged over the whole sample) does not receive corrections from interactions in arbitrary non - homogeneous equilibrium system. For completeness the alternative proof of this statement has been given in \ref{SectAppD}, where an alternative way of counting diagrams has been used.

\txt{Thus so far we have proven that the radiative  corrections to the total electric current vanish in the whole system that consists of the two parts (I) and (II). The electric field is constant but has opposite directions in those two regions. Besides, we obtained that the total electric current in the considered system can be expressed as
\begin {eqnarray}\label{IFIN}
{I}^k(\alpha) &=& \new{-}\int \frac{d^2 R }{S} \int_p Tr G_{\alpha,W}(R,p)\star  
     \frac{\partial}{\partial p_k} Q_{0,W}(R,p)\nonumber\\
&=&\new{-}\int\frac{d^2 R }{S} \int_p Tr G_{\alpha,W}(R,p)
  \frac{\partial}{\partial p_k} Q_{0,W}(R,p) 
\end{eqnarray}}

\txt{The total current of the whole system  (composed of regions (I) and (II)) along the x - axis can be written as
$$
I^1(\alpha) = I^{(I)1}_{persistent}(\alpha) + I^{(II)1}_{persistent} + \sigma^{(I)}_{xy}(\alpha) E - \sigma^{(II)}_{xy}(0) E
$$
Here $E$ is electric field directed along the $y$ axis in  region (I). In region (II) the electric field is given by $-E$. By $I^{(I,II)1}_{persistent}$ we denote persistent current that may appear in the system without external electric field. It is worth mentioning that according to the Bloch theorem in majority of systems such a current cannot appear. But anyway, its possible appearance in marginal cases does not affect our results on Hall conductivity.
	Recall that without interactions, in region (II) the conductivity  is given by Eq. (\ref{calM2d23c}) divided by $2\pi$. Since there are no interaction corrections to the total current of the whole system, and since this is valid for any value of $E$, we obtain:
$$
\sigma^{(I)}_{xy}(\alpha) = \sigma^{(I)}_{xy}(0)
$$
	 We come to conclusion that in the region (I) with interactions the total electric current is given by Eq. (\ref{IFIN}), where integration over $R$ is restricted to this region. It gives the Hall conductivity of this piece of material:
$
\sigma_{xy} = \frac{\cal N}{2 \pi},
$
${\cal N}$ is expressed by
Eq. (\ref{calM2d23c}) through the noninteracting Green function. But also we may substitute here the complete Green function with radiative corrections taken into account.}


\vspace{3mm}

\section{Conclusions and discussion}


In the present paper we \rvv{discuss} the influence of interactions on Integer Hall effect both in topological insulators (the intrinsic anomalous quantum Hall effect AQHE) and in the systems in the presence of external magnetic field (the conventional QHE). We consider a particular tight - binding model of the $2+1$ D topological insulator
discussed in \cite{tb2d,tb2d2,tb2d3,Z2016_1}.
The effect of Yukawa and Coulomb interactions is investigated in this case.
As expected, we obtain that in all considered cases the Hall conductivities are still given by expressions discussed in \cite{Z2016_1} in terms of the two-point Wigner-transformed Green functions of the interacting systems.
For the \rvv{2d} topological insulators the given expressions are topological invariants,
which remain unchanged when the system is modified smoothly.
Effect of interactions appears through the modification of certain parameters. For example, at the one - loop order, the corrections to Hall conductivities
comes from the renormalization of mass parameter.  The other corrections do not appear.

In the case of the \rvv{2d} topological insulators the interactions at the one-loop level
renormalize the mass  parameter of the considered tight - binding models.
If the strength of the interaction is larger than \rvv{a} certain threshold,
the system can be driven to the phase with the value of Hall conductivity
different from that of the non - interacting model.
Otherwise, if  the strength is lower than \rvv{the} certain threshold, the
Hall conductivity remains the same.
As for the effects of higher order, we took Yukawa interaction
as an example, and
corrections due to Yukawa interactions are considered to all orders in perturbation theory.
(Generalization to exchange by any bosonic excitations can be easily made.) In the latter consideration
we used an original method to prove diagrammatically that in the presence of interactions the Hall conductivity is given by the topological quantity expressed through the full fermion propagator.
 The essence of the proof is to construct the bubble-like
Feynman diagram (progenitor) for a group of diagrams, which contribute to the
Hall current. This  method is somehow similar to the progenitor approach
used by Coleman and Hill in QED$_3$ \cite{parity_anomaly}.
However, unlike \cite{parity_anomaly} we consider the more complicated model without relativistic invariance.

We also consider Hall current in the presence of magnetic field.
In the presence of non-uniform magnetic field, an electric potential of impurities, uniform external electric field
and Coulomb interactions, the Hall conductivity (averaged over the system area)
is proportional to the topological invariant in phase space of Eq. (\ref{calM2d23c}).
The present derivation of Eq. (\ref{calM2d23c}) (see also \cite{Zubkov+Wu_2019,FZ2019}
where this derivation has been given in the absence of interactions) is valid for the gauge field potential that varies slowly at the distances of the order of lattice spacing. This corresponds to the values of magnetic field much smaller than thousands Tesla
and the {typical wavelength } much larger than several Angstroms.
In the region of analyticity in $\alpha$,
 the Hall conductivity does not depend on $\alpha$ at all and is still given by the same expression as without Coulomb interactions! To the best of our knowledge this result has been obtained for the first time for the systems in the presence of varying magnetic field in \cite{Zhang_2019_JETPL}. Previously the non - renormalization by interactions of the TKNN expression for $\sigma_H$ was proved for the case of the constant magnetic field only. This proof of the absence of radiative corrections to Hall conductivity is somehow similar to that of \cite{parity_anomaly}).

In the absence of disorder there are two equivalent ways to understand Hall conductivity: physics in the bulk and physics at the edge (for the detailed explanation of this bulk - boundary correspondence see \cite{Hatsugai}).
The discussed method of the calculation of Hall conductivity, in principle, is able to unify the two approaches.
Namely, both varying magnetic field and  varying electric potential enter the expression for $Q_W$
on the same grounds as the components of {vector potential $A_\mu$}.
Varying electric potential, in principle is able to reflect both the electric field of impurities
and the confining potential at the boundary.
We expect that in the presence of disorder, although the current density is carried  mainly by the boundary,
the electric conductivity averaged over the whole area of the system is still
given by ${\cal N}/(2\pi)$ with $\cal N$ of Eq. (\ref{calM2d23c}).
However, the detailed consideration of this issue remains out of the scope of the present paper.

Thus, we conclude, that the total (averaged) Hall conductivity is proportional to that of Eq. (\ref{calM2d23c}) and is not affected neither by the smooth change of $\alpha$ nor by the weak disorder.

It is worth mentioning that  the mentioned proof of the absence of radiative corrections to the Hall conductivity
may be also  generalized to the other types of interactions  and to the $3+1$ D systems as well (see, e.g. \cite{Miransky:2015ava,V1,V2,V3,V4,V5,V6,V7}). It would be interesting to consider the generalization of this approach to the case, when elastic deformations are present (see, e.g. \cite{FZ2019}). In particular, in \cite{NV2018} it has been shown that the response of $\sigma_H$ to elastic deformations is quantized for the $3+1$D intrinsic AQHE in topological insulators. The influence of interactions on this response is worth to be considered.

The authors are grateful for useful discussions to I.Fialkovsky, M.Suleymanov, Xi Wu, M.Lewkowicz, and C.Banerjee.



\appendix

\section{Calculation of ${\cal N}$ for the $2+1$ D systems}
\label{AppendixA}

Here we repeat for completeness the calculation presented in Appendix C of \cite{Z2016_1}.
We calculate  the topological invariant ${\cal N}$  in the case, when the Green function has the form
 \begin{equation}
 {\cal G}^{-1}(p) = i\sigma^3\Big(\sum_{k}\sigma^{k} g_{k}(p) - i g_4(p)\Big)\label{G12d}
 \end{equation}
 where $\sigma^k$ are Pauli matrices while $g_k(p)$ and $g_4(p)$ are the real - valued functions, $k = 1,2,3$. Let us define
\begin{equation}\label{H_matrix}
{\cal H}(p) = \Big(\sum_{k}\sigma^{k} \hat{g}_{k}(p) - i \hat{g}_4(p)\Big)
\end{equation}
where $\hat{g}_k = g_k/g$, and $g = (\sum_{k=1}^{4} g_k^2)^{1/2}$. Then
\begin{eqnarray}\label{N3AH}
{\cal N}
&=&  -\frac{1}{24 \pi^2} {\rm Tr}\,
\int_{} \, {\cal G}^{-1} d {\cal G} \wedge d {\cal G}^{-1} \wedge d {\cal G}   \nonumber\\
&=& -\frac{1}{24 \pi^2} {\rm Tr}\,
\int_{} \, {\cal H} d {\cal H}^{\dagger} \wedge d {\cal H} \wedge d {\cal H}^{\dagger} .
\end{eqnarray}
Replacing ${\cal H}$ by Eq.(\ref{H_matrix})
and after some algebraic calculations, one finds that
\begin{eqnarray}\label{H_matrix_2}
{\cal H} d {\cal H}^{\dagger}
&=& (\hat{g}_i d \hat{g}_i + \hat{g}_4 d \hat{g}_4)
   + i(\epsilon^{ijk} \hat{g}_i d \hat{g}_j +  \nonumber \\
   && \hat{g}_c d \hat{g}_4 - \hat{g}_4 d \hat{g}_c)\sigma_c    \\
d{\cal H} \wedge d {\cal H}^{\dagger}
&=&  i(\epsilon^{ijk} d\hat{g}_i \wedge d \hat{g}_j + 2 d\hat{g}_c \wedge d \hat{g}_4) \sigma_c, \nonumber \\
\end{eqnarray}
in which $\sigma_a \sigma_b= \delta_{ab}+\epsilon_{abc}\sigma_c $ and
$d\theta \wedge d\theta=0$ have been applied. Then we obtain
\begin{eqnarray}
{\cal N} &=&  \frac{1}{12 \pi^2} \epsilon_{ijkl}\, \int_{} \, \hat{g}_i d \hat{g}_j \wedge d \hat{g}_k \wedge d \hat{g}_l
\end{eqnarray}
Let us introduce the parametrization
\begin{equation}
\hat{g}_4 = {\rm sin}\,\alpha, \quad \hat{g}_i = v_i\,{\rm cos}\,\alpha
\end{equation}
where $i=1,2,3$ while $\sum_{i}v_i^2=1$
with $v_i= g_i/(g_1^2+ g_2^2+ g_3^2)^{1/2}$,
and $\alpha \in [-\pi/2,\pi/2]$.
Let us suppose, that $\hat{g}_4(p)=0$ on the boundary of momentum space $p\in \partial {\cal M}$
with ${\cal M}=\{ (p_1,p_2, \omega) \mid p_1,p_2\in(-\pi,\pi],  \omega \in R \}$.
This gives
\begin{eqnarray}
{\cal N} &=&  \frac{1}{4 \pi^2} \epsilon_{ijk}\,
\int_{\cal M} \, {\rm cos}^2 \alpha\, v_i\,  d\,\alpha  \wedge d v_j \wedge d v_k \nonumber \\
&=&  \frac{\epsilon_{ijk}}{4 \pi^2} \,
\int_{\cal M}  v_i\,  d(\frac{\alpha}{2}+\frac{{\rm sin}\,2\alpha}{4})  \wedge d v_j \wedge d v_k \nonumber\\
&=& - \frac{\epsilon_{ijk}}{4 \pi^2} \,
\sum_l\int_{Y_l} v_i (\frac{\alpha}{2}+ \frac{{\rm sin}\,2\alpha}{4} )   d v_j \wedge d v_k ,  \nonumber
\end{eqnarray}
where $Y_l=\partial{\Omega(y_l)}$, $\Omega(y_l)$ is the small vicinity of point $y_l \in {\cal M}$,
and $y_l$'s are singular points of $v_i$'s.
The absence of the singularities of ${g}_k$ ($k=1,2,3$) implies that
$g_1^2+ g_2^2+ g_3^2=0$
and $\alpha \rightarrow \pm\pi/2$ at such points.

This gives
\begin{eqnarray}\label{N&Res}
{\cal N} &=& -  \frac{1}{2}\sum_l \, {\rm sign}(g_4(y_l)) \,{\rm Res}\,(y_l)
\end{eqnarray}
We use the notation:
\begin{eqnarray}
{\rm Res}\,(y) &=&  \frac{1}{8 \pi} \epsilon^{ijk}\, \int_{\partial \Omega(y)} \, v_i  d v_j \wedge d v_k
\end{eqnarray}
It is worth mentioning, that this symbol obeys { $\sum_l {\rm Res}\,(y_l)=0$}.

Let us illustrate the above calculation by the consideration of the  particular example of the system with the Green function ${\cal G}^{-1} = i \omega - H(p)$, where the Hamiltonian has the form
\begin{eqnarray}
H &=& {\rm sin}\,p_1\, \sigma^2 - {\rm sin}\, p_2 \, \sigma^1 - \nonumber \\
  && (m + 2 - {\rm cos}\,p_1 -{\rm cos}\,p_2) \, \sigma^3   \nonumber
\end{eqnarray}
This gives
\begin{eqnarray}
-i\sigma^3{\cal G}^{-1} &=& {\rm sin}\,p_1\, \sigma^1 + {\rm sin}\, p_2 \, \sigma^2 +  \omega \, \sigma^3 - \nonumber \\
 && i (m + 2 - {\rm cos}\,p_1 -{\rm cos}\,p_2) \nonumber
\end{eqnarray}
The boundary of momentum space corresponds to $\omega = \pm \infty$. We have
$$\hat{g}_4(p) = \frac{M}
{\sqrt{M^2+ {\rm sin}^2\,p_1+ {\rm sin}^2\,p_2 + \omega^2}} \nonumber $$
with $M=m + 2 - {\rm cos}\,p_1 -{\rm cos}\,p_2$
For example, for $m \in (-2,0)$ we have
\begin{eqnarray}
\hat{g}_4(p) & = & 0, \quad p\in \partial{\cal M}\nonumber\\
\hat{g}_4(p) & = & -1, \quad  \hat{g}_i(p)  = 0,\quad  p = (0,0,0), \quad {\rm Res}=1   \nonumber\\
\hat{g}_4(p) & = & 1,\quad  \hat{g}_i(p)  = 0,\quad  p = (0,\pi,0), \quad {\rm Res}=-1 \nonumber\\
\hat{g}_4(p) & = & 1,\quad  \hat{g}_i(p)  = 0,\quad  p = (\pi,0,0), \quad {\rm Res}=-1\nonumber\\
\hat{g}_4(p) & = & 1,\quad  \hat{g}_i(p)  = 0,\quad  p = (\pi,\pi,0), \quad {\rm Res}=1, \nonumber
\end{eqnarray}
where $i=1,2,3$ and $p$ is the triplet $(p_1,p_2,\omega)$
Therefore, from Eq.(\ref{N&Res}), we get immediately
\begin{eqnarray}
{\cal N}=  -\frac{1}{2}(-1-1-1+1) =  1
\end{eqnarray}
In the similar way ${\cal N}  = -1 $ for $m \in (-4,-2)$ and ${\cal N}  = 0 $ for $m \in (-\infty,-4)\cup (0,\infty)$.

\section{Comparison of the two expansions of Wigner-transformed Green function}
\label{AppB}

The Wigner-transformed Green function $G_W(x,p)$ satisfies the
so-called Greonewold equation:
\begin{eqnarray}\label{Greonewold}
Q_W(x,p)\star G_W(x,p)=1,
\end{eqnarray}
in which $\star=exp(i\overleftrightarrow{\Delta}/2)$,
with $\overleftrightarrow{\Delta}=
\overleftarrow{\partial_x}\overrightarrow{\partial_p}-
\overleftarrow{\partial_p}\overrightarrow{\partial_x}$.

Function $Q_W$ is related to the Hamiltonian of the
system. Here we consider the special case $Q_W(x,p)={\cal Q}(p-A(x))$,
with $A(x)$ linear in $x$.

\subsection{Expansion in partial derivatives}
\label{Sect_1}

In this subsection expansion $G_W=G^{(0)}_W + G^{(1)}_W + G^{(2)}_W +...$
corresponds to representation $\star=1+i\overleftrightarrow{\Delta}/2+...$.
From equations
\begin{eqnarray}\label{Greonewold_1}
Q_W(x,p)G^{(0)}_W(x,p)=1, \\
Q_W G^{(1)}_W+ \frac{i}{2}Q_W\overleftrightarrow{\Delta}G^{(0)}_W=0
\end{eqnarray}
we obtain $G^{(0)}_W(x,p)=Q_W^{-1}={\cal Q}(p-A(x))^{-1}$, which is a function of
$p-A(x)$. Furthermore,
$G^{(1)}_W=- \frac{i}{2}Q_W^{-1}[Q_W\overleftrightarrow{\Delta}G^{(0)}_W]$
is also a function of $p-A(x)$ given that $A(x)$ is linear in $x$.
Higher-order terms  $G^{(n)}_W$'s are also functions of $p-A(x)$, which may be shown via
mathematical induction.

We consider the case when the field strength $\partial_i A_j = const$ does not depend on coordinates. Then
one can easily show that if $U(x,p)$ and $V(x,p)$ are functions of $p-A(x)$,
and satisfy periodic boundary condition in $p$ (or approach fast to $0$ at the boundaries),
then
\begin{eqnarray}\label{star_and_dot}
{\rm tr} \int U \star V dp = {\rm tr}\int U  V dp={\rm tr} \int V \star U dp.
\end{eqnarray}
Here and below integral is over the whole momentum space.

Further, if $Q_W$ is a function of $p-A(x)$, then
\begin{eqnarray}\label{star_and_dot_2}
{\rm tr} \int G_W \star \frac{\partial Q_W}{\partial p_k} dp = {\rm tr}
\int G_W  \frac{\partial Q_W}{\partial p_k} dp,
\end{eqnarray}
and also ${\rm tr} \int {\rm} G_W \star \frac{\partial Q_W}{\partial p_k} dp = {\rm tr}
\int {\rm} \frac{\partial Q_W}{\partial p_k} \star G_W dp.$

Notice that because of the presence of $\overleftrightarrow{\Delta}$,
$G^{(1)}_W$ is proportional to field strength $F_{ij}=\partial_i A_j-\partial_j A_i$.

\subsection{Expansion in powers of a small parameter}
\label{Sect_2}

Let us consider expansion in powers of small parameter $\epsilon$, when
$Q_W(x,p)={\cal Q}(p-{A}(x)-\epsilon B(x))$.
Using Greonewold equation, Eq.\ref{Greonewold},
we expand $G_W(x,p)$ as
$G_W=G^{(0)}_W + \epsilon G^{(1)}_W + \epsilon^2 G^{(2)}_W +...$

We find iteratively:
\begin{eqnarray}\label{Greonewold_2}
Q_W^{(0)}\star G^{(0)}_W=1, \\
Q_W^{(0)}\star G^{(1)}_W+ Q_W^{(1)}\star G^{(0)}_W=0.
\end{eqnarray}
Suppose, we obtain the leading order term $G^{(0)}_W$
from Eq. (\ref{Greonewold_2}), the next-to-leading-order term
will be given by $G^{(1)}_W= - G^{(0)}_W\star Q_W^{(1)}\star G^{(0)}_W$.
{ Notice that such an expansion can't guarantee $G^{(n)}_W$'s
are functions of $p-A(x)$ at each order. Therefore, the considerations of the previous subsection cannot be applied for each order in $\epsilon$.}

Inserting the expansions into the current density $J_k(x)={\rm tr} \int_p G_W \partial_{p_k} Q_W$,
and keeping the terms up to $O(\epsilon)$,
we obtain the current density to the order $O(\epsilon)$
\begin{eqnarray}\label{current_density}
&& \int_x J^{(1)}_k(x)  = {\rm Tr}\int_x \int_p G^{(1)}_W\star \partial_{p_k} Q^{(0)}_W + G^{(0)}_W \star \partial_{p_k} Q^{(1)}_W \nonumber \\
            &&= {\rm Tr}\int_x \int_p - G^{(0)}_W\star Q_W^{(1)}\star G^{(0)}_W\star \partial_{p_k} Q^{(0)}_W + G^{(0)}_W \star \partial_{p_k} Q^{(1)}_W \nonumber \\
&&= {\rm Tr}\int_x \int_p( -  Q_W^{(1)}\star G^{(0)}_W\star \partial_{p_k} Q^{(0)}_W\star G^{(0)}_W + G^{(0)}_W \star \partial_{p_k} Q^{(1)}_W )\nonumber \\
&&= {\rm Tr}\int_x \int_p   Q_W^{(1)}\star  \partial_{p_k} G^{(0)}_W  + G^{(0)}_W \star \partial_{p_k} Q^{(1)}_W \nonumber \\
&&= {\rm Tr}\int_x \int_p  \partial_{p_k}[ Q_W^{(1)}\star   G^{(0)}_W  ] \nonumber \\
&&= 0.
\end{eqnarray}
Recall that here $\int_p = \int \frac{d^3 p}{(2\pi)^3}$. We also denote $\int_x = \int d^3 x$.
It is worth mentioning, that without integration over $x$ we cannot come to the similar result. At the same time, in the case considered in the previous subsection such an integral is divergent because the field $A$ is linear in $x$ and thus grows infinitely at large $x$. This does not allow to use Eq. (\ref{current_density}). As a result $\int_x J^{(1)}_k(x)$ does not vanish in this case unlike the case, when $A(x)$ is bounded everywhere.



\section{Alternative derivation of the current non - renormalization by interactions}

\label{SectAppD}

Here we represent an alternative proof of the
non-renormalization of average current $I$
to all orders, i.e. $\delta I = 0$,
where
\begin{eqnarray}
\delta I^i =\frac{1}{S\beta}\int_{p,x} {\rm Tr}\, \bar{\bf \Sigma}(x,p)\star\frac{\partial G_0}{\partial p_i}(x,p)\label{eq1}
\end{eqnarray}
with
$\int_{p,x} = \int \frac{d^3 x d^3 p}{(2 pi)^3}$. $G_0$ is the (Wigner transformed) Green function without interactions.
By $\bar{\bf \Sigma}$ we denote
$$
\bar{\bf \Sigma} = \Sigma + \Sigma \star G_0 \star \Sigma + \Sigma \star G_0 \star \Sigma \star G_0 \star \Sigma + ...
$$
while $\Sigma$ is the irreducible self - energy.
We represent $\delta I = \alpha\delta I_1 + \alpha^2\delta I_2+...$, where $\alpha$ is the coupling constant.
If for arbitrary positive integer $n$ there is $\mathcal{A}_n$,
such that
\begin{eqnarray}
\delta I_n = \int_{p,x} \frac{\partial \mathcal{A}_n}{\partial p_x},
\end{eqnarray}
then $\delta I =0$.

\begin{figure}[h]	\centering  \includegraphics[width=9cm]{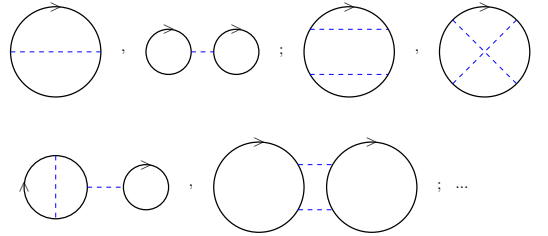}  	\caption{The elements in the bubble set $\mathcal{B}$ up to the second order.}  	\label{fig_BubbleSet}   \end{figure}

Let us consider the bubble set $\mathcal{B}$ (shown in
Fig.\ref{fig_BubbleSet} up to the second order). We define $s(b)$ for $b \in \mathcal{B}$ as the symmetry factor of the bubble diagram $b$,
and $[b]$ is the diagram itself (the result of the integration of the corresponding sequence of propagators) corresponding to bubble $b$.
We define
\begin{eqnarray}
B = \int_{x,p} Tr \bar{\bf \Sigma} \star G_0  = \alpha B_1 + \alpha^2 B_2 +...,
\end{eqnarray}
which is closely related to $\delta I$. This definition corresponds to the definition of $B$ of Sect. 6.4.3.
With the help of the bubble set $\mathcal{B}$,
$B$ can be expressed as
\begin{eqnarray}
B = \sum_{b\in \mathcal{B}} [b],
\end{eqnarray}
in which only topologically different diagrams contribute,
and each one appears once, because the self-energy functions
do not have the symmetry factor (i.e. their symmetry factor is 1).

\begin{figure}[h]
	\centering  %
	\includegraphics[width=9cm]{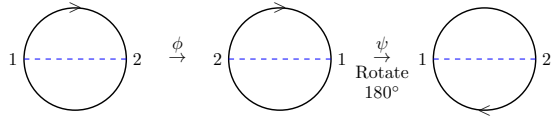}  %
	\caption{The example to illustrate the symmetry factor.}  %
	\label{fig_bubble_LO}   %
\end{figure}

To make the matter clear, we introduce the
concept "labelled bubble diagram" $\bar{b}$, obtained from
the unlabelled bubble diagram $b$.
In the language of graphs, for a bubble $b$ with $2n$ vertices,
the formation of  $\bar{b}$ from  $b$
is to label the vertices of the bubble $b$ by
 e.g. $1, 2,..., 2n$.
Suppose, $\phi$ is a permutation operation of these vertices,
i.e. $\phi\in\mathcal{S}_{2n}$ where $\mathcal{S}_{2n}$ is
the $2n$-th symmetric group. After applying the
operation to the graph $\bar{b}$, one obtains another
diagram, denoted by $\phi(\bar{b})$,  which is also a bubble,
but with the vertex labels changed.
If $\phi(\bar{b})$ is topologically equivalent to $\bar{b}$
(i.e. the two graphs can overlap each other by moving the vertices and the lines),
denoted by $\phi(\bar{b})\sim \bar{b}$, then
we call $\phi$ a {\bf symmetry operation} of the bubble $b$.
The symmetry factor $s(b)$ counts the number of these symmetry operations.
In the language of set theory, given a bubble diagram $b$ with $2n$ vertices,
if set $\mathcal{H}= \{ \phi\in \mathcal{S}_{2n}|
\phi(\bar{b})\sim \bar{b}\}$, then $s(b)=\#\mathcal{H} $ (i.e. the number of elements of set $\mathcal{H}$).
As a simple example, the bubble in
Fig.\ref{fig_bubble_LO} has symmetry factor 2.

{\bf Lemma} 1

Suppose $\bar{b}$ is a labelled {\bf connected} bubble diagram,
formed by directed lines (fermion propagators) and undirected dashes (boson propagators)
and a permutation mapping $\phi$ of the vertices of $\bar{b}$ is one of
the symmetry operations of the graph $\bar{b}$,  i.e. $\phi(\bar{b})\sim \bar{b}$. If there is a vertex $P$ of $\bar{b}$ such that $\phi(P)=P$,
then $\phi$ is identity mapping.

\begin{figure}[h]
	\centering  %
	\includegraphics[width=9cm]{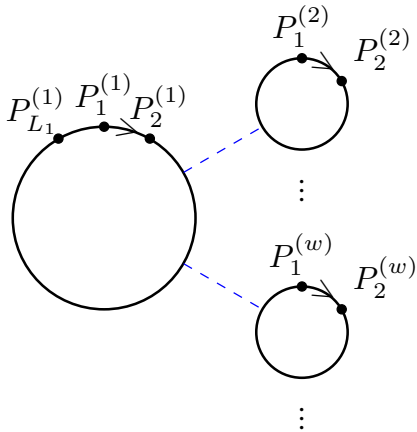}  %
	\caption{A schematic diagram to show the vertex-labels.}  %
	\label{fig_multiCirc}   %
\end{figure}

Proof:

Without loss of generality, we assume that $b$ has $2n$ vertices,
$n$ dashes and  $m$ fermion loops, and we  label its vertices in the following way.
First, fix a fermion loop of $b$ (just choose one if there are many), label its vertices
along the line following the  direction of arrows on the fermion line,
 by $P^{(1)}_1, P^{(1)}_2, ..., P^{(1)}_{L_1} $,
with $ P^{(1)}_{L_1 +1}=P^{(1)}_1 $, if there are $L_1$  vertices on this
loop; and then take another fermion loop, label its vertices
similarly by $P^{(2)}_1, P^{(2)}_2, ..., P^{(2)}_{L_2} $; ...
until all of the $2n$ vertices of $b$ are labelled.
Now we obtain the set of labeled  vertices
 $\{ P^{(1)}_1, ..., P^{(1)}_{L_1} , ... , P^{(N)}_1, ..., P^{(N)}_{L_N} \}$,
with $\sum_{w=1}^{m} L_w =2n$, which is shown schematically
in Fig.\ref{fig_multiCirc}.

The above mentioned (in the condition of the lemma) vertex $P$ is one of the elements of
this vertex set, say,  $P=P^{(w)}_i$. Because $\phi$ is a
 symmetry operation of the graph $\bar{b}$, $\phi(\bar{b}) \sim \bar{b}$,
$\phi(P^{(w)}_i)=P^{(w)}_i$ implies that
for the next vertex $P^{(w)}_{i+1}$ along the directed fermion loop,
$\phi(P^{(w)}_{i+1})=P^{(w)}_{i+1}$ still holds.
For the same reason, $\phi(P^{(w)}_{j})=P^{(w)}_{j}$ holds for any $1\leq j \leq L_w$,
which means all of the vertices on the $w$-th Fermi circle remain the same, i.e. the map $\phi$
does not change them.

Furthermore, because the bubble $\bar{b}$  is connected, the relation
$\phi(P^{(w)}_{j})=P^{(w)}_{j}$  can be extended to the other fermion
loops through the dashes. Therefore, we can finally show that none of the
vertices are changed by  $\phi$, which implies that $\phi$ is
identity.

Inserting
 derivative $\partial/\partial p_i$ into the
integrand of the expression of a bubble diagram
we generate the sum of self-energy functions contributing
to the current $I$.
It has been shown graphically in figures of Sect.4 that
the action of $\partial/\partial p^i$ corresponds
to cutting one fermion segment along the fermion line.
For  a bubble diagram $b$ formed by one fermion loop,
its value is given by
\begin{eqnarray}
[b] =\int_{p,x}\int_{q_1,...,q_n} {\rm Tr}\, G_0(x,p_1)\star \circ G_0(x,p_2)\star ... \star \circ G_0(x,p_{2n}) D(x,q_1)... D(x,q_n)
\end{eqnarray}
in which $p_1=p$, $p_2=p_1+k_1$, ..., $p_l=p_{l-1}+k_{l-1}$ ($1<l<2n$),
...,$p_{2n}=p_{2n-1}+k_{2n-1}$, and $p_{1}=p_{2n}+k_{2n}$,
implying that $k_1+k_2+...+k_{2n} =0$. We denote $\int_{q_1,...,q_n}=\int \frac{d^3 q_1 ... d^3 q_{2n}}{(2\pi)^{6n}}$.
$D(x,q_i)$'s are Wigner transformed bosonic propagators. Momentum $k_i$ is given by $-q_j$ if at the $i$-th point the bosonic line with propagator $D(q_j)$ ends, and it is given by $+q_j$ if such a bosonic line starts at this point.   Therefore,
$k_i \in \{ \pm q_1, \pm q_2, ... \pm q_n \}$. We also add the circle products that act both on bosonic and fermionic propagators according to the rules of Sect. 5. We write them symbolically in above expressions by symbols $\circ$.
Now let us consider the action of $\partial/\partial p_i$
on the bubble $b$, which generates the sum
\begin{eqnarray}
[X(\bar{b})] &=& \int_{x,p}\int_{q_1, ..., q_n} \frac{\partial}{\partial p^i} {\rm Tr}\, G_0(x,p_1)\star \circ G_0(x,p_2)\star ...G_0(x,p_{2n}) D(x,q_1)... D(x,q_n)
\\
&=& \sum_{l=1}^{2n} W_l,
\end{eqnarray}
with
\begin{eqnarray}
W^i_l &=& \int_{x,p}\int_{q_1, ..., q_n}   {\rm Tr}\, G_0(x,p_1)\star \circ ...G_0(x,p_{l-1}) \star \circ\nonumber \\&& \frac{\partial G_0(x,p_{l})}{\partial p_{l}^i}
G_0(x,p_{l+1})\star \circ ...G_0(x,p_{2n}) D(x,q_1)... D(x,q_n)
   \nonumber \\
&=& \int_{x,p}\int_{q_1, ..., q_n}   {\rm Tr}\, \frac{\partial G_0(x,p_{l})}{\partial p_l^i}
G_0(x,p_{l+1})\star \circ \nonumber \\&& ...G_0(x,p_{2n}) \star \circ G_0(x,p_1)...G_0(x,p_{l-1}) D(x,q_1)... D(x,q_n)
\end{eqnarray}
One can move function $\partial G_0(x,p_{l})/\partial p_l^i$,
because of the trace.
After changing variable $p_l \to p$, the above quantity
can be transformed into the form of the right-hand side of Eq. (\ref{eq1}),
and
\begin{eqnarray}
\int_{q_1,q_2,...,q_n}
G_0(x,p_{l+1})\star \circ ...G_0(x,p_{2n})\star \circ G_0(x,p_1)\star \circ ...G_0(x,p_{l-1}) D(x,q_1)... D(x,q_n)
\end{eqnarray}
is a contribution to $\bar{\bf \Sigma}$.
Keep in mind that $[X(\bar{b})]$ is an integral of
 full derivative. Therefore, if the above quantity
is equal to or proportional to $[X(\bar{b})]$,
we can say that the former is equal to zero.

For the diagram $b'$ with more than one fermion loops
(e.g. $m$ loops: the $w$-th loop has $L_w$ vertices, and
$\sum_w L_w= 2n$),
its value can be written as
\begin{eqnarray}
[b'] &=&-(-1)^m\int_{q_1 ... q_n} \prod_{w=1}^{m}\Big(\int_{x^{(w)},p^{(w)}} {\rm Tr}\, [ G_0(x^{(w)}, p^{(w)}_1)\star \circ ...G_0(x^{(w)},p^{(w)}_{L_w})D(x^{(w)},q_1)...D(x^{(w)},q_{L_w - r_w})\Big)\nonumber\\&&
D(x^{(1)},x^{(2)})...
\end{eqnarray}
Here propagators $D(x^{(1)},x^{(2)})$ connect points of the first and the second loops. At the places, where the corresponding dashed line starts (there are $r_w$ such points on the $w$ - th loop) there should be the appropriate circle product (see Sect. 5). For the bubbles with several fermion loops the $X$ operation contains sum over the fermionic loops with the derivatives with respect to momentum circulating within these loops.

Above we used symbol $[X(\bar{b})]$ rather than $[X(b)]$.
In the general case
not all of the  $2n$
contributions to $\bar{\Sigma}$ generated from the bubble are
different from each other, and they may be repeated
several times, i.e. there may be $W_l$'s  with different $l$
that are equivalent.
However, if we start from the labelled bubble $\bar{b}$ (with $2n$ vertices and $n$ dashes),
and then
cut(delete) each fermion segment,
we will obtain $2n$
{\bf different} Feynman diagrams.
All of them are different from each other because of the presence of
the vertex labels.
In the language of graphs, if we treat $X(b)$ and  $X(\bar{b})$
as sets of diagrams generated by the bubble $b$,
$X(b)$  only considers the topological different ones, while
 $X(\bar{b})$ takes into account their reiterations and includes
all of the $2n$ diagrams (from  $X(\bar{b})$ to $X(b)$,
one just needs to erase the vertex labels).
Therefore, we can write
\begin{equation}
[X(\bar{b})]=
\sum_{\bar{\sigma}\in X(\bar{b})}
\int {\rm Tr} [\bar{\sigma}]\star
\frac{\partial G_0(x,p)}{\partial p_i}  \frac{d^3 p d^3 x}{(2\pi)^3}\label{eee}
\end{equation}
This equation is important, because it is a bridge between
integration formulae (LHS) and the set of graphs (RHS).

There is a relation between the sets $X[b]$ and  $[X(\bar{b})]$,
which will be discussed below.

{\bf Lemma} 2

Consider bubble $b\in \mathcal{B}$, its symmetry factor
is $s(b)$ and $\bar{b}$ is the corresponding diagram with
vertex-labels, then
\begin{eqnarray}
\sum_{\bar{\sigma}\in X(\bar{b})}[\bar{\sigma}]
=s(b) \sum_{\sigma\in X(b)}[\sigma]
\end{eqnarray}

Proof:

Adopting the vertex labels introduced in the previous Lemma,
we introduce the labels for the fermionic
segments of $b$ on the basis of
the vertex labels:
$l^{(w)}_i ={\rm seg} (P^{(w)}_i, P^{(w)}_{i+1})$. We assume that it
follows the direction of the arrow along the fermionic loop,
with $1\leq i \leq L_w$, and $1\leq w \leq m$.
Assuming $b$ has $2n$ vertices, it has 2n fermionic segments. We denote this
$\# \mathcal{L} = 2n$, where $\mathcal{L}$ the set of fermionic segments.

Suppose $s(b)=r$, then there will be $r$ different permutations
acting on the vertices of $\bar{b}$ but leaving the graph $\bar{b}$ unchanged,
and  the symmetry group of $b$ can be written as
$\mathcal{H} = \{ \phi_1 =e, \phi_2, ... , \phi_r \}$,
in which each $\phi_k$ satisfies $\phi_k(\bar{b})\sim\bar{b}$.
Furthermore, the mapping $\phi_k$ of vertices induces
the mapping for the segments
$\phi_k(l^{(w)}_i )=
{\rm seg} (\phi_k(P^{(w)}_i), \phi_k(P^{(w)}_{i+1}))$.
Because $\phi_k(\bar{b}) \sim \bar{b}$ for $1\leq k \leq r$, the two segments
${\rm seg} (P^{(w)}_i, P^{(w)}_{i+1})$ and
${\rm seg} (\phi_k(P^{(w)}_i), \phi_k(P^{(w)}_{i+1}))$ of the same graph $\bar{b}$
are equivalent.
 The latter means that if one deletes the segment ${\rm seg} (P^{(w)}_i, P^{(w)}_{i+1})$
or deletes the segment
${\rm seg} (\phi_k(P^{(w)}_i), \phi_k(P^{(w)}_{i+1}))$,
the resulting contributions to $\bar{\bf \Sigma}$ are topologically equivalent.

Let us construct the set of segments
$\mathcal{L}_1= \{  \phi_k(l^{(1)}_1) | k=1,2,...,r  \} $,
which is a subset of $\mathcal{L}$, i.e. $\mathcal{L}_1\subset \mathcal{L}$.
The number of its elements should not be bigger than $r$.
From Lemma 1, one knows that $\# \mathcal{L}_1=r$,
because vertices $\phi_1(P^{(1)}_1),\phi_2(P^{(1)}_1),
..., \phi_r(P^{(1)}_1)$ are all different from each other.
The $r$ generated diagrams from this subset $\mathcal{L}_1$
(by cutting one segment in $\mathcal{L}_1$)
are topologically equivalent (regardless of the labels), and all them correspond to
the same element of $X[b]$.

Next, one can construct $\mathcal{L}_2$ based on a remaining
segment from $\mathcal{L}-\mathcal{L}_1$, and then continue this
procedure on $\mathcal{L}-\mathcal{L}_1-\mathcal{L}_2$...
Finally, we obtain the segment sets $\mathcal{L}_1, \mathcal{L}_2, ..., \mathcal{L}_q$,
such that $\# \mathcal{L}_i =r$ for each $i$,
$\mathcal{L}_i\cap \mathcal{L}_j=\emptyset$ for $i\not= j$,
$\mathcal{L}_1\cup \mathcal{L}_2...\cup \mathcal{L}_q = L$,
and $\forall \mathcal{L}_i$, all the segments in $\mathcal{L}_i$
are equivalent to each other.

Therefore, in order to construct $X(b)$,
one just needs to choose one representative
segment in each $\mathcal{L}_i$ set to cut.
There will be $q$ representative segments.
Let us denote this set of representative segments
by $\mathcal{L}^*$ ($\#\mathcal{L}^*=q$).
The set $X(b)$ will be obtained by cutting
the segments in $\mathcal{L}^*$.
Furthermore, because $\# \mathcal{L}_i =r$ for each $i$,
each element of $X[b]$ will be repeated
$r$ times in $X[\bar{b}]$. It implies that
\begin{eqnarray}
\sum_{\bar{\sigma}\in X(\bar{b})}[\bar{\sigma}]
=r \sum_{\sigma\in X(b)}[\sigma]
\end{eqnarray}
and noticing that $s(b)=r$, the lemma has been proven.

Now, let us return to current $I$. Corrections to it
 can be expressed as
\begin{eqnarray}
\delta I^i = \frac{1}{S \beta}
\sum_{b\in \mathcal{B}}
\sum_{\sigma\in X(b)}
\int_{x,p} {\rm Tr} [\sigma] \star
\frac{\partial G_0(x,p)}{\partial p_i}
\end{eqnarray}
Applying the above proven lemma to the right-hand side of the above equation,
we obtain
\begin{eqnarray}
\delta I^i
&=& \frac{1}{S \beta}
\sum_{b\in \mathcal{B}} \frac{1}{s(b)}
\sum_{\bar{\sigma}\in X(\bar{b})}
\int_{x,p} {\rm Tr} [\bar{\sigma}]\star
\frac{\partial G_0(x,p)}{\partial p_i}  \nonumber \\
&=& \frac{1}{S \beta}\sum_{b\in \mathcal{B}} \frac{1}{s(b)} [X(\bar{b})]
\end{eqnarray}
where we used definition of $ [X(\bar{b})]$ of Eq. (\ref{eee}).
We mentioned above that $ [X(\bar{b})]$ is an integral of a full derivative
and, therefore, it is equal to zero. We obtain $\delta I =0$.


\end{document}